# STUDY OF THE NEUTRON AND PROTON CAPTURE REACTIONS $^{10,11}B(n, \gamma)$, $^{11}B(p, \gamma)$, $^{14}C(p, \gamma)$, AND $^{15}N(p, \gamma)$ AT THERMAL AND ASTROPHYSICAL ENERGIES

SERGEY DUBOVICHENKO[*,†],
ALBERT DZHAZAIROV-KAKHRAMANOV[*,†]

[*] *V. G. Fessenkov Astrophysical Institute "NCSRT" NSA RK, 050020, Observatory 23, Kamenskoe plato, Almaty, Kazakhstan*
[†] *Institute of Nuclear Physics CAE MINT RK, 050032, str. Ibragimova 1, Almaty, Kazakhstan*
[*] *dubovichenko@mail.ru*
[†] *albert-j@yandex.ru*

We have studied the neutron-capture reactions $^{10,11}B(n, \gamma)$ and the role of the $^{11}B(n, \gamma)$ reaction in seeding *r*-process nucleosynthesis. The possibility of the description of the available experimental data for cross sections of the neutron capture reaction on $^{10}B$ at thermal and astrophysical energies, taking into account the resonance at 475 keV, was considered within the framework of the modified potential cluster model (MPCM) with forbidden states and accounting for the resonance behavior of the scattering phase shifts. In the framework of the same model the possibility of describing the available experimental data for the total cross sections of the neutron radiative capture on $^{11}B$ at thermal and astrophysical energies were considered with taking into account the 21 and 430 keV resonances. Description of the available experimental data on the total cross sections and astrophysical *S*-factor of the radiative proton capture on $^{11}B$ to the ground state of $^{12}C$ was treated at astrophysical energies. The possibility of description of the experimental data for the astrophysical *S*-factor of the radiative proton capture on $^{14}C$ to the ground state of $^{15}N$ at astrophysical energies, and the radiative proton capture on $^{15}N$ at the energies from 50 to 1500 keV was considered in the framework of the MPCM with the classification of the orbital states according to Young tableaux. It was shown that, on the basis of the *M*1 and the *E*1 transitions from different states of the $p^{15}N$ scattering to the ground state of $^{16}O$ in the $p^{15}N$ channel, it is quite succeed to explain general behavior of the *S*-factor in the considered energy range in the presence of two resonances.

## 1. Introduction

This review is the logical continuation of works devoted to the radiative proton and neutron capture on light nuclei that were published in the International Journal of Modern Physics E **21** (2012) 1250039, **22** (2013) 1350028, and **22** (2013) 1350075.

### 1.1. *Astrophysical aspects of the review*

As we know, light radioactive nuclei play an important role in many astrophysical environments. In addition, such parameter as cross section of the capture reactions as a function of energy is very important for investigation of many astrophysical problems such as primordial nucleosynthesis of the Universe, main trends of stellar evolution, novae and super-novae explosions, X-ray bursts etc. The continued interest in the study of processes of radiative neutron capture on light nuclei at thermal and astrophysical energies is caused by several reasons. Firstly, this process plays a significant part in the study of many fundamental properties of nuclear reactions, and secondly, the data on the capture cross sections are widely used in a various applications of nuclear physics and nuclear astrophysics, for example, in the process of studying of the primordial nucleosynthesis reactions.

The study of the reaction $^{11}$B(n, γ)$^{12}$B, from the astrophysical point of view, may play a definite role in reaction chains in the so-called inhomogeneous Big Bang models[1,2,3,4,5] that allow to the synthesis of heavy elements via the next chain of neutron capture reactions:

$$^{1}\mathrm{H}(n,\gamma)^{2}\mathrm{H}(n,\gamma)^{3}\mathrm{H}(^{2}\mathrm{H},n)^{4}\mathrm{He}(^{3}\mathrm{H},\gamma)^{7}\mathrm{Li}(n,\gamma)^{8}\mathrm{Li}(^{4}\mathrm{He},n)\underline{^{11}\mathrm{B}(n,\gamma)^{12}\mathrm{B}}(\beta^{-})^{12}\mathrm{C}(n,\gamma)$$
$$^{13}\mathrm{C}(n,\gamma)\ ^{14}\mathrm{C}(n,\gamma)^{15}\mathrm{C}(\beta^{-})^{15}\mathrm{N}(n,\gamma)^{16}\mathrm{N}(\beta^{-})^{16}\mathrm{O}(n,\gamma)^{17}\mathrm{O}(n,^{4}\mathrm{He})^{14}\mathrm{C}\ldots \quad (1)$$

While, the study of the reaction $^{10}$B(n, γ)$^{11}$B, from the astrophysical point of view, is not so interesting because this capture channel will be negligible, as opposed to $^{7}$Li(α, γ)$^{11}$B, for production of $^{11}$B for reaction chains of Eq. (1).

However, it seems to us that the study of this reaction is also interesting, even though that it has been impossible for us to find any similar theoretical calculations for the reaction $^{10}$B(n, γ)$^{11}$B in thermal and astrophysical energy range. In addition, $^{10}$B is a very good absorber of neutrons that it is used in control rods in nuclear reactors. This property also makes it useful for construction of neutron detectors. Boron is used to make windows that are transparent to infrared radiation, for high-temperature semiconductors, and for electric generators of a thermoelectric type.[6]

Let us note that usually supposed[7] that proton capture reaction on $^{11}$B at the astrophysical energies ($E_{\mathrm{p}} < 100$ keV) leads to the small cross sections of the formation of $^{12}$C, because of large Coulomb barrier. The proton capture reaction on $^{11}$B is also neglected in the primordial nucleosynthesis and formation of $^{12}$C nucleus at the neutron capture on $^{11}$B with the following beta decay of $^{12}$B to $^{12}$C is supposed. The density of $^{4}$He nuclei in the star nucleosynthesis, produced in the p-p chain, is sufficiently large and namely triple alpha process is responsible for synthesis of $^{12}$C. However, as we know now, it is impossible to neglect by the neutron capture on $^{11}$B in full.[7] Besides, the following nuclear reaction chain was found to be very important in nucleosynthesis processes of proton-rich nuclei:[2]

$$^{7}\mathrm{Be}(p,\gamma)^{8}\mathrm{B}(\alpha,\gamma)^{11}\mathrm{C}(e^{+},n)\underline{^{11}\mathrm{B}(p,\gamma)^{12}\mathrm{C}}. \quad (2)$$

The measurements of the reaction rate for $^{11}$B(p,γ) process at low energies, generally, are focused in the narrow resonance range at $E_{\mathrm{p}} = 163$ keV with small width. Study of this resonance with the help of the $^{11}$B(p,$\gamma_{0+1}$) reaction[8] allows one to obtain width of $\Gamma = 6.7$ keV and cross section of $\sigma_{\gamma} = 125$ μb in the resonance range, which differ from the assumed values[9] ($\Gamma = 5.3$ keV and $\sigma_{\gamma} = 158$ μb) more than by 20%. The values $\Gamma = 5.4$ keV and $\sigma_{\gamma} = 130$ μb were obtained relatively recently in Ref. 10, and was drawn the conclusion that this resonance plays the key role in the determination of the reaction rate for the $^{11}$B(p,γ) process at low and astrophysical energies. Therefore, this reaction can play the certain role in different thermonuclear processes in the Universe at synthesis of $^{12}$C.

Recently, for example in Ref. 11, it was supposed that baryon number fluctuations in the early Universe lead to the formation of high-density proton-rich and low-density neutron-rich regions. This might be the result of the nucleosynthesis of elements with mass $A \geq 12$ in the neutron-rich regions of the early Universe.[12,13] The special interest is the nucleus $^{14}C$,[14] which is produced by successive neutron capture[1,2,3,4,5]

$$\ldots\ ^{12}C(n,\gamma)^{13}C(n,\gamma)^{14}C\ \ldots\ . \tag{3}$$

The nucleus $^{14}C$ has a half-life about 6000 years and is stable on the time scale of the Big-Bang nucleosynthesis. Therefore the synthesis of elements with mass $\geq 14$ depends on the rate of the neutron, alpha, and proton capture reactions on $^{14}C$. Because the cross section for neutron capture on $^{14}C$ at thermal energies is very small ($\sigma \leq 1$ μb, Ref. 15) and it is located at the level of 5–15 μb in the region of 100–1000 keV, it is assumed[12] that the alpha capture reaction is dominant. However, the proton capture on $^{14}C$ might be of equal importance, because it depends on the proton abundance and density of their distribution in certain regions of the early Universe.

Note furthermore that the results of new studies of the reaction $^{14}C(p,\gamma)^{15}N$ in the nonresonance energy range[14] lead to the cross sections that higher on the order and more than the cross sections obtained earlier in Ref. 16. It allows one to obtain higher rate of the $^{14}C(p,\gamma)^{15}N$ reaction at lower temperatures, notably, lower than $0.3T_9$, that is essentially rise the role of this reaction for synthesis of heavier elements in the low energy range at the different stages of formation and development our Universe.[12]

Continuing the study of the radiative proton capture reactions on light atomic nuclei, which are the part of different thermonuclear processes,[17] let us stop on the capture reaction $p^{15}N \to {}^{16}O\gamma$ at astrophysical energies. This process is a part of basic chain of thermonuclear reactions of the CNO cycle,[18] which determine the formation of the Sun and stars at early stages of their evolution[18]

$$^{12}C(p,\gamma)^{13}N(\beta^+\nu)^{13}C(p,\gamma)^{14}N(p,\gamma)^{15}O(\beta^+\nu)^{15}N(p,\alpha)^{12}C. \tag{4}$$

This chain has alternative reaction channel, which begins from the last reaction of the proton interaction with $^{15}N$ (see Ref. 19)

$$\underline{^{15}N(p,\gamma)^{16}O}(p,\gamma)^{17}F(\beta^+\nu)^{17}O(p,{}^4He)^{14}N. \tag{5}$$

At stellar temperatures the $^{14}N(p,\gamma)^{15}O$ reaction is the slowest process in the cycle, defining the time scale and the overall energy production rate.[20,21,22] This reaction is therefore of importance for the interpretation of CNO burning. The proton capture by $^{15}N$ is relevant, as it is a branch point linking the first CNO or CN cycle with the second CNO or NO cycle.

Evidently, the considered reactions can play a certain role in some models of the Universe,[1-5] when the number of forming nuclei, perhaps, is dependent on the presence of dark energy and its concentration,[23] on the rate of growth of baryonic matter perturbations,[24] or on the rotation of the early Universe.[25] However, perturbations in the primordial plasma not only stimulate the process of nucleosynthesis,[26] but also kill it, for example, through the growth of the perturbations of non-baryonic matter of the Universe[27] or because of the oscillations of cosmic strings.[28]

### 1.2. *Nuclear aspects of the review*

One extremely successful line of development of nuclear physics in the last 50-60 years has been the microscopic model known as the Resonating Group Method (RGM, see, for example, Refs. 29, 30, 31, 32, and 33). And the associated with it models, for example, Generator Coordinate Method (see, particularly, Refs. 33 and 34]) or algebraic version of RGM.[35,36] However, the rather difficult RGM calculations are not the only way in which to explain the available experimental facts. But, the possibilities offered by a simple two-body potential cluster model (PCM) have not been studied fully up to now, particularly if it uses the concept of forbidden states (FS).[37] The potentials of this model for discrete spectrum are constructed in order to correctly reproduce the main characteristics of the bound states (BSs) of light nuclei in cluster channels, and in the continuous spectrum they directly take into account the resonance behavior of the elastic scattering phase shifts of the interactive particles at low energies.[38,39] It is enough to use the simple PCM with FSs taking into account the described methods of construction of potentials and classification of the orbital states according to Young tableaux for consideration many problems of nuclear physics of low energy and nuclear astrophysics. Such a model can be called a modified PCM (MPCM). In many cases, such an approach, as has been shown previously, allows one to obtain adequate results in the description of many experimental studies for the total cross sections of the thermonuclear reactions at low and astrophysical energies.[37,38,39]

Therefore, in continuing to study the processes of radiative capture,[38,39] we will consider the $n + {}^{10,11}B \to {}^{11,12}B + \gamma$, $p + {}^{11}B \to {}^{12}C + \gamma$, $p + {}^{14}C \to {}^{15}N + \gamma$, and $p + {}^{15}N \to {}^{16}O + \gamma$ reactions within the framework of the MPCM at low and thermal energies. The resonance behavior of the elastic scattering phase shifts of the interacting particles at low energies will be taken into account. In addition, the classification of the orbital states of the clusters according to the Young tableaux allows one to clarify the number of FSs and allowed states (ASs), i.e., the number of nodes of the wave function of the relative motion of the cluster. The potentials of the $n^{10}B$ interaction for scattering processes will be constructed based on the reproduction of the spectra of resonance states for the final nucleus in the $n^{10}B$ channel. The $n^{10}B$ potentials are constructed based on the description both of the binding energies of these particles in the final nucleus and of certain basic characteristics of these states; for example, the charge radius and the asymptotic constant (AC) for the BS or the ground state (GS) of ${}^{11}B$, formed as a result of the capture reaction in the cluster channel, which coincide with the initial particles.[39]

## 2. Model and calculation methods

Let us give firstly the basic expressions for the phase shift analysis of the $p^{14}C$ elastic scattering and note that earlier we already have performed the phase shift analysis in systems $p^6Li$, $n^{12}C$, $p^{12}C$, ${}^4He^4He$, ${}^4He^{12}C$, $p^{13}C$, and $n^{16}O$,[40,41,42,43,44,45,46] meanwhile, essentially at astrophysical energies.[39]

The nuclear scattering phases obtained on the basis of the experimental differential cross sections allow one to extract certain information about the structure of resonance states of light atomic nuclei.[47] In this case, the processes of the elastic scattering of particles with total spin of 1/2 on the nucleus with zero spin take place in nuclear systems like $N^4He$, $^3H^4He$, $N^{12}C$, $N^{16}O$, $N^{14}C$ etc. The cross section of the elastic scattering of such particles is presented in the simple form[48]

$$\frac{d\sigma(\theta)}{d\Omega} = |A(\theta)|^2 + |B(\theta)|^2 , \tag{6}$$

where

$$A(\theta) = f_c(\theta) + \frac{1}{2ik}\sum_{L=0}^{\infty}\{(L+1)S_L^+ + LS_L^- - (2L+1)\}\exp(2i\sigma_L)P_L(\cos\theta) ,$$

$$B(\theta) = \frac{1}{2ik}\sum_{L=0}^{\infty}(S_L^+ - S_L^-)\exp(2i\sigma_L)P_L^1(\cos\theta) ,$$

$$f_c(\theta) = -\left(\frac{\eta}{2k\sin^2(\theta/2)}\right)\exp\{i\eta\ln[\sin^{-2}(\theta/2)] + 2i\sigma_0\} \quad . \tag{7}$$

Here $S_L^{\pm} = \eta_L^{\pm}\exp(2i\delta_L^{\pm})$ – scattering matrix, $\delta_L^{\pm}$ – required scattering phase shifts, $\eta_L^{\pm}$ – inelasticity parameters, and signs "±" correspond to the total moment of system $J = L \pm 1/2$, $k$ – wave number of the relative motion of particles $k^2 = 2\mu E/\hbar^2$, $\mu$ – reduced mass, $E$ – the energy of interacting particles in the center-of-mass system, $\eta$ – Coulomb parameter.

The multivariate variational problem of finding these parameters at the specified range of values appears when the experimental cross sections of scattering of nuclear particles and the mathematical expressions, which describe these cross sections with certain parameters $\delta_L^J$ – nuclear scattering phase shifts, are known. Using the experimental data of differential cross-sections of elastic scattering, it is possible to find a set of phase shifts $\delta_L^J$, which can reproduce the behavior of these cross-sections with certain accuracy. Quality of description of experimental data on the basis of a certain theoretical function or functional of several variables of Eqs. (6) and (7) can be estimated by the $\chi^2$ method, which is written as

$$\chi^2 = \frac{1}{N}\sum_{i=1}^{N}\left[\frac{\sigma_i^t(\theta) - \sigma_i^e(\theta)}{\Delta\sigma_i^e(\theta)}\right]^2 = \frac{1}{N}\sum_{i=1}^{N}\chi_i^2 , \tag{8}$$

where $\sigma^e$ and $\sigma^t$ are experimental and theoretical, i.e., calculated for some defined values of the scattering phase shifts cross-sections of the elastic scattering of nuclear particles for $i$-angle of scattering, $\Delta\sigma^e$ – the error of experimental cross-sections at these angles, $N$ – the number of measurements. The details of the using by us searching method of scattering phase shifts were given in Ref. 48 and in our works Refs. 39 and 43.

The nuclear part of the intercluster interaction potential, for carrying out calculations of photonuclear processes in the considered cluster systems, has the form:

$$V(r) = -V_0 \exp(-\alpha r^2), \tag{9}$$

with point-like Coulomb term of the potential.

The potential is constructed completely unambiguously with the given number of BSs and with the analysis of the resonance scattering when in the considered partial wave at energies up to 1 MeV where there is a rather narrow resonance with a width of about 10–50 keV. Its depth is unambiguously fixed according to the resonance energy of the level at the given number of BS, and the width is absolutely determined by the width of such resonance. The error of its parameters does not usually exceed the error of the width determination at this level and equals 3–5%. Furthermore, it concerns the construction of the partial potential according to the phase shifts and determination of its parameters according to the resonance in the nuclear spectrum.

Consequently, all potentials do not have ambiguities and allow correct description of total cross sections of the radiative capture processes, without involvement of the additional quantity – spectroscopic factor $S_f$.[49] It is not required to introduce additional factor $S_f$ under consideration of capture reaction in the frame of PCM for potentials that are matched, in continuous spectrum, with characteristics of scattering processes that take into account resonance shape of phase shifts, and in the discrete spectrum, describing the basic characteristics of nucleus BS.

All effects that are present in the reaction, usually expressed in certain factors and coefficients, are taken into account at the construction of the interaction potentials. It could be possible, exactly because they are constructed and take into account FS structure. On the basis of description of observed, i.e., experimental characteristics of interacting clusters in the initial channel and formed, in the final state, a certain nucleus that has a cluster structure consisting of initial particles. In other words, the presence of $S_f$, is apparently taken into account in the BS wave functions of clusters, determining the basis of such potentials due to solving the Schrödinger equation.[50]

The AC for any GS potential was calculated using the asymptotics of the wave function (WF) having a form of exact Whittaker function[51]

$$\chi_L(r) = \sqrt{2k}\,C_W W_{-\eta_L+1/2}(2kr), \tag{10}$$

where $\chi_L$ is the numerical wave function of the bound state obtained from the solution of the radial Schrödinger equation and normalized to unity; $W$ is the Whittaker function of the bound state which determines the asymptotic behavior of the WF and represents the solution of the same equation without nuclear potential, i.e., long distance solution; $k$ is the wave number determined by the channel binding energy; $\eta$ is the Coulomb parameter that is equal to zero in this case; $L$ is the orbital moment of the bound state.

The total radiative capture cross sections $\sigma(NJ,J_f)$ for the $EJ$ and $MJ$ transitions in the case of the PCM are given, for example, in Ref. 49 or Refs. 38, 39, 52, and 53 are written as:

$$\sigma_c(NJ,J_f) = \frac{8\pi K e^2}{\hbar^2 q^3}\frac{\mu}{(2S_1+1)(2S_2+1)}\frac{J+1}{J[(2J+1)!!]^2}$$

$$\times A_J^2(NJ,K)\sum_{L_i,J_i} P_J^2(NJ,J_f,J_i) I_J^2(J_f,J_i) \qquad (11)$$

where $\sigma$ – total radiative capture cross section; $\mu$ – reduced mass of initial channel particles; $q$ – wave number in initial channel; $S_1$, $S_2$ – spins of particles in initial channel; $K$, $J$ – wave number and momentum of $\gamma$-quantum in final channel; $N$ – is the $E$ or $M$ transitions of the $J$ multipole ordered from the initial $J_i$ to the final $J_f$ nucleus state.

The value $P_J$ for electric orbital $EJ(L)$ transitions has the form[38-49]

$$P_J^2(EJ,J_f,J_i) = \delta_{S_iS_f}\left[(2J+1)(2L_i+1)(2J_i+1)(2J_f+1)\right](L_i 0 J 0 \mid L_f 0)^2 \begin{Bmatrix} L_i & S & J_i \\ J_f & J & L_f \end{Bmatrix}^2 ,$$

$$A_J(EJ,K) = K^J \mu^J \left(\frac{Z_1}{m_1^J} + (-1)^J \frac{Z_2}{m_2^J}\right), \qquad I_J(J_f,J_i) = \langle \chi_f | R^J | \chi_i \rangle . \qquad (12)$$

Here, $S_i$, $S_f$, $L_f$, $L_i$, $J_f$, and $J_i$ – total spins, angular and total moments in initial ($i$) and final ($f$) channels; $m_1$, $m_2$, $Z_1$, $Z_2$ – masses and charges of the particles in initial channel; $I_J$ –integral over wave functions of initial $\chi_i$ and final $\chi_f$ states, as functions of cluster relative motion of n and $^{10}$B particles with intercluster distance $R$.

For consideration of the $M1(S)$ magnetic transition, caused by the spin part of magnetic operator,[54] it is possible to obtain an expression[38,39] using the following:[55]

$$P_1^2(M1,J_f,J_i) = \delta_{S_iS_f}\delta_{L_iL_f}\left[S(S+1)(2S+1)(2J_i+1)(2J_f+1)\right] \begin{Bmatrix} S & L & J_i \\ J_f & 1 & S \end{Bmatrix}^2 ,$$

$$A_1(M1,K) = \frac{e\hbar}{m_0 c} K\sqrt{3}\left[\mu_1 \frac{m_2}{m} - \mu_2 \frac{m_1}{m}\right] , \quad I_J(J_f,J_i) = \langle \chi_f | R^{J-1} | \chi_i \rangle , \quad J = 1. \qquad (13)$$

Here, $m$ is the mass of the nucleus, and $\mu_1$ and $\mu_2$ are the magnetic moments of the clusters, the values of which are taken from Refs. 56 and 57.

The construction methods used here for intercluster partial potentials at the given orbital moment $L$, are expanded in Refs. 38, 39, 58 and here we will not discuss them further. The next values of particle masses are used in the given calculations: $m_n$ = 1.00866491597 amu,[59] $m_p$ = 1.00727646577 amu,[60] $m(^{10}B)$ = 10.012936 amu,[61] $m(^{11}B)$ = 11.0093052 amu,[62] $m(^{14}C)$ = 14.003242 amu,[63] $m(^{15}N)$ = 15.000108 amu,[64] and constant $\hbar^2/m_0$ is equal to 41.4686 MeV fm$^2$.

## 3. Neutron-capture reaction $^{10}$B(n, $\gamma$)$^{11}$B

### 3.1. *Structure of cluster states*

We regard the results of the classification of $^{11}$B by orbital symmetry in the n$^{10}$B channel as qualitative, because there are no complete tables of Young tableaux productions for systems with more than eight nucleons,[65] which have been used in earlier similar calculations.[38,39] At the same time, simply based on such a classification, we succeeded in describing the available experimental data on the radiative capture of neutrons and charged particles for a wide range of reactions.[38,39,66,67,68] This is why the classification procedure by orbital symmetry given above was used here for the determination of the number of FSs and ASs in partial intercluster potentials and, consequently, to the specified number of nodes of the wave function of the relative motion of the cluster for the case of neutrons and $^{10}$B.

Furthermore, we will suppose that it is possible to assume the orbital Young tableau in the form {442} for $^{10}$B; therefore, for the n$^{10}$B system, we have $\{1\} \times \{442\} \rightarrow \{542\} + \{443\} + \{4421\}$.[65] The first of the obtained tableaux is compatible with orbital moments $L = 0$, 2, 3, and 4, and is forbidden because it contains five nucleons in the *s*-shell. The second tableau is allowed and is compatible with orbital moments $L = 1$, 2, 3, and 4, and the third is also allowed and is compatible with $L = 1$, 2, and 3.[69] As mentioned before, the absence of tables of Young tableaux productions for when the number of particles is 10 and 11 prevents the exact classification of the cluster states in the considered system of particles. However, qualitative estimations of the possible Young tableaux for orbital states allow us to detect the existence of the FSs in the *S* and *D* waves and the absence of FSs for the *P* states. The same structure of FSs and ASs in the different partial waves allows us to construct the potentials of intercluster interactions required for the calculations of the total cross sections for the considered radiative capture reaction. Thus, by limiting our consideration to only the lowest partial waves with orbital moment $L = 0$, and 1, it could be said that for the n$^{10}$B system (for $^{10}$B it is known $J^{\pi},T = 3^{+},0$; Ref. 70), the only allowed state exists in the *P* wave potentials and the FS is in the *S* waves. The state in the $^{6}P_{3/2}$ wave (representation in $^{(2S+1)}L_{J}$) corresponds to the GS of $^{11}$B with $J^{\pi},T = 3/2^{-},1/2$ and is at the binding energy of the n$^{10}$B system of -11.4541(2) MeV.[9] Let us note that some n$^{10}$B scattering states and BSs can be mixed by isospin with $S = 5/2$ ($2S+1 = 6$) and $S = 7/2$ ($2S + 1 = 8$).

The spectrum of $^{11}$B for excited states (ESs), bound in the n$^{10}$B channel, shows that at the energy of 2.1247 MeV above the GS or -9.3329 MeV (see Ref. 9) relative to the threshold of the n$^{10}$B channel, the first ES can be found, bound in this channel with the moment $J^{\pi} = 1/2^{-}$, which can be compared with the $^{6}F_{1/2}$ wave with an FS. However, we will not consider it, because of the large value of the angular momentum barrier. The second ES at the energy 4.4449 MeV (see Ref. 9) above the GS or -7.0092 MeV relative to the threshold of the n$^{10}$B channel has the moment $J^{\pi} = 5/2^{-}$, and it can be compared with the mixture of the $^{6}P_{5/2}$ and $^{8}P_{5/2}$ waves without FSs. Furthermore, the unified potential of such a mixed $P_{5/2}$ state will be constructed, because the model used does not allow one to divide states with different spin clearly. The wave function obtained with this potential at the calculation of the Schrödinger equation and, in principle, consisting of two components for different spin channels, does not divide into these components in the explicit form, i.e., we are using the total form of the WF in all calculations. The third ES at the energy of 5.0203 MeV (see Ref. 9) relative to the GS or -6.4338 MeV relative to the channel threshold has the moment $J^{\pi} = 3/2^{-}$, and it can be matched to the $^{6}P_{3/2}$ wave without an FS. The fourth ES at the energy of 6.7429 MeV (see Ref. 9) relative to the GS or -4.7112 MeV relative to the channel threshold has the moment $J^{\pi} = 7/2^{-}$, and it can be matched to the mixture of the $^{6}P_{7/2}$ and $^{8}P_{7/2}$ waves without an FS. In addition, it is possible to consider the ninth ES at the energy of 8.9202 MeV with the moment $5/2^{-}$, i.e., at the energy of -2.5339 MeV relative to the n$^{10}$B threshold, which can be matched to the mixture of the $^{6}P_{5/2}$ and $^{8}P_{5/2}$ states without FSs.

Consider now the resonance states in the n$^{10}$B system, i.e., states at positive energies. The first resonance state of $^{11}$B in the n$^{10}$B channel, located at the energy 0.17 MeV, has the neutron width of 4 keV and the moment $J^{\pi} = 5/2^{+}$.[9] It is possible to compare this state with the $^6S_{5/2}$ scattering wave with an FS. We have not succeeded in the construction of the potential with such small width; therefore, we will consider this scattering wave as nonresonance, which leads to zero scattering phase shifts. The second resonance state has the energy of 0.37 MeV – its neutron width equals 0.77 MeV and the moment $J^{\pi} = 7/2^{+}$;[9] therefore, it is possible to compare this with the $^8S_{7/2}$ scattering wave with an FS. Because of the large width of resonance (two times greater than its energy), we will use nonresonance values of the parameters for the potential to coincide with the previous $^6S_{5/2}$ potential. The third resonance state has the energy of 0.53 MeV in the laboratory system (l.s.) – its neutron width equals 0.031 MeV in the center-of-mass system (c.m.) and the moment $J^{\pi} = 5/2^{-}$.[9] Therefore, it can be compared with the mixed $^6P_{5/2} + {}^8P_{5/2}$ scattering waves without FSs. These characteristics of the resonance are given in Table 11.11 of Ref. 9, and in the note to this table are given the energy and the width values equal to 0.495(5) MeV and 140(15) keV, respectively, with reference to Ref. 71. At the same time, the value of 0.475(17) MeV (l.s.) with the total width 200(20) keV (c.m.) is given in Table 11.3 of Ref. 9 for this resonance. Furthermore, under the construction of this potential, we will proceed from two variants of data, notably, the first and the last given above with the width of 31 keV (c.m.), but at the energy of the resonance of 475 keV (l.s.), which follows from the level spectrum.[65]

The next resonance is at the energy above 1 MeV and we will not consider it (see Table 11.11, Ref. 9). There are no resonance levels lower than 1 MeV in the spectrum of $^{11}B$ that can be matched to the $^6P_{3/2}$ and $^{6+8}P_{7/2}$ states.[9] Therefore, their phase shifts are taken as equal to zero, and as far as there are no FSs in the $P$ waves, by way of the first variant, such potentials can be simply equalized to zero.[38,39] We ought to note here that there are more up-to-date values for all these states[72] – the results from this review do not differ for the ESs (see Table 11.18, Ref. 72), but do have slightly different values for resonance states. Particularly, the excited energy of 11.893(13) MeV with the adjusted total width of 194(6) keV, which gives 483 keV (l.s.) for the resonance energy, is given for the state $J^{\pi} = 5/2^{-}$, which can be matched to the $^{6+8}P_{5/2}$ scattering waves without FSs. The width equal to 1.34 MeV, which is given for the state with $J^{\pi} = 7/2^{+}$, compares with the $^8S_{7/2}$ scattering wave with an FS – it is twice that given in Ref. 9.

Continuing to the analysis of possible electromagnetic $E1$ and $M1$ transitions, let us note that we will consider only transitions to the GS and to four (2nd, 3rd, 4th, and 9th) ESs from the $S$ and $P$ scattering waves. As the GS is matched with the $^6P_{3/2}$ level, it is possible to consider $E1$ transitions from the $^6S_{5/2}$ scattering wave to the GS of $^{11}B$.

No. 1. $^6S_{5/2} \rightarrow {}^6P_{3/2}$.

In addition, it is possible to consider $E1$ transitions from the $^6S_{5/2}$ and $^8S_{7/2}$ scattering waves to the second ES of $^{11}B$, which is the mixture of two $P$ states:

No. 2. $\begin{array}{l} ^6S_{5/2} \rightarrow {}^6P_{5/2} \\ ^8S_{7/2} \rightarrow {}^8P_{5/2} \end{array}$.

Because here we have transitions from the initial $S$ states that differ by spin to the different parts of one total WF of the BS, which, evidently, have no essential difference, the cross section of these transitions will sum up, i.e., $\sigma = \sigma(^6S_{5/2} \rightarrow {}^6P_{5/2}) + \sigma(^8S_{7/2} \rightarrow {}^8P_{5/2})$. The third ES is matched with the $^6P_{3/2}$ level as the GS, and it is possible to consider the $E1$ transitions from the $^5S_{5/2}$ scattering waves to this ES of $^{11}$B.

No. 3. $^6S_{5/2} \rightarrow {}^6P_{3/2}$.

Another $E1$ transition is possible from the $^6S_{5/2}$ and $^8S_{7/2}$ scattering waves of the fourth ES of $^{11}$B at $J^\pi = 7/2^-$:

No. 4. $\begin{matrix} ^6S_{5/2} \rightarrow {}^6P_{7/2} \\ ^8S_{7/2} \rightarrow {}^8P_{7/2} \end{matrix}$.

The cross section of these two transitions will also sum up. The last of considered $E1$ transitions is the capture from the $^6S_{5/2}$ and $^8S_{7/2}$ scattering waves to the ninth ES of $^{11}$B at $J^\pi = 5/2^-$:

No. 5. $\begin{matrix} ^6S_{5/2} \rightarrow {}^6P_{5/2} \\ ^8S_{7/2} \rightarrow {}^8P_{5/2} \end{matrix}$.

The cross section of these transitions will also sum up, as given in the case of reaction 2.

Furthermore, it is possible to consider $M1$ transitions to the GS from the resonance scattering wave $^6P_{5/2}$ at 0.475(17) MeV, and from the nonresonance $^6P_{3/2}$ wave.

No. 6. $\begin{matrix} ^6P_{5/2} \rightarrow {}^6P_{3/2} \\ ^6P_{3/2} \rightarrow {}^6P_{3/2} \end{matrix}$.

As will be shown later, the cross section of the transition $^6P_{3/2} \rightarrow {}^6P_{3/2}$ for the first variant of the $^6P_{3/2}$ scattering potential with zero depth will stay at the level 1–2 μb, and the other transitions from the nonresonance waves will not be considered. Furthermore, it is possible to consider $M1$ transitions to the second ES $^6P_{5/2}$ and $^8P_{5/2}$ from the resonance $^6P_{5/2}$ and $^8P_{5/2}$ scattering waves.

No. 7. $\begin{matrix} ^6P_{5/2} \rightarrow {}^6P_{5/2} \\ ^8P_{5/2} \rightarrow {}^8P_{5/2} \end{matrix}$.

Because this is the transition from the mixed-by-spin $P_{5/2}$ scattering wave to the mixed-by-spin second ES, the cross section will be averaged according to transitions given above, i.e., $\sigma = 1/2\ \{\sigma(^6P_{5/2} \rightarrow {}^6P_{5/2}) + \sigma(^8P_{5/2} \rightarrow {}^8P_{5/2})\}$. The possible transition $^{6+8}P_{7/2} \rightarrow {}^{6+8}P_{5/2}$ from the nonresonance $^{6+8}P_{5/2}$ wave to the second ES is not taken into account here.

It is possible to consider the $M1$ transitions to the third ES $^6P_{3/2}$ from the resonance $^6P_{5/2}$ scattering wave

No. 8. ${}^6P_{5/2} \to {}^6P_{3/2}$.

The *M*1 transitions are feasible to the fourth ES ${}^6P_{7/2} + {}^8P_{7/2}$ from the resonance ${}^6P_{5/2} + {}^8P_{5/2}$ scattering wave

No. 9. $\begin{matrix} {}^6P_{5/2} \to {}^6P_{7/2} \\ {}^8P_{5/2} \to {}^8P_{7/2} \end{matrix}$.

This cross section will be averaged over two transitions, as was given in the case of reaction 7.

Finally, we can consider the *M*1 transitions to the ninth ES ${}^6P_{5/2} + {}^8P_{5/2}$ from the resonance ${}^6P_{5/2} + {}^8P_{5/2}$ scattering wave, because there are experimental data of Ref. 73 for this.

No. 10. $\begin{matrix} {}^6P_{5/2} \to {}^6P_{5/2} \\ {}^8P_{5/2} \to {}^8P_{5/2} \end{matrix}$.

This cross section will also be averaged over the two transitions to the ES given here.

### 3.2. *Interaction potentials*

For all partial waves of the $n^{10}B$ interaction potentials, i.e., for each partial wave with the given *L*, we used the Gaussian potential of the form of Eq. (9).

Here, as mentioned before, we will not consider the influence of the first resonance at 0.17 MeV in the ${}^6S_{5/2}$ wave; therefore, we will use the potential with FSs leading to the zero scattering phase

$$V_0 = 160.5 \text{ MeV}, \ \ \alpha = 0.5 \text{ fm}^{-2}. \tag{14}$$

The ${}^6S_{5/2}$ scattering phase shift of this potential at energy up to 1.0 MeV is less than 0.5°. The same parameters we be used for the ${}^8S_{7/2}$ scattering wave, also ignoring the resonance.

The following parameters were obtained for the third resonance state ${}^6P_{5/2} + {}^8P_{5/2}$ at 0.475(17) MeV:

$$V_0 = 106.615 \text{ MeV}, \ \ \alpha = 0.4 \text{ fm}^{-2}. \tag{15}$$

Such potential leads to resonance, i.e., the scattering phase shift equals 90.0°(1), at 475(1) keV (l.s.) with a width of 193(1) keV (c.m.), which is in good agreement with the data of reviews of Refs. 9 and 72.

For the potential of the pure-by-spin GS of ${}^{11}B$ in the $n^{10}B$ channel, where the ${}^6P_{3/2}$ wave is used, the following parameters were obtained:

$$V_0 = 165.3387295 \text{ MeV}, \ \ \alpha = 0.45 \text{ fm}^{-2}. \tag{16}$$

We have obtained the value of the dimensionless AC = 1.53(1) in the range of 3–10 fm, the charged radius of 2.44 fm, and the mass radius of 2.39 fm at the binding energy of -11.454100 MeV with the accuracy of the finite-difference method (FDM), used for the calculation of the binding energy at $\varepsilon = 10^{-6}$ MeV.[74] The AC error is connected with its averaging over the above-mentioned range of distances. The phase shift for such potential decreases smoothly until a value of 179° when the changes from zero to 1.0 MeV. The generalized Levinson theorem[37] is used for the determination of the value of scattering phase shift at zero energy.

The AC value equal to 1.72 $\mathrm{fm}^{-1/2}$ was obtained in Ref. 75 for the GS of $^{11}$B in the cluster channel n$^{10}$B, where the coefficient of neutron identity was assigned (see expression 83b in Ref. 76). In this work, a slightly different definition of AC was used,

$$\chi_L(r) = C \cdot W_{-\eta L+1/2}(2k_0 r) \tag{17}$$

(with the Coulomb parameter $\eta$ equals zero, in this case), which is different from our previous works of Refs. 38 and 39 to the value $\sqrt{2k_0}$ that equals 1.19 $\mathrm{fm}^{-1/2}$ for the GS; therefore, the AC value equals 1.44 in the dimensionless form. The improved value of 1.82(15) $\mathrm{fm}^{-1/2}$ is given in the latest results for this AC,[77] and after re-computation, it gives 1.52(12) in the dimensionless form and agrees absolutely with the value for the GS potential of Eq. (16) obtained here.

The parameters of the GS potential and any BSs in the considered channel at the given number of the bound, allowed or forbidden states in the partial wave, are fixed quite unambiguously by the binding energy, the charge radius, and the asymptotic constant. The accuracy of the determination of the BS potential parameters is connected with the accuracy of the AC, which is usually equal to 10% to 20%. There are no another ambiguities in this potential, because the classification of the states according to the Young tableaux allows us unambiguously to fix the number of BSs in this partial wave, which defines its depth completely, and the width of the potential depends wholly on the values of the charge radius and the AC.

The next parameters were obtained for the parameters of the $^6P_{5/2} + {}^8P_{5/2}$ potential without FSs for the second ES of $^{11}$B in the n$^{10}$B channel with $J^\pi = 5/2^-$:

$$V_0 = 151.61181 \text{ MeV}, \ \ \alpha = 0.45 \text{ fm}^{-2}. \tag{18}$$

This potential leads to the binding energy of -7.0092 MeV at $\varepsilon = 10^{-4}$, which is completely coincident with the experimental value[9,72] for the charge radius of 2.44 fm, and the AC of 1.15(1) at the range of 3–13 fm.

The next parameters were obtained for the potential without FSs for the third ES $^6P_{3/2}$ pure-by-spin with $J^\pi = 3/2^-$:

$$V_0 = 149.70125 \text{ MeV}, \ \ \alpha = 0.55 \text{ fm}^{-2}. \tag{19}$$

These parameters lead to the binding energy of -6.4338 MeV at $\varepsilon = 10^{-4}$ coinciding with the experimental value[9,72] for the AC equals 1.10(1) at the range of 3–13 fm, and the charge and mass radii are equal to 2.44 fm and 2.41 fm, respectively. The scattering phase shift for this potential decreases until 178° at the energy of 1.0 MeV.

The next parameters were obtained for the $^6P_{7/2} + {}^8P_{7/2}$ potential without FSs for the fourth ES of $^{11}$B in the n$^{10}$B channel with $J^{\pi} = 7/2^{-}$:

$$V_0 = 143.72353 \text{ MeV}, \ \alpha = 0.45 \text{ fm}^{-2}. \tag{20}$$

The binding energy of -4.7112 MeV at $\varepsilon = 10^{-4}$, which absolutely coincides with the experimental value[9,72] for the charge radius of 2.44 fm, and dimensionless AC equal to 0.94(1) at the range of 3–15 fm, was obtained with this potential. The scattering phase shift for this potential decreases smoothly until 178° at the energy of 1.0 MeV.

These parameters were obtained for the $^6P_{5/2} + {}^8P_{5/2}$ potential without FSs for the ninth ES of $^{11}$B in the n$^{10}$B channel with $J^{\pi} = 5/2^{-}$:

$$V_0 = 135.39620 \text{ MeV}, \ \alpha = 0.45 \text{ fm}^{-2}. \tag{21}$$

This potential leads to the binding energy of -2.5339 MeV at $\varepsilon = 10^{-4}$, which completely coincides with the experimental value[9,72] for the charge radius of 2.44 fm, and dimensionless AC equal to 0.70(1) at the range of 3–24 fm. The scattering phase shift for this potential decreases smoothly until 178° at the energy of 1.0 MeV.

### 3.3. *The total cross section of the radiative neutron capture on $^{10}$B*

The next experimental data were used for the comparison of the calculation results given in Figs. 1 and 2. The black points (●) show the total summed capture cross section from Ref. 73 at 23, 40, and 61 keV. The triangle (▲) represents the cross section of 500(200) μb from Ref. 78 at the energy of 25 meV, and the open reverse triangle (∇) shows the new results for the cross section of 305(16) μb at 25 meV from Ref. 79, given in review Ref. 72. It should be noted that other data for 390(11) μb, obtained in Ref. 80 and also shown in Figs. 1 and 2 by the open reverse triangle (∇), were published later – reference to these results is also given in review Ref. 72. The experimental measurements of Ref. 73 for the transitions to different ESs of $^{11}$B are shown in Figs. 1 and 2: open circles (o) represent the total capture cross section to the GS $^6P_{3/2}$, open squares (□) represent the total capture cross section to the second ES $^{6+8}P_{5/2}$, black squares (■) represent the total capture cross section to the fourth ES $^{6+8}P_{7/2}$, and open triangles (Δ) represent the total capture cross section to the ninth ES $^{6+8}P_{5/2}$. Furthermore, in Figs. 3 and 4, only part of these experimental results is given.

The $E1$ transition $^6S_{5/2} \rightarrow {}^6P_{3/2}$ from the $S$ scattering wave with potential of Eq. (14) to the GS with potential of Eq. (16) was considered initially in our calculations, and the obtained capture cross section is shown in Fig. 1, represented by the short dashed line. The general dashed line shows the capture cross section to the second ES for potential of Eq. (18), identified in section 2 as No. 2. The dotted line shown at the bottom of Fig. 1 denotes the cross section of the transition $^6S_{5/2} \rightarrow {}^6P_{3/2}$ to the third ES of Eq. (19). The dot-dashed line shows the cross section of the transition to the fourth ES of Eq. (20), identified in section 2 as No. 4. The dot-dot-dashed line, which is almost superimposed with the dotted line, shows the transition from the $S$ scattering waves to the ninth ES with potential of Eq. (21). The solid line gives the total summed cross section of all the above considered transitions, which largely describes the experimental data for the total summed cross sections from Refs. 73 and 78 at the energy range from 25 meV to 61 keV correctly.

Let us note that in the measurements of Ref. 73 the transition to the third ES is not taken into account, and as seen in Fig. 1, this leads to nearly the same cross section of the transition to the third ES and the ninth ES – dotted and dot-dot-dashed lines. Therefore, probably, it is necessary to add the cross section of the transition to the ninth ES to the total cross sections from Ref. 73 to obtain summed cross sections that are more correct, and this will be equivalent to taking into account the transition to the third ES. Such cross sections are shown in Figs. 1–4 by the open rhombus – this account influences weakly the total cross sections, which also agree with the results of our calculations.

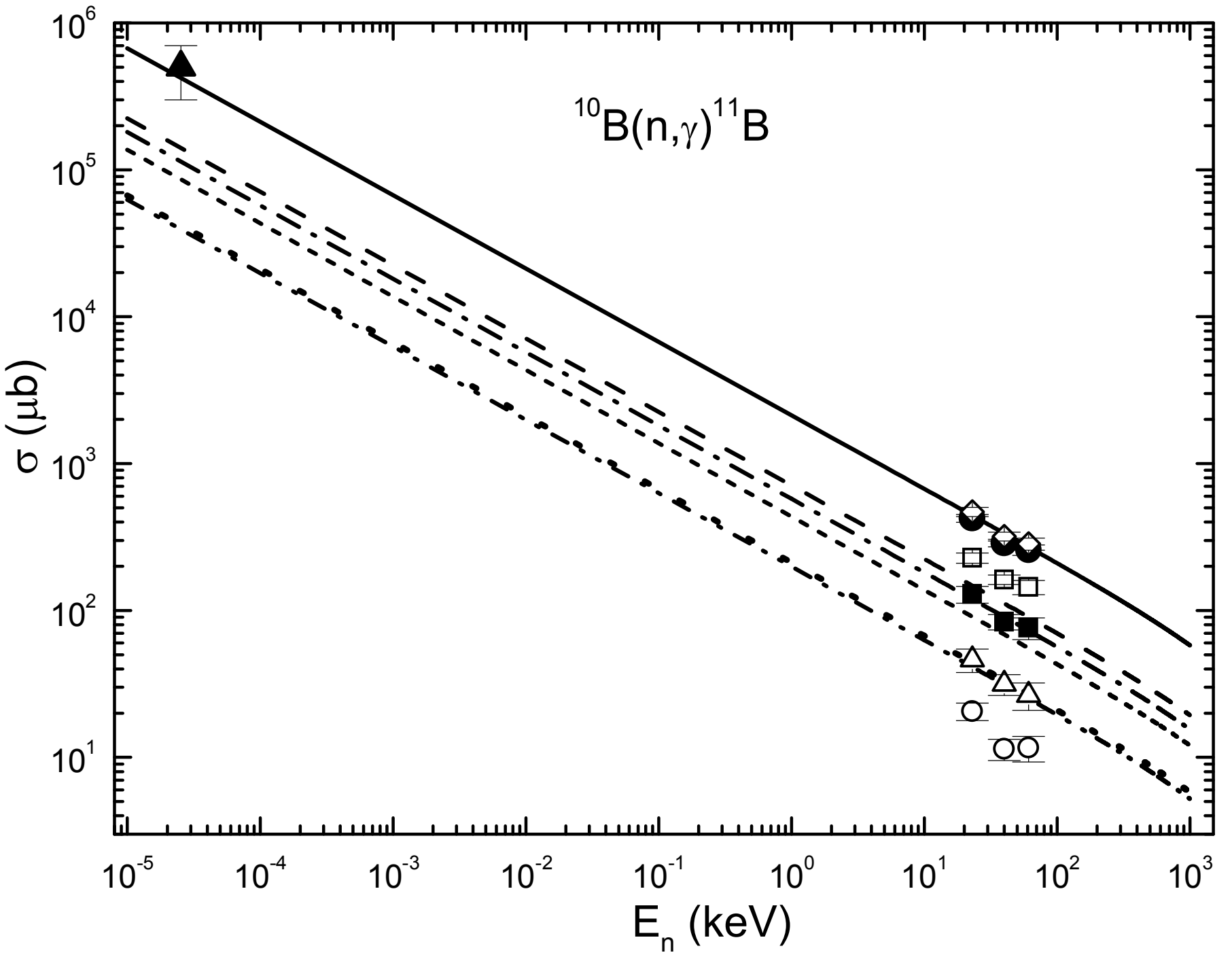

Fig. 1. The total cross sections of the radiative neutron capture on $^{10}$B. Experimental data: black triangle (▲) – the capture cross section at 25 meV from Ref. 78, points (●) – the total summed cross section of the neutron capture on $^{10}$B from Ref. 73, circles (o) – the total capture cross section to the GS, open squares (□) – the total capture cross section to the second ES, black squares (■) – the total capture cross section to the fourth ES, and open triangles (Δ) – the total capture cross section to the ninth ES from Ref. 73, open reversed triangles (∇) – the capture cross section at 25 meV from Refs. 79 and 80, open rhombus (◊) – the summed total capture cross section from Ref. 73 taking into account the transition to the third ES. Lines: the short dashed line is the cross section of the $E1$ transition ${}^6S_{5/2} \to {}^6P_{3/2}$ from the $S$ scattering wave with potential of Eq. (14) to the GS with potential of Eq. (16), the general dashed line is the capture cross section to the second ES of Eq. (18), the dotted line is the capture cross section of the transition ${}^6S_{5/2} \to {}^6P_{3/2}$ to the third ES of Eq. (19), the dot-dashed line is the cross section of the transition to the fourth ES of Eq. (20), the dot-dot-dashed line is the cross section of the transition from the $S$ scattering waves to the ninth ES with potential of Eq. (21), the solid line is the total summed cross section of all considered transitions.

As can be seen from the obtained results, the calculated line for the transition to the fourth ES is in a good agreement with the given black squares (experimental data).[73] The good agreement of the calculation, shown by the dot-dot-dashed line, can also be observed for the transition to the ninth ES, the experiment for which is shown by the open triangles.[73] The measurements for the transition to the second ES, shown in Fig. 1 by the open squares,[73] lie appreciably higher than the corresponding calculated line, shown by the dashed line. The measurements of the cross section for the transition to the GS, shown by the open circles,[73] lie much lower than the calculated line, shown by the short dashed line. Thereby, only two calculations conform to the experimental results for the transitions to the fourth and ninth ESs,[73] although the total summed cross sections, shown by the black points or rhombus, are described completely by the calculated line – the solid line in Fig. 1.

Because, we do not know the AC value for the second ES, it is always possible to construct the potential correctly describing the capture cross sections to this state, shown in Figs. 1 and 2 by the open squares.[73] For example, it is possible to use the potential with the parameters:

$$V_0 = 108.37443 \text{ MeV}, \;\; \alpha = 0.3 \text{ fm}^{-2}, \tag{22}$$

which leads to the binding energy of -7.0092 MeV, the charged radius of 2.44 fm, and the value of the AC equal to 1.45(1) at the range of 4–13 fm. The calculation results of the capture cross sections to this state from the $S$ scattering waves are shown in Fig. 2 by the dashed line, which is in a quite agreement with the experimental data of Ref. 73 shown by the open squares.

At the same time, the other variant of the GS potential that describes the total capture cross sections to the GS correctly, shown in Figs. 1 and 2 by the open circles, will not agree with the known AC or that given above for the GS. For example, the parameters

$$V_0 = 602.548373 \text{ MeV}, \;\; \alpha = 2.0 \text{ fm}^{-2} \tag{23}$$

allow one to describe reasonably the available experimental cross section measurements of the transition,[73] as is shown in Fig. 2 by the short dashed line. However, although this potential leads to the correct binding energy of -11.454100 MeV and describes reasonably the charged radius of 2.43 fm, the value of the AC is equal to 0.71(1) at the range of 2–8 fm, which is half that of the results from other experimental data from Refs. 75 and 77. This result can be explained by the imperfection of the MPCM used here; however, on such occasions, the MPCM led to the correct description of the cross sections both to the transitions to the GS and to the total summed cross section of the capture processes of Refs. 38, 39, 58, and 66. Therefore, it could be supposed that the experimental measurements for transitions to different ESs of $^{11}$B at the radiative neutron capture on $^{10}$B should be improved in the future; it will also be interesting to obtain new data in the range of possible resonances from 100 to 600 keV.

Reverting to the calculation results given in Fig. 1, we note that at the energies from 10 meV to 10 keV, the calculated cross section is almost a straight line, and it can be approximated by a simple function of the form:

$$\sigma_{\text{ap}} = \frac{A}{\sqrt{E_{\text{n}}}} \,. \tag{24}$$

The value of the given constant $A = 2123.4694$ μb·keV$^{1/2}$ was determined from a single point of the cross-sections (solid line in Fig. 1) at a minimal energy of 10 meV. The absolute value

$$M(E) = \left|[\sigma_{\rm ap}(E) - \sigma_{\rm theor}(E)] / \sigma_{\rm theor}(E)\right| \qquad (25)$$

of the relative deviation of the calculated theoretical cross-sections ($\sigma_{\rm theor}$), and the approximation of this cross-section ($\sigma_{\rm ap}$) by the expression given above in the energy range until 10 keV, is at the level of 0.2%. It is supposed that this form of total cross-section dependence on energy will be conserved at lower energies. In this case, the estimation of the cross-section value, for example, at the energy of 1 μ keV, gives the value of 67.2 b. The coefficient for the solid line in Fig. 2 in the expression of Eq. (24) given above for the approximated calculation results for cross sections, is equal to 2150.3488 $\mu$b·keV$^{1/2}$.

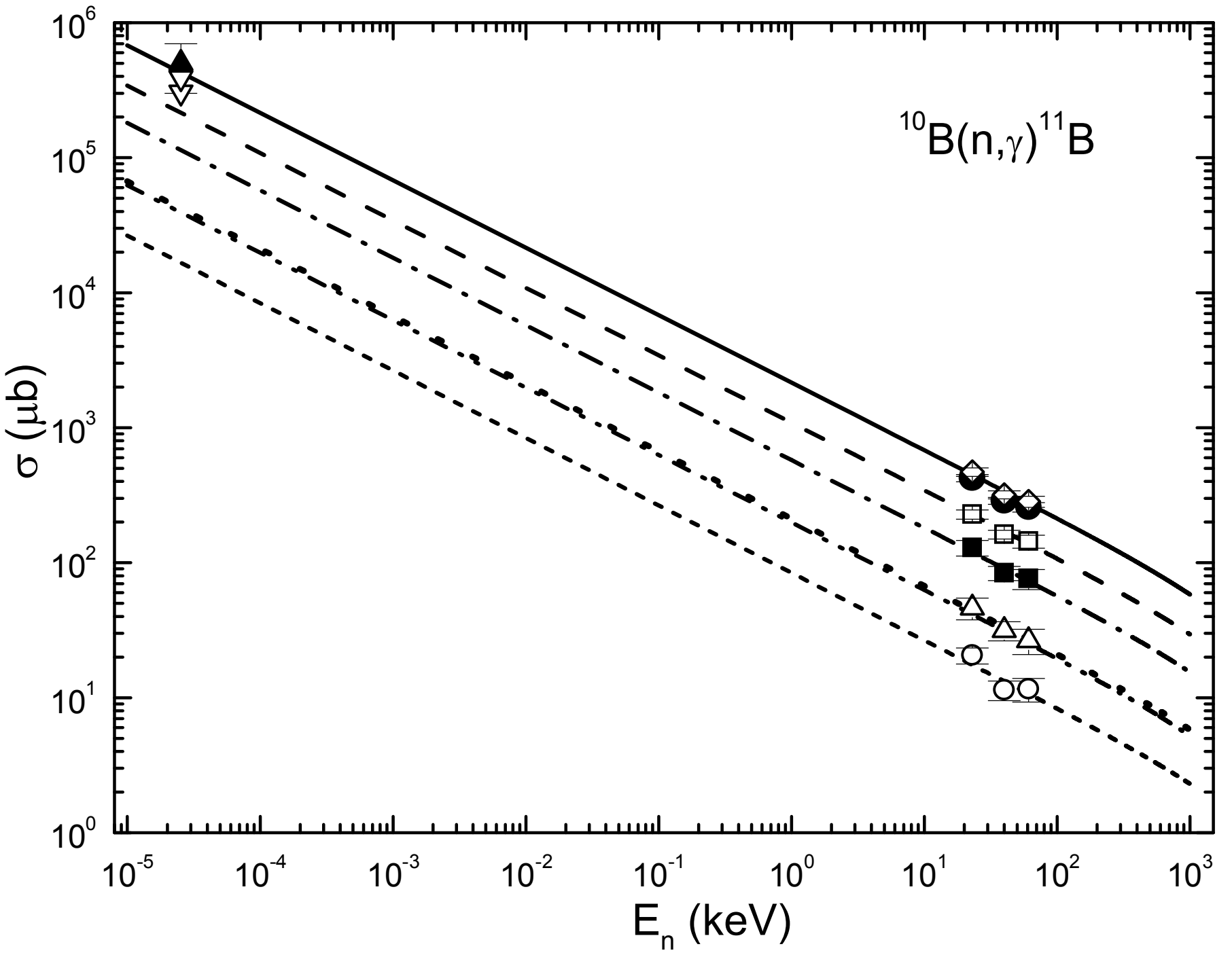

Fig. 2. The total cross sections of the radiative neutron capture on $^{10}$B. Experimental data: the same as in Fig. 1. Lines: the short dashed line is the cross section of the *E*1 transition $^6S_{5/2} \to {}^6P_{3/2}$ from the *S* scattering wave with potential of Eq. (14) to the GS with potential of Eq. (23), the general dashed line is the capture cross section to the second ES of Eq. (22), the dotted line is the capture cross section of the transition $^6S_{5/2} \to {}^6P_{3/2}$ to the third ES of Eq. (19), the dot-dashed line is the cross section of the transition to the fourth ES of Eq. (20), the dot-dot-dashed line is the cross section of the transition from the *S* scattering waves to the ninth ES with potential of Eq. (21), the solid line is the total summed cross section of all considered transitions.

Furthermore, the considered *M*1 transitions to the GS and to the different ESs are shown in Fig. 3, together with the summed cross section for the *E*1 processes, which is shown by the dashed line (it is represented by the solid line in Fig. 1). The dotted line at the top of the figure shows the cross section of the *M*1 transition to the GS with potential of Eq. (16) from the resonant $^6P_{5/2}$ scattering wave for potential of Eq. (15), identified in section 3.1 as No. 6. The dot-dashed line it is the transition from the $P_{5/2}$ scattering wave of Eq. (8) to the second ES with potential of Eq. (18), identified in section 2 as No. 7. The dot-dot-dashed line shows the cross section of the $M$1 transition ${}^6P_{5/2} \to {}^6P_{3/2}$ to the third ES with potential of Eq. (19) in Fig. 3.

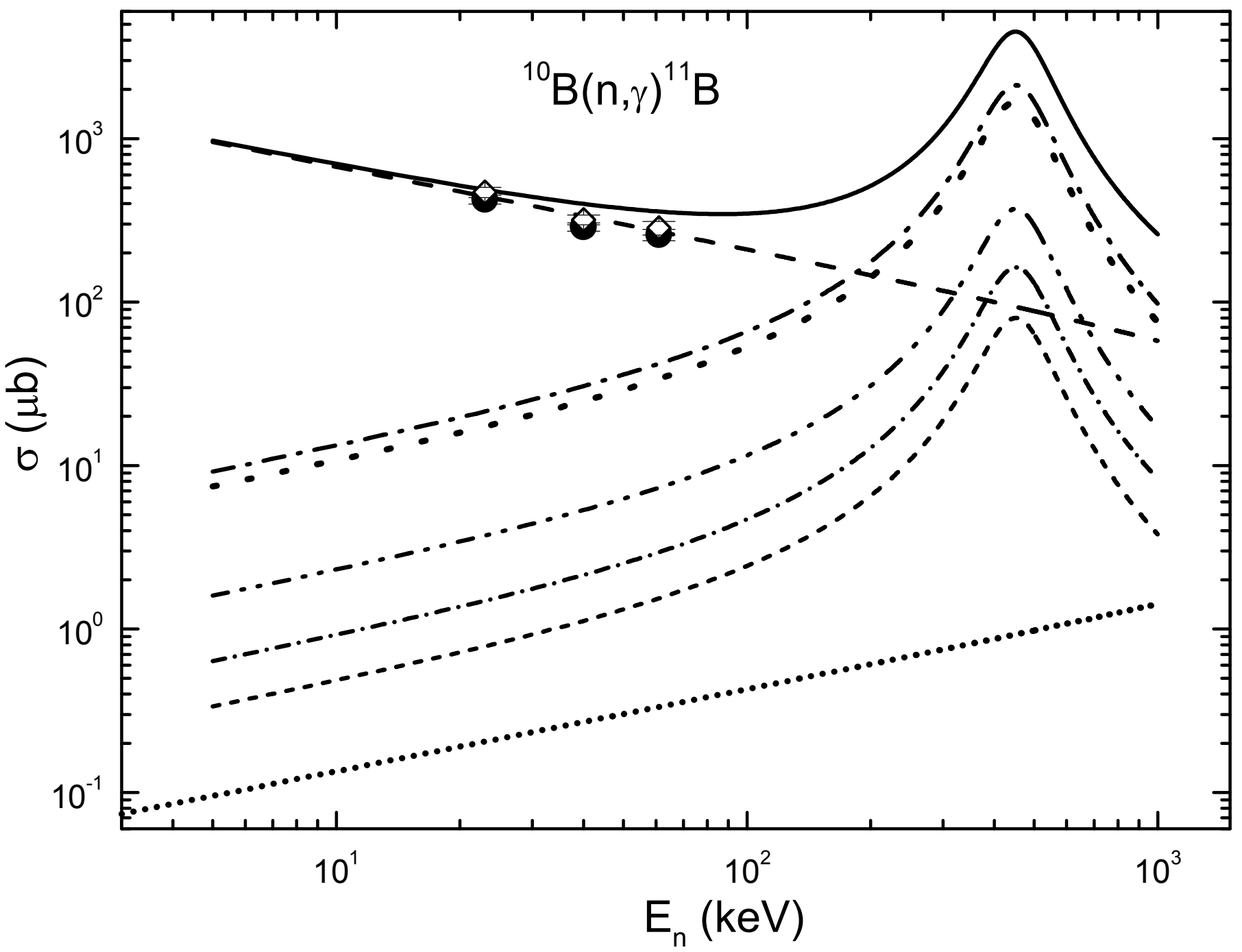

Fig. 3. The total cross sections of the radiative neutron capture on $^{10}$B. Experimental data: points (●) – the total summed cross section of the neutron capture on $^{10}$B from Ref. 73, open rhombus (◊) – the summed total capture cross section from Ref. 73 taking into account the transition to the third ES. Lines: the dashed line is the summed cross section of the $E$1 transitions, shown in Fig. 1 by the solid line, the dotted line gives the cross section of the $M$1 transition to the GS of Eq. (16) from the resonant ${}^6P_{5/2}$ scattering wave for potential of Eq. (15), the dot-dashed line is the cross section of the transition from the resonant ${}^6P_{5/2}$ scattering wave to the second ES of Eq. (18), the dot-dot-dashed line shows the cross section of the $M$1 transition ${}^6P_{5/2} \to {}^6P_{3/2}$ to the third ES of Eq. (19), from the $S$ scattering waves to the ninth ES with potential of Eq. (21), the short dashed line is the $M$1 transition to the fourth ES of Eq. (21), the dot-dashed line with closely placed dashes shows the ninth ES of Eq. (21), the dotted line with closely placed dots at the bottom of the figure shows the $M$1 transition from the nonresonance ${}^6P_{3/2}$ scattering wave to the GS, the solid line shows the sum of the $E$1 and $M$1 transitions considered above.

Another possible $M$1 transition to the fourth ES of Eq. (20), identified in section 2 as No. 9, has the form of the cross section shown in Fig. 3 by the short dashes. In addition, the $M$1 transition to the ninth ES of Eq. (21) is possible; it is identified in section 2 as No. 10, and shown in Fig. 3 by the dot-dashed line with closely placed dashes (the third line at the bottom), which lies slightly higher than the short dashed line. In addition, the $M$1 transition to the GS from the nonresonance ${}^6P_{3/2}$ scattering wave was considered with the potential of zero depth – the second transition under No. 6 in section 2. The result is shown by the dotted line with closely placed dots in the bottom of Fig. 3. Its value at the maximum is about 1.5 μb and it has practically no influence on the calculated cross sections in the range of the resonance at 475 keV, almost reaching to 4.5 mb.

The sum of all the $E1$ and $M1$ transitions described above is shown in Fig. 3 by the solid line, which gives a suitable description of the given experimental data. The small overshoots of the calculated cross sections over the experimental one at 40 and 61 keV can be used to argue that the used potential of Eq. (15) leads to the overestimated value of the resonance width in the $P_{5/2}$ scattering wave of 193 keV. As mentioned before, some values for the energy and width of this resonance are given in review Ref. 9, and it will be possible to construct new potentials, which will be matched with the width 31 keV, as shown in Table 11.11, Ref. 9. The use of such potentials can change the results for the resonance cross sections, reducing their influence to the total summed calculated cross sections in the energy range 40–60 keV. We will use the resonance potential of the $P_{5/2}$ scattering wave in the form

$$V_0 = 3555.983 \text{ MeV}, \ \alpha = 13.0 \text{ fm}^{-2}. \qquad (26)$$

This potential, as before, leads to the resonance at 475 keV, but its width is reduced until 32 keV (c.m.), in full accordance with the data listed in Table 11.11 of Ref. 9.

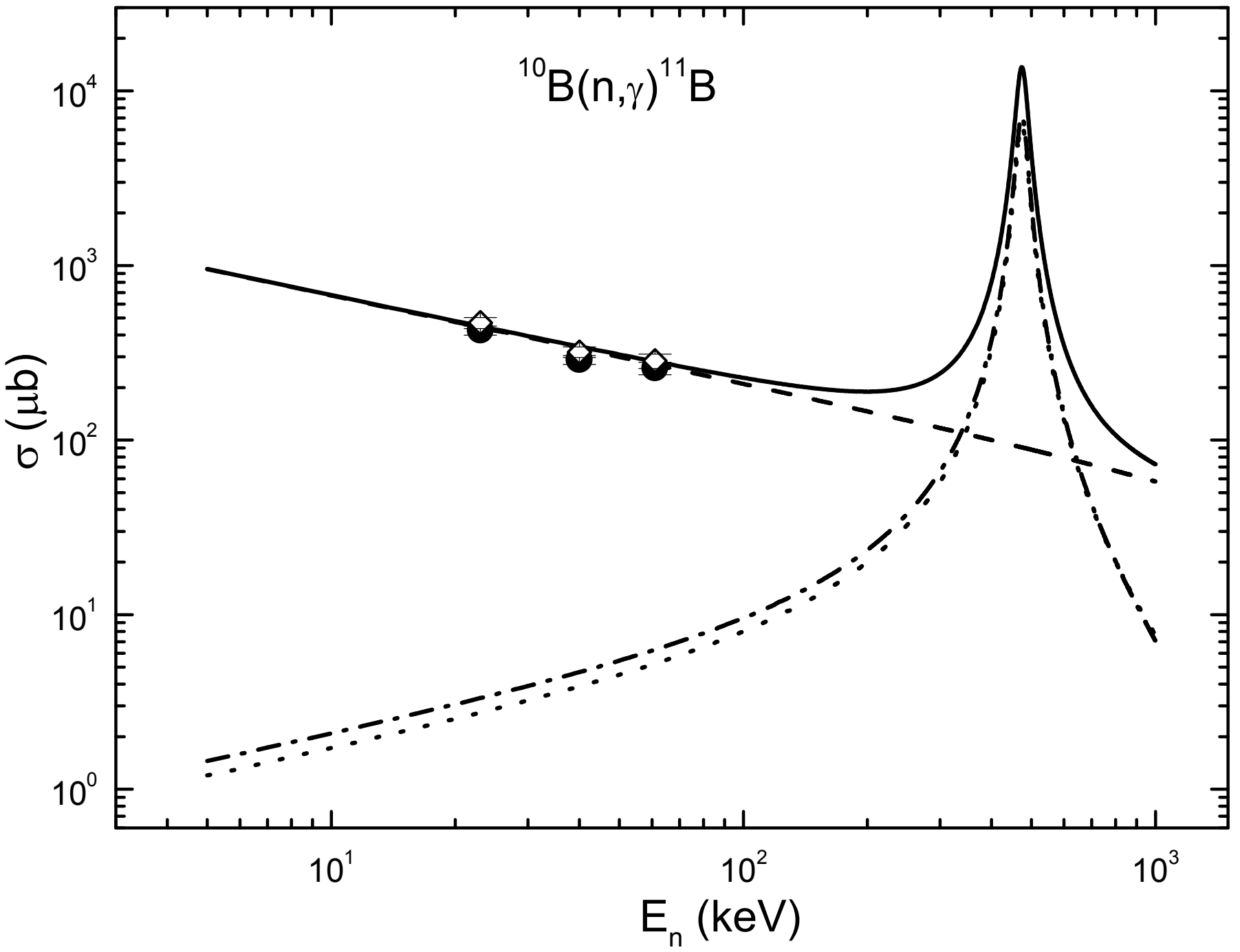

Fig. 4. The total cross sections of the radiative neutron capture on $^{10}B$. Experimental data: the same as in Fig. 3. Lines: the dashed line is the capture cross section for the E1 processes shown in Fig. 1 by the solid line, the dotted line shows the cross section of the $M1$ transition to the GS of Eq. (16) from the resonance $^6P_{5/2}$ scattering wave for potential of Eq. (26), the dot-dashed line is the cross section of the transition from the resonance $P_{5/2}$ scattering wave of Eq. (26) to the second ES with potential of Eq. (18), the solid line shows the total summed cross section.

The calculation results of the total cross sections with potential of Eq. (26) are shown in Fig. 4. The dashed line is the summed cross section for the $E1$ processes, shown in Fig. 1 by the solid line. The cross section of the $M1$ transition to the GS with potential of Eq. (16) from the resonance $^6P_{5/2}$ scattering wave for potential of Eq. (26) is shown by the dotted line, and the dot-dashed line shows the cross section from the resonance $P_{5/2}$ scattering wave of Eq. (26) to the second ES with potential of Eq. (18). The solid line is the summed cross sections for all considered transitions. All other transitions shown in Fig. 3 lead to the cross sections that do not provide an essential contribution to the total summed cross sections at the resonance energy. As can be seen from Fig. 4, the resonance part of the calculated cross section does not really change the total summed cross sections at 61 keV, which now are in a good agreement with the available experimental data of Ref. 73.

## 4. Neutron-capture reaction $^{11}$B(n, $\gamma$)$^{12}$B

Furthermore, we will consider the reaction n$^{11}$B $\rightarrow \gamma^{12}$B at thermal and astrophysical energies in the framework of the MPCM with FSs, which plays a certain role in the solar nucleosynthesis reactions.[81] In this case the potentials for intercluster interactions for scattering processes are constructed on the basis of description of the spectrum structure of the resonance states of the n$^{11}$B system. The intercluster potentials are constructed on the basis of description the binding energy of these particles in the final nucleus and certain basic characteristics of such states[39] for the bound and ground states of nuclei, generating as a result of the capture reaction in the cluster channel, which coincide with the initial particles.

### 4.1. *Structure of states in the n$^{11}$B system*

In view of absence of the complete tables for the products of Young tableaux for the systems with a number of nucleons more than eight,[65] the presented below results might be consider as the qualitative estimation of the possible orbital symmetries in the ground state of $^{12}$B for considering the n$^{11}$B channel.

At the same time, just basing on such classification we succeeded to reproduce, and, what is more important, to explain available experimental data on the radiative nucleon capture reactions in channels p$^{13}$C,[82] n$^{14}$C, and n$^{14}$N,[83] as well as a greater number of neutron radiative capture processes.[39,66] So, it is reasonable to apply our approach basing on the classification of cluster states by orbital symmetry what leads to appearing of a number of FSs and ASs in partial two-body potentials, and as a consequence relative motion WFs (in present case those refer to the neutron and $^{11}$B system) have a set of nodes.

Let us assume the {443} orbital Young tableau for $^{11}$B, then within the 1$p$-shell treating of n$^{11}$B system one has $\{1\} \times \{443\} \rightarrow \{543\} + \{444\} + \{4431\}$.[65,69] The first tableau in this product is compatible with the orbital momenta $L = 1, 2, 3, 4$, it is forbidden, as no more than four nucleons may be on $s$-shell. Second tableau is allowed and compatible with the orbital momenta $L = 0, 2, 4$, as for the third one also allowed the corresponding momenta are $L = 1, 2, 3$.[69]

It should be noted that even such qualitative analysis of orbital symmetries allow to define that there are forbidden states in $P$ and $D$ waves, and no ones in the $^3S_1$ state (here the notation $^{(2S+1)}L_J$ is used). Actually, such a structure of FSs and ASs allows one to construct two-body interaction potentials required for the calculations of total cross sections for the treating reaction.

Thereby, restricting the problem by the lower partial waves with orbital momenta

$L = 0$, 1 and 2, it can be said that for the n$^{11}$B system ($^{11}$B in the GS has quantum numbers $J^\pi$, $T$ = 3/2$^-$, 1/2)[9] interactive potential for $^3S_1$ wave should include only AS, as far as potentials for $^3P$ waves they should have both AS and FS. The AS, namely $^3P_1$, corresponds to the GS of $^{12}$B with quantum numbers $J^\pi$, $T$ = 1$^+$, 1 with the binding energy of -3.370 MeV in the n$^{11}$B channel.[9] Generally speaking, some scattering states in the n$^{11}$B channel, as well as the BS, may be mixed by channel spin $S$ = 1 or 2, but here we are assuming the spin states as pure triplet.

Let us list now excited but bound states in $^{12}$B nuclei in n$^{11}$B channel.[9] Notice, after the level energy relatively the GS, the energy relatively the threshold of the n$^{11}$B channel is pointed in brackets.

1. The first ES with $J^\pi$ = 2$^+$; 0.95 MeV (-2.4169 MeV). It can be associated with triplet the $^3P_2$ wave with the FS.

2. The second ES with $J^\pi$ = 2$^-$; 1.67 MeV (-1.6964 MeV). It can be associated with the triplet $^3D_2$ wave with the FS.

3. The third ES with $J^\pi$ = 1$^-$; 2.62 MeV (-0.7492 MeV). It can be associated with the triplet $^3S_1$ wave without the FS.

4. The fourth ES with $J^\pi$ = 0$^+$; 2.72 MeV (-0.647 MeV). It can be associated with the triplet $^3P_0$ wave with the FS.

Notice, if the bound state has a binding energy less than 1 MeV, then one may neglect the most strong $E1$ transition, cause of its minor input to the total cross section.

Consider now the spectrum of resonance states (RS) in the n$^{11}$B system,[9] i.e., states with positive energies.

1. The first RS at 20.8(5) keV with $J^\pi$ = 3$^-$ has the width less than 1.4 keV. It can be associated with the $^3D_3$ scattering wave with the bound FS.

2. The second RS at 430(10) keV with $J^\pi$ = 2$^+$ has the width of 37(5) keV. It can be associated with the $^3P_2$ scattering wave with the bound FS.

3. The third RS at 1027(11) keV with $J^\pi$ = 1$^-$ has the width of 9(4) keV. It can be associated with the $^3S_1$ scattering wave without the bound FS.

The third and following resonances lay higher than 1 MeV, so we will not consider them. There are no resonance states which may be associated with $^3S_1$ scattering wave below 1 MeV.[9] So, its phase shift is close or equal zero, and as there is no FS in this wave, the interaction potential may be regarded zero.[39]

Thus, we will consider the following electromagnetic transitions to the GS and four ESs:

$$
\begin{array}{llll}
(a) & {}^3S_1 \xrightarrow{E1} {}^3P_1 & (g) & {}^3S_1 \xrightarrow{E1} {}^3P_0 \\
(b) & {}^3D_3 \xrightarrow{E1} {}^3P_2 & (h) & {}^3P_2 \xrightarrow{E1} {}^3D_2 \\
(c1) & {}^3P_2 \xrightarrow{E1} {}^3S_1 & (i) & {}^3P_2 \xrightarrow{M1} {}^3P_1 \\
(c2) & {}^3P_1 \xrightarrow{E1} {}^3S_1 & (j) & {}^3P_2 \xrightarrow{M1} {}^3P_2 \\
(c3) & {}^3P_0 \xrightarrow{E1} {}^3S_1 & (k) & {}^3S_1 \xrightarrow{M1} {}^3S_1 \\
(f) & {}^3S_1 \xrightarrow{E1} {}^3P_2 & (l) & {}^3D_3 \xrightarrow{M1} {}^3P_2
\end{array}
$$

Detailed discussion on each transition will be given in Section 4.3.

### 4.2. *Interaction potentials*

The partial $n^{11}B$ interaction potential for each orbital momentum *L,* total momentum *J,* and parity $\pi$ as usual is presented in the Gaussian form of Eq. (9).

For the potential of the resonating $^3P_2$ wave with one BS, which is forbidden, the following parameters have been obtained basing on the $^{12}B$ spectrum as well as data on the elastic scattering in the $n^{11}B$ channel

$$V_0 = 11806.017 \text{ MeV}, \quad \alpha = 15.0 \text{ fm}^{-2}. \tag{27}$$

Within this potential the resonance energy level $E = 430$ keV was obtained and its width equals 37 keV, which completely coincide with experimental data of Ref. 9: at level energy phase shift turned to be 90.0°(1). To calculate the level width basing on the phase shift $\delta$ we used expression $\Gamma = 2(d\delta/dE)^{-1}$. Energy dependence of the $^3P_2$ phase shit is given in Fig. 5.

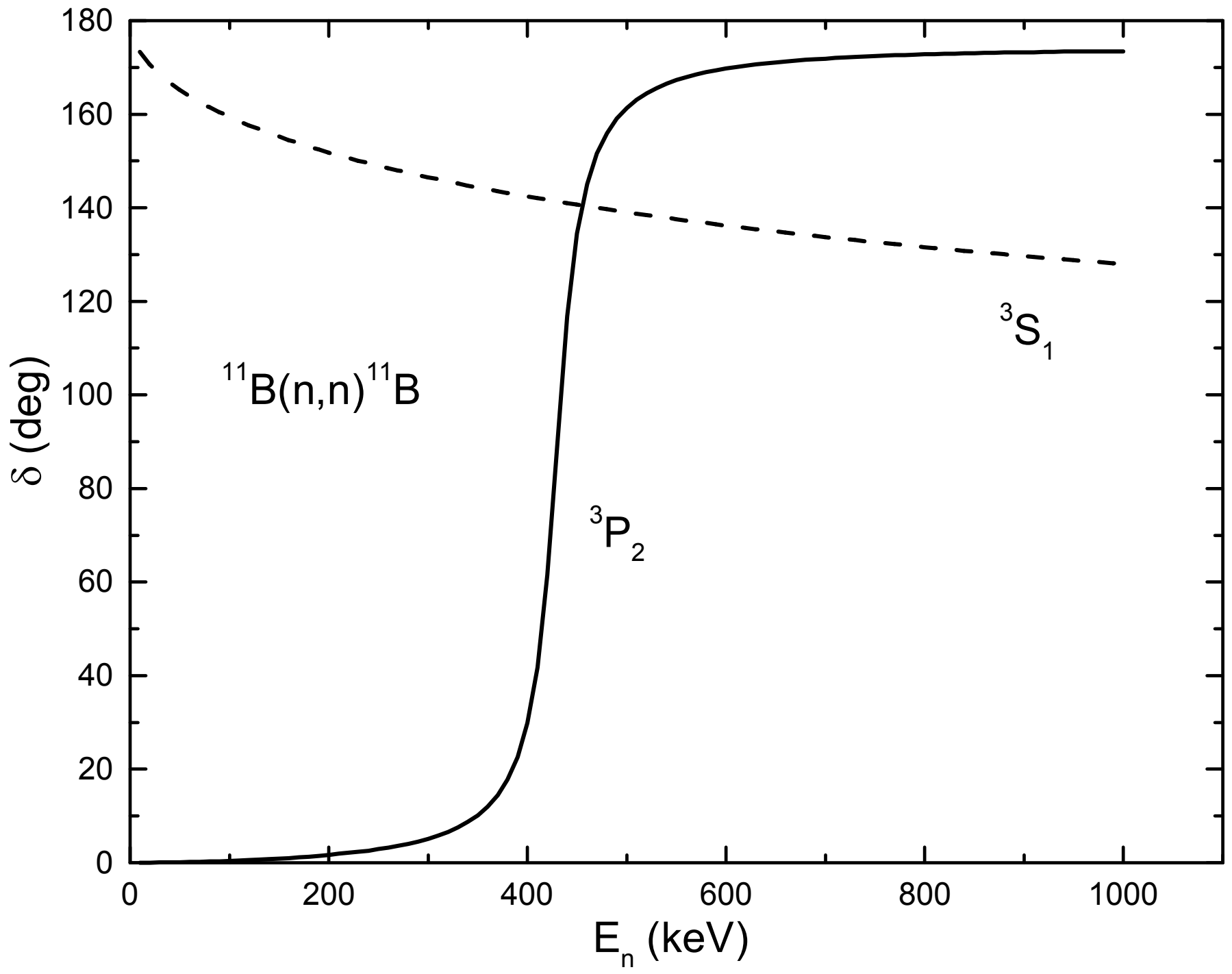

Fig. 5. The elastic scattering $^3P_2$ phase shift with resonance at 430 keV (solid line); the $^3S_1$ phase shift (dashed line).

Let us note, the analysis of resonance scattering bellow 1 MeV when the resonance width is about 10–50 keV, and number of BS is given, the interaction potential may be defined completely unambiguously. At given number of BSs the potential depth is fixed by the resonance energy of the level, and its width is defined by the resonance width.

The parameters error usually is not more than accuracy of the definition of the corresponding level and it is about 3–5%. Same remark refers to the reconstruction procedure of the partial potentials using scattering phase shifts and information of the resonances in spectrum of final nucleus.[39,58] However some ambiguity in such potential may exist.

Basing on the given above classification of states in treating system by Young tableaux the number of ASs and FSs may be defined only, but definite conclusion either AS is bound or not cannot be done. In particular, AS in $^3P_2$ scattering wave might not be necessarily bound. So, we consider all FS in scattering waves as bound, what allows exclude the pointed ambiguity.

We propose the following set of parameters for the nonresonance $^3P_0$ и $^3P_1$ scattering waves with bound FSs:

$$V_0 = 11000.0 \text{ MeV}, \ \ \alpha = 15.0 \text{ fm}^{-2}. \tag{28}$$

This potential gives scattering phase shits less than 0.5° in the energy region bellow 1.0 MeV.

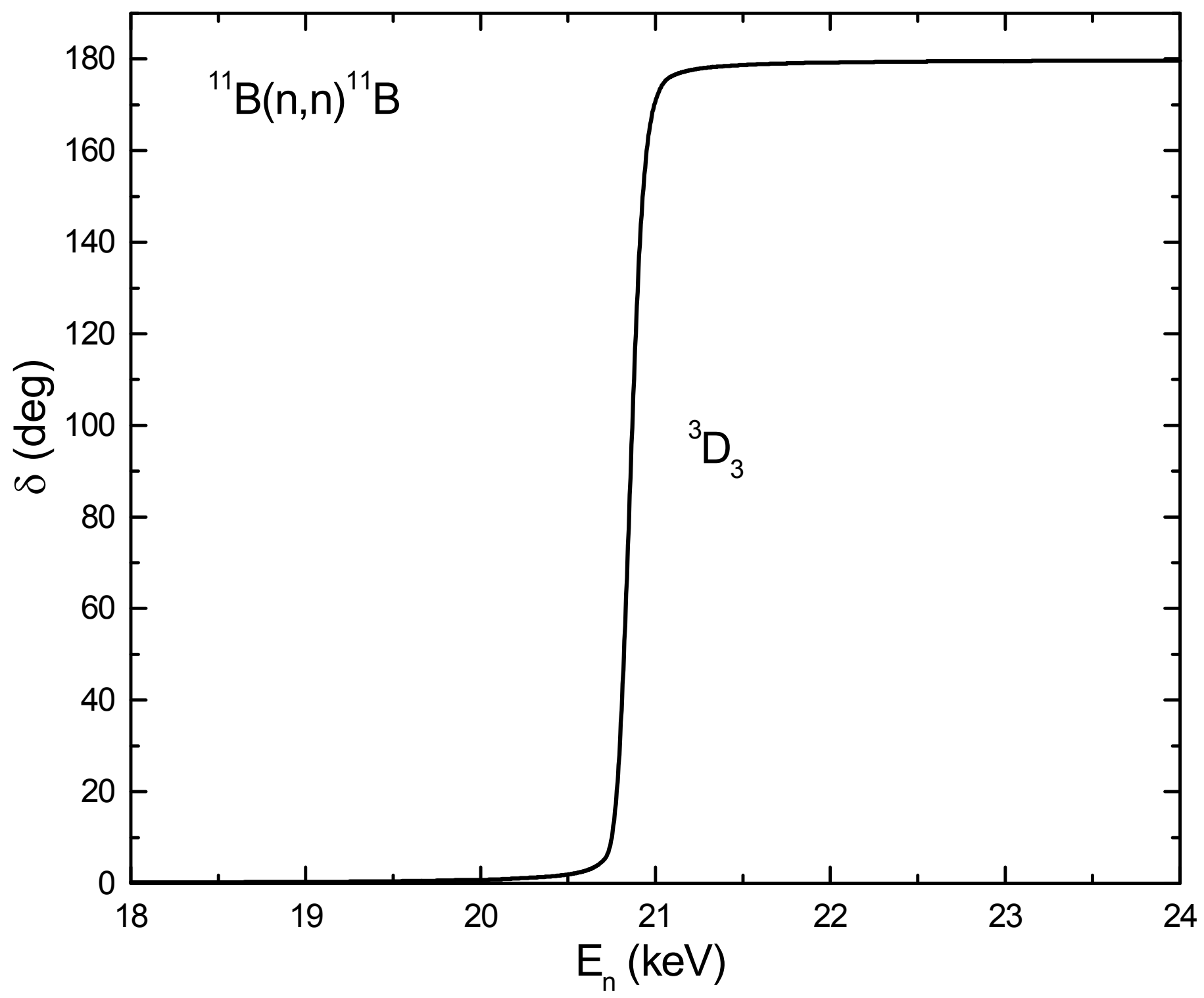

Fig. 6. The elastic scattering $^3D_3$ phase shift with resonance at 20.9(1) keV.

For the resonating $^3D_3$ wave with FS the following set of potential parameters was found:

$$V_0 = 129.305 \text{ MeV}, \ \ \alpha = 0.1 \text{ fm}^{-2}. \tag{29}$$

Fig. 6 shows corresponding phase shifts with resonance at 20.9(1) keV and width that less than 1 keV, what is in excellent coincidence with known data.[9]

One should keep in mind, that if potential contains $N + M$ forbidden and allowed states, then it obeys to the generic Levinson theorem, and the corresponding phase shifts at zero energy should start from $\pi\,(N + M)$.[37] However, on Figs. 5 and 6 for more conventional representation, the results for $P$ and $D$ phase shifts where bound FS appear are given from zero, but not from 180°. As far as the $S$ phase shift which has the bound AS is given from 180º according Levinson theorem.

The parameters set for the nonresonance $^3D_2$ and $^3D_1$ waves with FSs was used

$$V_0 = 78.0 \text{ MeV}, \ \alpha = 0.1 \text{ fm}^{-2}. \quad (30)$$

It led to the values of phase shifts less than 0.1º in the energy region bellow 1.0 MeV.

Now, let us turn to the description of the constructed for BS potentials, both for the GS, and all excited but bound states in the n$^{11}$B channel. For the ground $^3P_1$ state of $^{12}$B in the n$^{11}$B channel with FS we found:

$$V_{\rm BS} = 3183.6365 \text{ MeV}, \ \alpha = 4.0 \text{ fm}^{-2}. \quad (31)$$

This potential allows to obtain the mass radius of $R_{\rm m} = 2.36$ fm, charge radius of $R_{\rm ch} = 2.41$ fm, calculated binding energy of -3.3700 MeV (see Ref. 74) coincides with experimental value of -3.370 MeV.[9] The AC value of 0.43(1) written in the dimensionless form of Eq. (10) (see Ref. 51) was obtained over the range 2–14 fm. Error of the calculated constant is defined by its averaging over the mentioned above distance range. For the mass and charge radii of $^{11}$B value of 2.406(29) fm was used.[84] It seems that $^{12}$B radius should not differ greatly from those of $^{11}$B and $^{12}$C, for the latter it is also known and equals 2.4702(22) fm;[84] neutron mass radius was taken same as the proton one 0.8775(51) fm.[56] Note, in scattering channel the corresponding phase shift calculated with potential of Eq. (31) smoothly slows down to 178° at 1.0 MeV.

For the AC of $^{12}$B in the GS of the n$^{11}$B cluster channel the numerical value of 0.245 fm$^{-1}$ (or 0.495 fm$^{-1/2}$)[75] was obtained with accounting of identity of the nucleons (see Eq. 83b in Ref. 76). In Ref. 75 another definition of the AC was used (see Eq. (17)), what differs from the used here by a factor $\sqrt{2k_0}$ equals 0.88 fm$^{-1/2}$ for the GS, what gives $C_{\rm W} = 0.56$.

We suggest one more set of potential parameters for the GS which reproduces given above AC

$$V_{\rm BS} = 1606.331 \text{ MeV}, \ \alpha = 2.0 \text{ fm}^{-2}. \quad (32)$$

Within this potential we obtained AC = 0.55(1) over the range 2–16 fm, charge and mass radii of 2.37 fm and 2.41 fm at the binding energy of -3.3700 MeV obtained with the accuracy of 10$^{-4}$ MeV.[74] Scattering phase shift shows same behavior as in case of potential of Eq. (31).

For parameters of the $^3P_2$ potential with FS corresponding to the first ES of $^{12}$B ($J^\pi = 2^+$) in the n$^{11}$B channel lying at 0.95 MeV the following values have been found

$$V_{\rm BS} = 3174.75797 \text{ MeV and } \alpha = 4.0 \text{ fm}^{-2}. \quad (33)$$

This potential gives the binding energy of -2.4169 MeV with the accuracy of $\varepsilon = 10^{-4}$ quite well coinciding with the experimental value of -2.41686 MeV,[9] the charge radius of 2.41 fm and the AC = 0.38(1) over the range 2–14 fm. Ref. 75 reported the AC values of 0.098 fm$^{-1}$ or 0.313 fm$^{-1/2}$, and recalculation to the dimensionless AC with $\sqrt{2k_0} = 0.81$ fm$^{-1/2}$ it turned to be equal 0.386 what is in agreement with the AC obtained with potential of Eq. (33). Corresponding phase shift smoothly decreases up to 177° at 1.0 MeV.

For parameters of the $^3D_2$ potential with FS corresponding to the second ES of $^{12}$B ($J^\pi = 2^-$) in the n$^{11}$B channel lying at 1.67 MeV the following values have been found

$$V_{\rm BS} = 5187.0744 \text{ MeV}, \ \ \alpha = 4.0 \text{ fm}^{-2}. \qquad (34)$$

They give the binding energy of -1.6964 MeV with the accuracy of $\varepsilon = 10^{-4}$, what is in very good agreement with the experimental value of -1.69635 MeV (see Ref. 9), the charge radius of 2.41 fm and the mass radius of 2.33 fm, the AC = 0.033(1) over the range 2–12 fm. The phase shift is equal to 180° in the energy interval from zero up to 1.0 MeV.

It should be noted, this excited state may be associated also with the $^5S_2$ wave without FS but with the spin channel $S = 2$. We do not examine this state here, however the corresponding potential was found

$$V_{\rm BS} = 279.6746 \text{ MeV}, \ \ \alpha = 4.0 \text{ fm}^{-2}. \qquad (35)$$

This potential gives the characteristics of the level very close to those are coming from potential of Eq. (34): the binding energy of -1.6964 MeV with the accuracy of $\varepsilon = 10^{-4}$, the charge and the mass radii of 2.42 fm and 2.45 fm respectively, but the AC = 1.10(1) over the range 2–25 fm. The scattering phase shift is less than 0.1° up to 1.0 MeV. As it is clear these two options give considerably different values for the asymptotic constants. Thus, it is interesting proposal for the experimental measurement of AC (for example, as it was done in Ref. 75) with a view to conclude either this exited level belongs to the $^3D_2$ or to the $^5S_2$ waves.

For parameters of the $^3S_1$ potential without FS corresponding to the third ES of $^{12}$B ($J^\pi = 1^-$) in the n$^{11}$B channel lying at 2.62 MeV the following values have been found

$$V_{\rm BS} = 266.3015 \text{ MeV}, \ \ \alpha = 4.0 \text{ fm}^{-2}. \qquad (36)$$

They give the binding energy of -0.7492 MeV with the accuracy of $\varepsilon = 10^{-4}$, what coincides with the experimental value from Ref. 9, the charge radius of 2.43 fm, and the AC equals 1.07(1) over the range 2–30 fm. Fig. 5 shows the corresponding phase shift (dashed line).

For parameters of the $^3P_0$ potential with FS corresponding to the fourth ES of $^{12}$B ($J^\pi = 0^+$) in the n$^{11}$B channel the following values have been found

$$V_{\rm BS} = 3156.9385 \text{ MeV}, \ \ \alpha = 4.0 \text{ fm}^{-2}. \qquad (37)$$

It gives the binding energy of -0.6470 MeV with the accuracy of $\varepsilon = 10^{-4}$, what coincides with experimental value from Ref. 9, the charge radius 2.41 fm, and the AC = 0.25(1) over the range 2–24 fm. In the energy interval from zero up to 1.0 MeV corresponding phase shift is decreasing up to 175°.

### 4.3. *The total cross section of the radiative neutron capture on $^{11}B$*

First, the most strong dipole electric $E1$ transition $^3S_1 \xrightarrow{E1} {}^3P_1$ to the GS, i.e., the (n, $\gamma_0$) process, was treated with zero potential for the scattering $^3S_1$ wave and set of Eq. (31) for the final nucleus. The corresponding cross section is given in Fig. 7 by the solid line. It is seen that within this potential experimental data at 25 meV (closed triangles) and in the energy range 23–61 keV (open circles) have been reproduced reasonably well.

For comparison in Fig. 7 by the dashed line similar results for the (n, $\gamma_0$) process are given being obtained with the GS potential of Eq. (32), which gives the AC of 0.55 what coincides with the results of Ref. 75. In this case calculated cross sections more or less describe within the error bars experimental points at 23–61 keV, but fail in description of the crucial point at 25 meV.[73,79] Therefore, we concluded that the GS potential of Eq. (31) is more appropriate one for study of the (n, $\gamma_0$) cross section – its parameters were varied exclusively for correct description of the cross section for 25 meV.[79] With the same parameters, without any additional variations or adjustments, the measurements data for the transition to the GS at the energy range 23–61 keV were obtained.[73] In other words, just the GS potential of Eq. (31), because the $^3S_1$ potential equals zero, completely determines the slope and location of the line in Fig. 7, shown the calculated cross section of the $^3S_1 \rightarrow {}^3P_1$ transition. Difference in AC of Ref. 75 and those calculated with potential of Eq. (31) is not very considerable. Note, Ref. 75 was published 35 years ago, so it seems reasonable to refine a value of AC in the future.

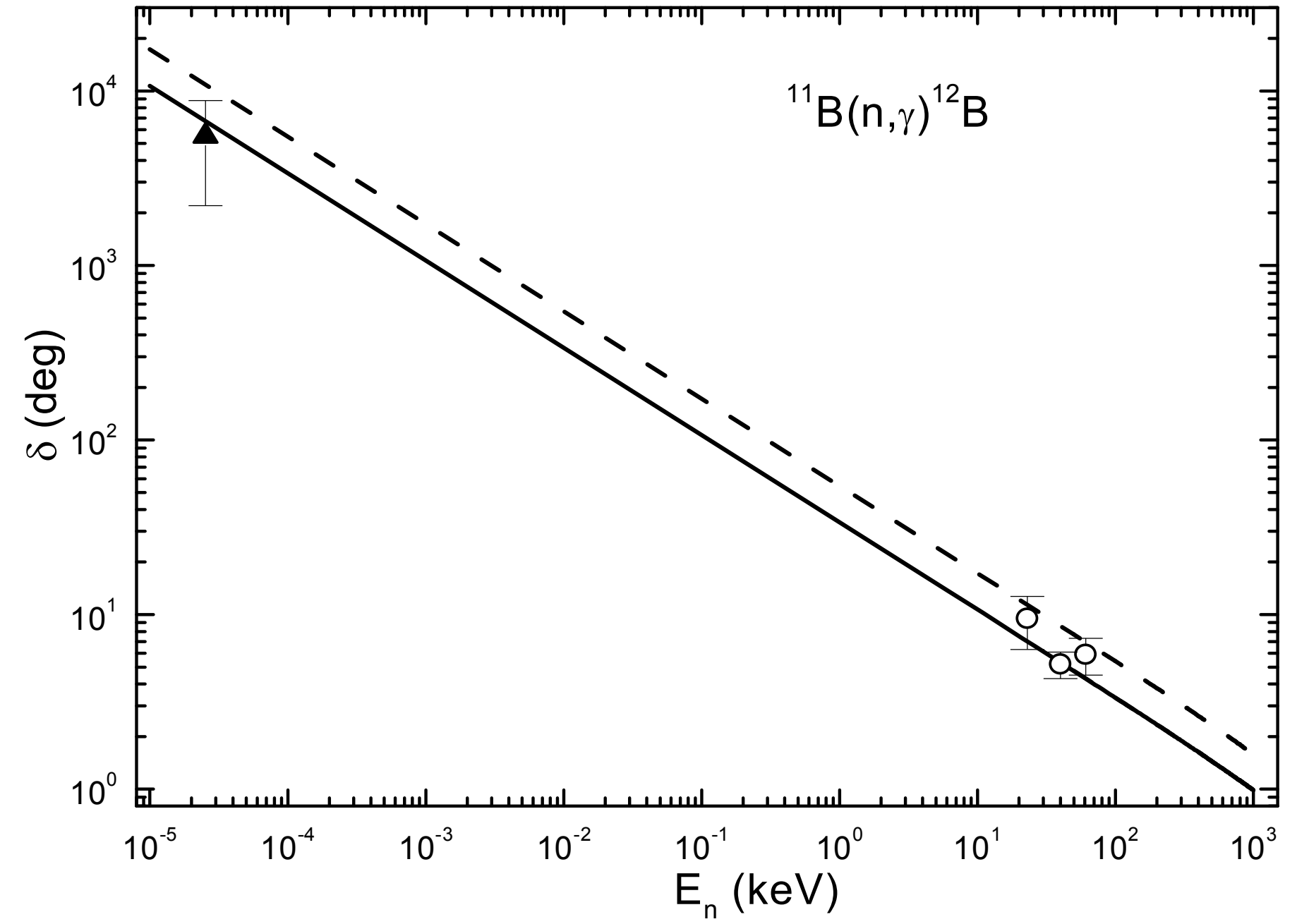

Fig. 7. The total cross sections of the $^{11}$B(n, $\gamma_0$)$^{12}$B reaction. Experimental data: ▲ – Ref. 79, o – Ref. 73. Lines correspond to the calculation of the total cross sections for the transitions to the GS with potentials given in Sect. 4: potential of Eq. (31) – solid line; potential of Eq. (32) – dashed line.

Let us turn now to analysis of the total cross sections of $^{11}$B(n, $\gamma$)$^{12}$B reaction in the energy range 1–1000 keV and pay special attention to the signature of the resonances at 21 and 430 keV. In Fig. 8 we compare the available experimental data and present calculations with account of $E1$ transition amplitudes ($a$)-($g$) listed in Sect. 4.1.

On the background of ${}^3S_1 \xrightarrow{E1} {}^3P_1$ transition to the GS (long dashes) fist resonance appears at 20.8 keV due to ${}^3D_3 \xrightarrow{E1} {}^3P_2$ transition from the resonating ${}^3D_3$ wave to the first ES ${}^3P_2$ (dot dashed line). For completeness let us remark that similar transitions from nonresonance ${}^3D_2$ and ${}^3D_1$ waves led to the cross sections less than $10^{-3}$ μb at 1.0 MeV according our calculations, so they give minor input on the whole.

Transition ${}^3P_2 \xrightarrow{E1} {}^3S_1$ to the third ${}^3S_1$ ES from the resonating ${}^3P_2$ scattering wave is leading to the cross section at 430 keV higher than 1mb (dot curve in Fig. 8). Input of nonresonance partial cross sections ${}^3P_1 \xrightarrow{E1} {}^3S_1$ and ${}^3P_0 \xrightarrow{E1} {}^3S_1$ is not essential (dot-dot-dashed line in Fig. 8).

Partial cross section corresponding to the ${}^3S_1 \xrightarrow{E1} {}^3P_2$ transition to the first ES (densely dots) is parallel to the line conforming to the ${}^3S_1 \xrightarrow{E1} {}^3P_1$ transition to the GS, and transition to the fourth ES ${}^3S_1 \xrightarrow{E1} {}^3P_0$ have been also calculated, but as it seen in Fig. 8 (short dashed line) it is practically insignificant.

Finally, the total sum of partial cross sections is shown by the solid line in Fig. 8. Experimental data for the transitions to the GS are taken from Ref. 73 (open circles), and data for the sum of partial cross sections are taken from Refs. 73, 85, and 86.

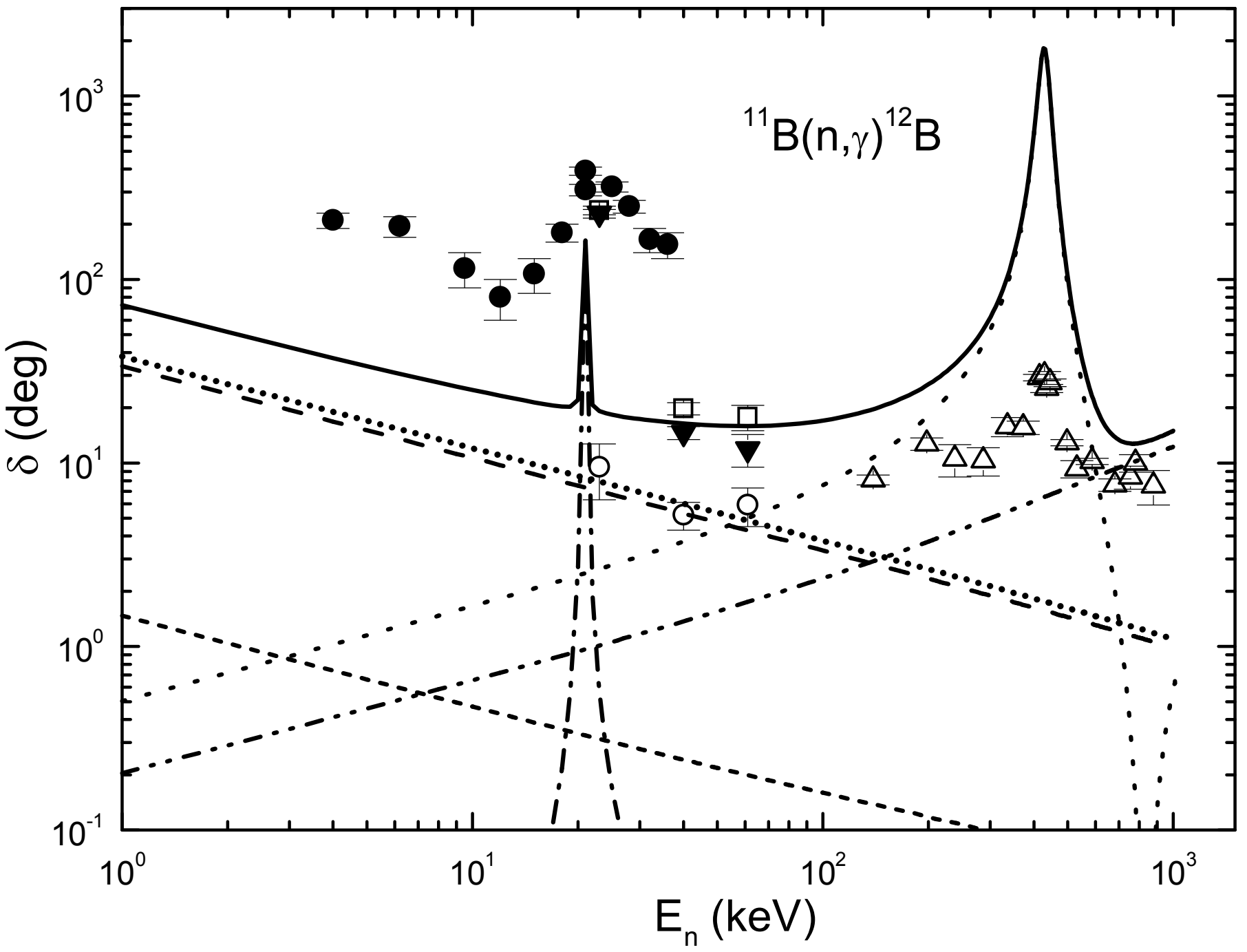

Fig. 8. The total cross sections of the $^{11}$B(n, γ)$^{12}$B reaction in the energy range 1–1000 keV. including ($a$)–($g$) amplitudes. Experimental data: ▼ – transition to the second ES lying at -1.6964 MeV, Ref. 73; ● – Ref. 85; □ – sum of partial cross sections from Ref. 73; Δ – Ref. 86; o – capture to the GS, Ref. 73. Commentaries for the calculated lines are given in the text.

Let us present another to somewhat extent alternative treatment of $^{11}$B(n, γ)$^{12}$B reaction in the energy range 1–1000 keV illustrated by Fig. 9. There transitions ($a$), ($b$), and ($f$) are preserved with same notations as in Fig. 8. Additional transitions ($h$) - ($l$) have been examined. So, instead of the process ($c1$) ${}^3P_2 \xrightarrow{E1} {}^3S_1$ transition ($h$) ${}^3P_2 \xrightarrow{E1} {}^3D_2$ was taken into account (dotted line in Fig. 9). Solid line is the result of the summation of all mentioned partial cross sections.

In Fig. 9 the transition ${}^3P_2 \xrightarrow{M1} {}^3P_1$ which contributes to the energy resonance region at 430 keV is shown by densely dashes, and gives 77.5 μb in maximum. For comparison our estimations for $M1$ capture from ${}^3P_1$ and ${}^3P_0$ waves with potentials of Eq. (28) give $10^{-2}$ μb what is negligibly small.

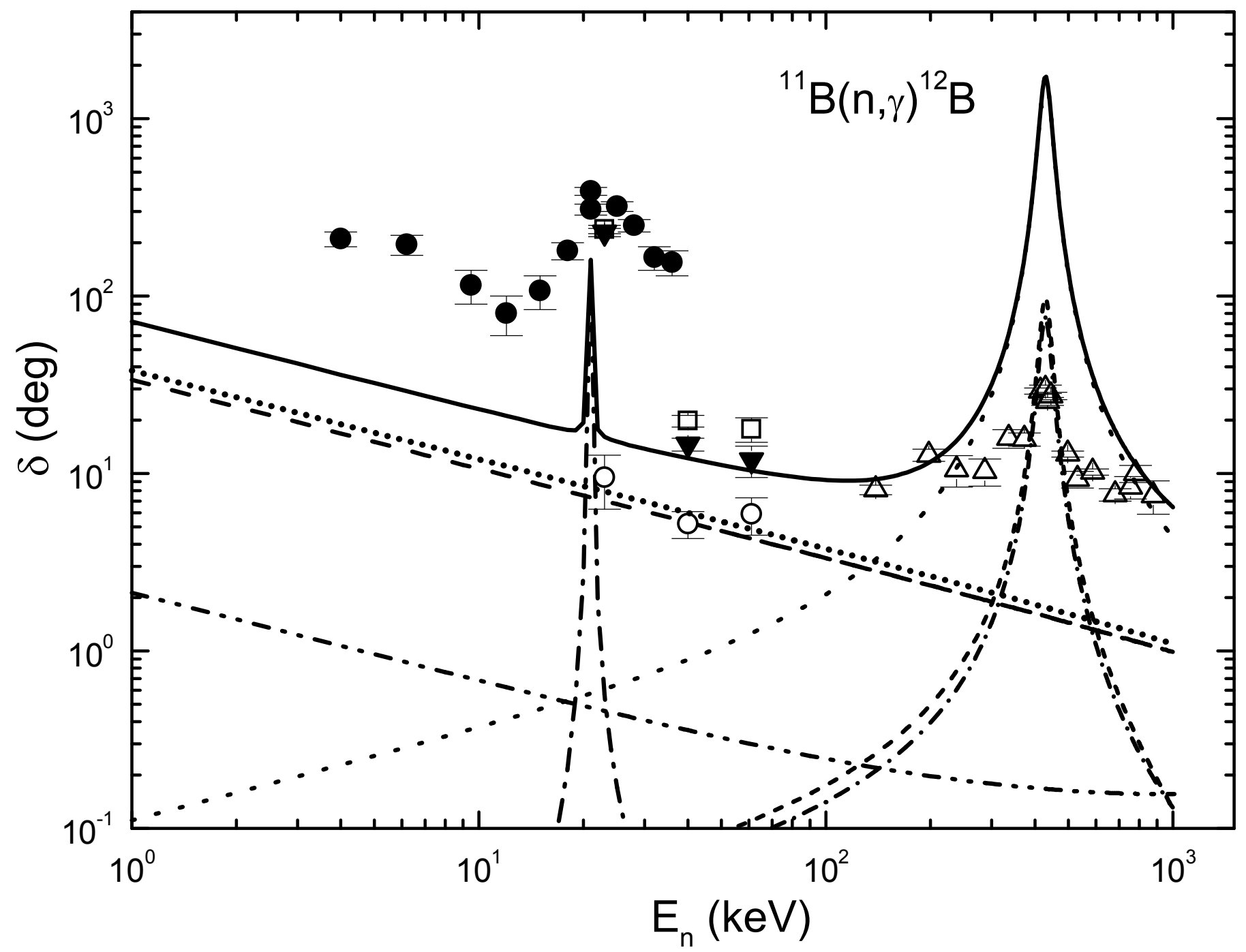

Fig. 9. The total cross sections of the $^{11}B(n, \gamma)^{12}B$ reaction in the energy range 1–1000 keV. including ($h$)–($l$) amplitudes. Experimental data: ▼ – transition to the second ES lying at -1.6964 MeV, Ref. 73; ● – Ref. 85; □ – sum of partial cross sections from Ref. 73; Δ – Ref. 86; o – capture to the GS, Ref. 73. Commentaries for the calculated curves are given in the text.

One more resonating transition ${}^3P_2 \xrightarrow{M1} {}^3P_2$ is observed in the region of 430 keV with value for the cross section 98 μb, what is in sum with the last transition amount less than 10% comparing the treated above $E1$ process ${}^3P_2 \xrightarrow{E1} {}^3D_2$.

We included also in consideration the ${}^3S_1 \xrightarrow{M1} {}^3S_1$ process (dot-dot-dashed line in Fig. 9) and found it to be near order less comparing the cross section of $E1$ transition from the same scattering wave to the GS.

Resonating capture ${}^3D_3 \xrightarrow{M1} {}^3P_2$ cross section is near 5 μb and does not change the picture at the resonance energy 20.8 keV, as the $E1$ process ($b$) gives for the corresponding cross section a value equals to 144 μb. In addition all $M1$ transitions from the nonresonance ${}^3D$ scattering waves to the second ES ${}^3D_2$ have been evaluated and turned to be less than $10^{-5}$ μb.

Summarizing the results of Section 5, we may conclude that according Fig. 7 that cross sections at 25 meV and 23–61 keV corresponding to the transitions onto the GS have been reproduced well and fit experimental data (Refs. 73 and 79). Same statement valid for the cross sections at 21 and 430 keV resonances, but as Fig. 8 shows in the nonresonance energy region experimental data of Ref. 85 are order of magnitude greater than the calculated ones. There it should be noted, that as AC (this is input information for the construction of interaction potentials) are known for the GS and the first ES only, then results obtained for the second, third, and fourth excited states might be regarded as preliminary ones. Moreover, all experimental investigations on this channel (Refs. 73, 85 and 86) were performed in 60-ies of the last century and, apparently, to be specified.

Finally, let us indicate one important predictive opportunity of developed here approach (see also Refs. 73, 85 and 86). As the calculated cross-section is practically the straight line in the energy range from 10 meV to 10 keV (see solid line in Fig. 7), then it may be approximated by the simple function of the form of Eq. (24). The given constant value $A = 33.8364$ μb·keV $^{1/2}$ has been defined over one point in the cross-sections at the minimal energy equals 10 meV.

Relative modulus deviation of Eq. (25) of the calculated theoretical cross-section ($\sigma_{theor}$) and approximation of this cross-section by the given above function ($\sigma_{ap}$) at the energies up to 10 keV is about 0.3%. If it is assumed, that this form of the energy dependence of the total cross-section will be also conserved at lower energies, then one may estimate the cross-section value which, for example, at the energy of 1 μeV is equal to 1.1 barn.

## 5. Proton-capture reaction $^{11}$B(p, γ)$^{12}$C

### 5.1. *Structure of states in the p$^{11}$B system*

In view of absence of the complete tables for the products of Young tableaux for systems with a number of nucleons more than eight,[65] the presented below results might be consider as the qualitative estimation of the possible orbital symmetries in the GS of $^{12}$C in the p$^{11}$B cluster channel.

At the same time, just basing on such classification we succeeded to reproduce, and, what is more important, to explain available experimental data on the radiative nucleon capture reactions in the p$^{12}$C (see Ref. 38) and p$^{13}$C (see Ref. 82) channels. So, it is reasonable to apply our approach basing on the classification of cluster states by orbital symmetry what leads to appearing of a number of FSs and ASs in partial two-body potentials, and as a consequence relative motion WFs (in present case those refer to the proton and the $^{11}$B system) have a set of nodes.

Let us assume the {443} orbital Young tableau for $^{11}$B, then within the 1*p*-shell treating of the p$^{11}$B system one has {1} × {443} → {543} + {444} + {4431}.[65] The first tableau in this product is compatible with the orbital momenta $L = 1, 2, 3, 4$, it is forbidden, as no more than four nucleons may be on *s*-shell. The second tableau is allowed and compatible with the orbital momenta $L = 0, 2, 4$, as for the third one also allowed the corresponding momenta are $L = 1, 2, 3$.[69]

It should be noted that even such qualitative analysis of orbital symmetries allow one to define that there are forbidden states in the *P* and *D* waves, and no ones in the *S* state. Actually, such a structure of FSs and ASs allows one to construct two-body interaction potentials required for the calculations of total cross sections for the treating reaction.

Thereby, just restricting by the lowest partial waves with orbital angular moments $L = 0$ and 1 one may state that for the p$^{11}$B system ($^{11}$B in the GS has quantum numbers $J^{\pi}$, $T = 3/2^{-}, 1/2$)[9] the $^3S_1$ potential (here the notation $^{(2S+1)}L_J$ is used) has the AS only, which may be not bound and lay in continuous spectrum, as far as FS it is absent. Each of $^3P$ waves has bound forbidden and allowed state. One of them corresponds to the $^3P_0$ GS of $^{12}$C with quantum numbers $J^{\pi}$, $T = 0^{+}$, 0 and the binding energy of -15.9572 MeV in the p$^{11}$B channel.[9] Others triplet $^3P$ states have bound FSs, but may also have ASs in continuous spectrum. Besides, there is mixture by spins $S = 1$ and 2 may also appear both in scattering and bound states in the p$^{11}$B system.

Besides, the GS let us treat two more resonance states appearing in the p$^{11}$B system at positive energies:

1. First resonance state appears at 162 keV (l.s.) or 148.6(4) keV in c.m. and has a width less than 5.3(2) keV in c.m., and quantum numbers $J^{\pi} = 2^{+}$ (Tables 12.11 and 12.6 in Ref. 9). It corresponds to 16.1058(7) MeV level in $^{12}$C and may be compare to mixed in spin $^{3+5}P_2$ wave with FS.

2. Second resonance state appearing at 675 keV (l.s.) has the width 300 keV in c.m. and quantum numbers $J^{\pi} = 2^{-}$.[9] It corresponds to the level with 16.576 MeV and might be associated with $^5S_2$ scattering wave without FS.

The next resonance is at 1.388 MeV, i.e., higher than 1 MeV, so it will not be considered. In the spectrum of $^{12}$C below 1 MeV there are no resonance levels which may be correlated to the $^3S$ or $^5S$ scattering resonances.[9] That is why the corresponding phase shifts may be taken close to zero. There are no FS in the $S$ wave, so the potentials for both spin channels $S = 1$ and $S = 2$ may be regarded zero also.[39] Thus, the minimum set of electromagnetic transitions to the GS will be considered in order to reproduce the energy dependence of the measured total cross sections.

So far as the GS of $^{12}$C is the $^3P_0$ level it is actual to include in consideration dipole electric $E1$ transition from the nonresonance $^3S_1$ scattering wave with zero interaction potential to the GS:

1. $^3S_1 \to {}^3P_0$.

Besides, the $E2$ transition from the triplet part of the $^3P_2$ scattering wave to the GS also should be evaluated as it has the resonance at 162 keV:

2. $^3P_2 \to {}^3P_0$.

There are a lot of details on the methods of calculation of cross sections in the framework of MPCM,[38,39,49,52,53] so we are dropping this part here. Same concerns the methods of construction of two-body interaction cluster potentials at the given orbital momentum $L$, so we refer to Refs. 38, 39 and 58.

## 5.2. *Interaction potentials*

Let us now give the corresponding parameters for the GS and two resonating states of

$^{12}C$ presented as the $p^{11}B$ system. For all $p^{11}B$ potentials the Gaussian form of Eq. (9) was used. The following parameters have been found for the resonating $^{3+5}P_2$ wave with FS and $J = 2^+$

$$V_0 = 24.38058 \text{ MeV}, \ \alpha = 0.025 \text{ fm}^{-2}. \qquad (38)$$

Within this potential the resonance energy level of $E$ = 162.0(1) keV (l.s.) and its width equals 0.8(1) keV (c.m.), which, in general, coincide with the experimental data of Ref. 9 (it should be noted that this ref. gives the proton width of this level of 0.0217(18) keV (l.s.) data have been reproduced exactly: at level energy phase shift turned to be 90.0°(1). To calculate the level width basing on the phase shift we used expression $\Gamma = 2(d\delta/dE)^{-1}$. The energy dependence of the $^{3+5}P_2$ phase shift is given in Fig. 10.

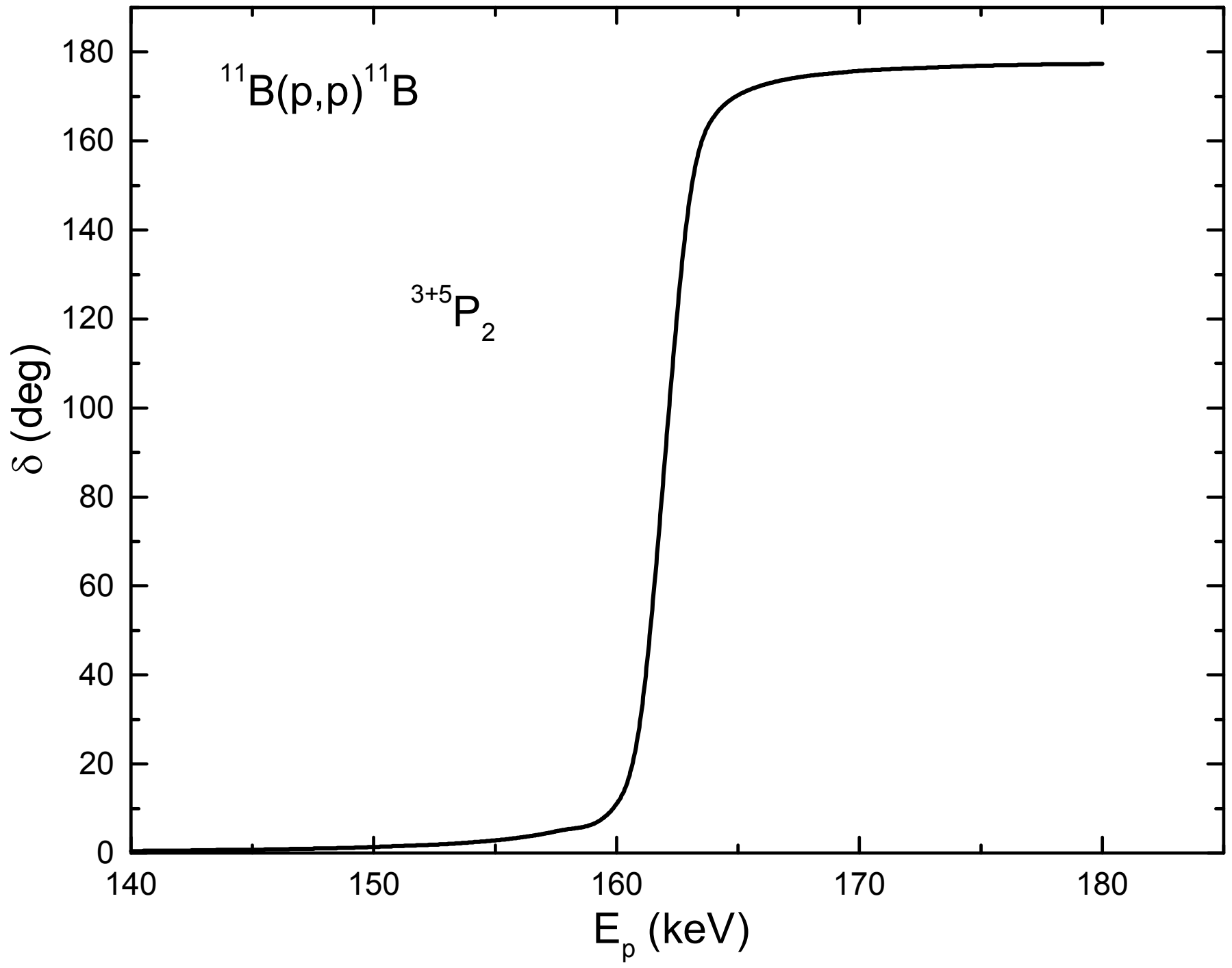

Fig. 10. The $p^{11}B$ elastic scattering $^5P_2$ phase shift with the resonance at 162 keV.

One should keep in mind, that if potential contains $N + M$ forbidden and allowed states, then it obeys to the generic Levinson theorem,[37] and the corresponding phase shifts at zero energy should start from $\pi\cdot(N + M)$ or from 180º in present case as the FS is bound, but the AS is not.

However, in Fig. 10 for more conventional representation, the results for $^{3+5}P_2$ phase shifts are given from zero, but not from 180º. Let us note, that using model does not allow to separate $^3P$ and $^5P$ parts of such a potential, and that is why the $P_2$ potential with spin mixture has been obtained. Further down we are going to use the potential of Eq. (38) both for $^3P_2$ and $^5P_2$ scattering states in the calculations of $M1$ or $EJ$ transitions.

Let us note, the analysis of resonance scattering bellow 1 MeV, when the number of BS is given, the interaction potential may be defined completely unambiguously. At the given number of BSs the potential depth is fixed by the resonance energy of the level, and its width is defined by the resonance width. The parameters error usually not more than accuracy of the definition of the corresponding level and it is about 3–5%. Same remark refers to the reconstruction procedure of the partial potentials using scattering phase shifts and information of the resonances in spectrum of final nucleus.[39]

For the nonresonance $^3P_0$ and $^{3+5}P_1$ scattering waves with FSs the following parameters are suggested

$$V_0 = 60.0 \text{ MeV}, \alpha = 0.1 \text{ fm}^{-2}. \tag{39}$$

This potential leads to the phase shifts close to 180(1)º in the energy range from zero up to 1.0 MeV and has one bound FS.

The interaction potential of the nonresonance process may be also performed unambiguously basing on the data for the scattering phase shifts and number of the allowed and forbidden states in the bound channels. The accuracy of the definition of the potential parameters depends on the accuracy of the corresponding phase shifts extracted from the scattering data and may be of 20–30%. But such a potential has no ambiguities as its depth is fixed by the number of BSs coming from the classification by Young tableaux, and its width is defined by the phase shifts shape.

While constructing the nonresonance scattering potential basing on the data for the nuclear energy spectra it is much more difficult to estimate the accuracy of its parameters definition even when the number of BSs is fixed. Usually such a potential should give the scattering phase shift close to zero bellow 1 MeV, or should give smoothly drop shape of the phase, because there are no resonance levels in the nucleus spectrum.[39]

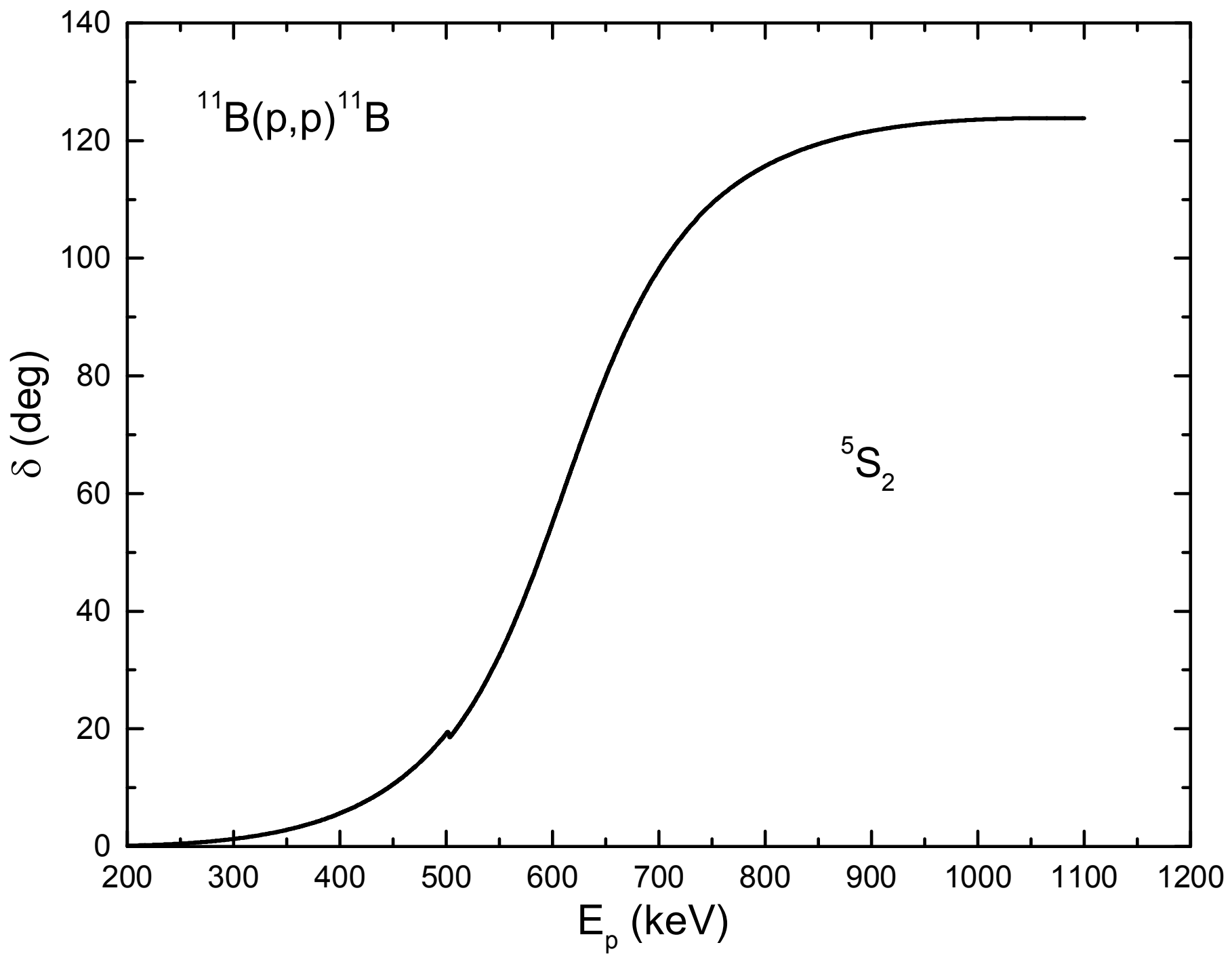

Fig. 11. The p$^{11}$B elastic scattering $^5S_2$ phase shift with the resonance at 675 keV.

For the resonating $^5S_2$ scattering wave without FS with $J = 2^-$ the following parameters are suggested:

$$V_0 = 10.9256\ \text{MeV},\ \ \alpha = 0.08\ \text{fm}^{-2}, \tag{40}$$

which lead to the phase shifts presented in Fig. 11. This potential reveals the resonance at 675(1) keV (l.s.) with the width of 289(1) keV (c.m.) what is in agreement of data in Ref. 9, especially if keep in mind that given their value for the proton width is 150 keV (l.s.)

For the potential corresponding to the $^3P_0$ GS of $^{12}$C with FS in the p$^{11}$B cluster channel the following parameters were found:

$$V_0 = 142.21387\ \text{MeV},\ \alpha = 0.1\ \text{fm}^{-2}. \tag{41}$$

This potential allows to obtain the mass radius of $R_m = 2.51$ fm, the charge radius of $R_{ch} = 2.59$ fm, the calculated binding energy of -15.95720 MeV with an accuracy of $\varepsilon = 10^{-5}$ MeV.[74] For the AC written in the dimensionless form of Eq. (10) (see Ref. 51) value of 23.9(2) was obtained over the range 7–13 fm. Error of the calculated constant is defined by its averaging over the mentioned above distance range. The value of the mass and charge radii of $^{11}$B equals 2.406(29) fm (see Ref. 84) was used, the radius of $^{12}$C was taken as 2.4702(22) fm, the charge and mass proton radius is equal to 0.8775(51) fm.[87]

Scattering phase shift for such a potential does not exceed 0.1° at 1.0 MeV. As there is no any information on the AC in the p$^{11}$B channel in the GS potential of Eq. (41) was constructed basing on the demands of reproducing of the nonresonance part of the total cross sections exclusively. It is clear that we should take into account the corresponding error bars, so in this very case the accuracy for the parameter definition pretends onto 10% near. Let us emphasized, that in discussed case no independent data on ACs have been found.

### 5.3. *Total cross sections*

First, the dipole electric $E1$ transition $^3S_1 \rightarrow {}^3P_0$ was treated when capture occurs from the $^3S_1$ scattering wave with zero interaction central potential to the GS $^3P_0$ state constructed with potential of Eq. (41). The calculated cross section is given by the dashed line in Fig. 12 in the energy range 50 keV–1.5 MeV. It fits well the nonresonance part of experimental data from Refs. 7, 88, 89, and 90 known in the range 80–1500 keV.

Second, the quadrupole electric $E2$ transition from the resonating $^3P_2$ wave at 162 keV to the GS was calculated. The sum of these two partial cross sections is shown by the solid line in Fig. 12, and well reproducing of the experimental data is obviously seen. At a time, it should be noted that calculated value of the cross section at the resonance energy of 162 keV equals 101 $\mu b$ is essentially higher the measured one equals 5.5 $\mu b$. However, experimental data have been obtained at 163 keV, and as the resonance has very narrow width then the energy change even within one keV may lead to such an essential difference.

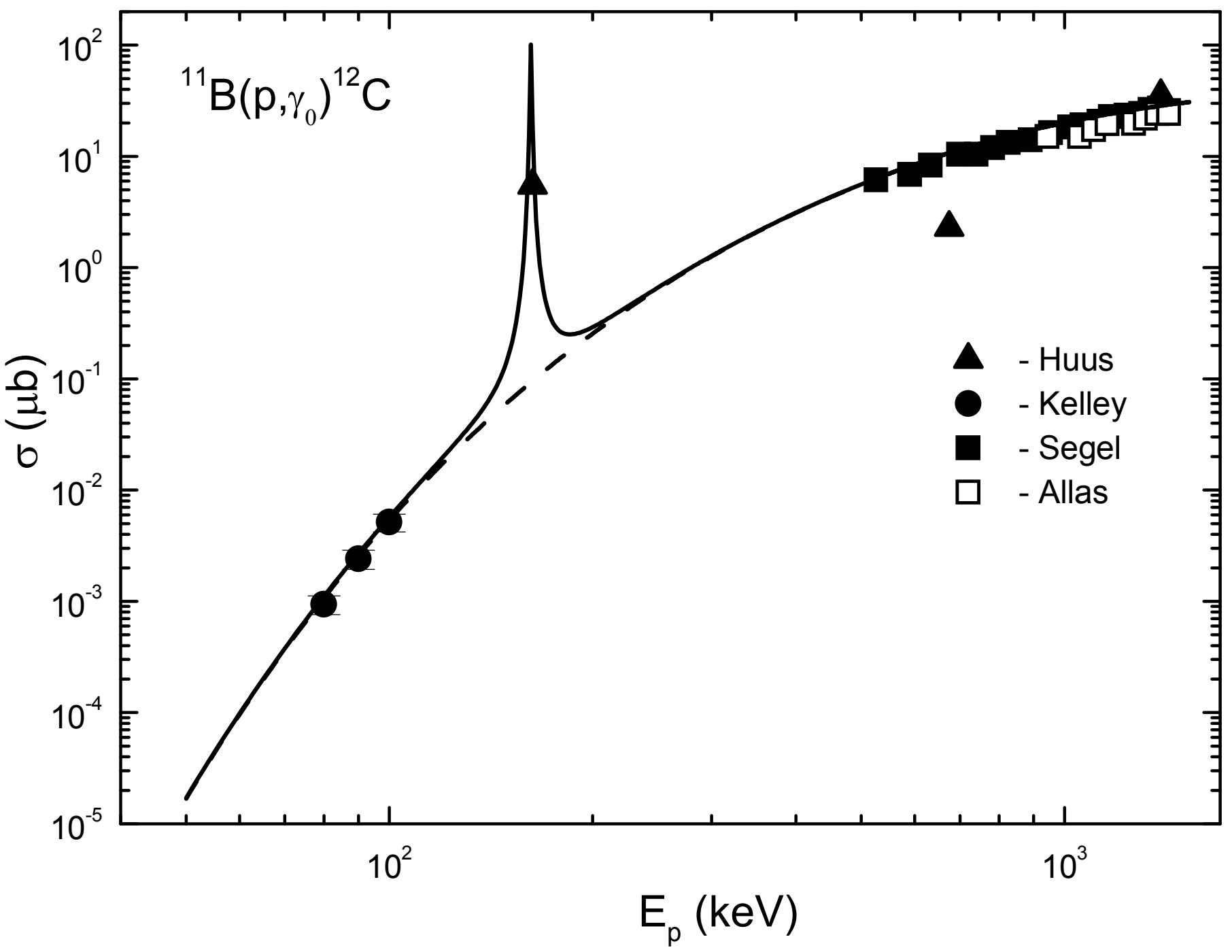

Fig. 12. The total cross sections of the proton radiative capture on $^{11}$Be to the GS in the energy range 50–1.5 $10^3$ keV. Experiment: black triangles (▲) – Ref. 88, dots (●) – Ref. 7, open squares (□) – sum of cross sections from Ref. 90, black squares (■) – Ref. 89.

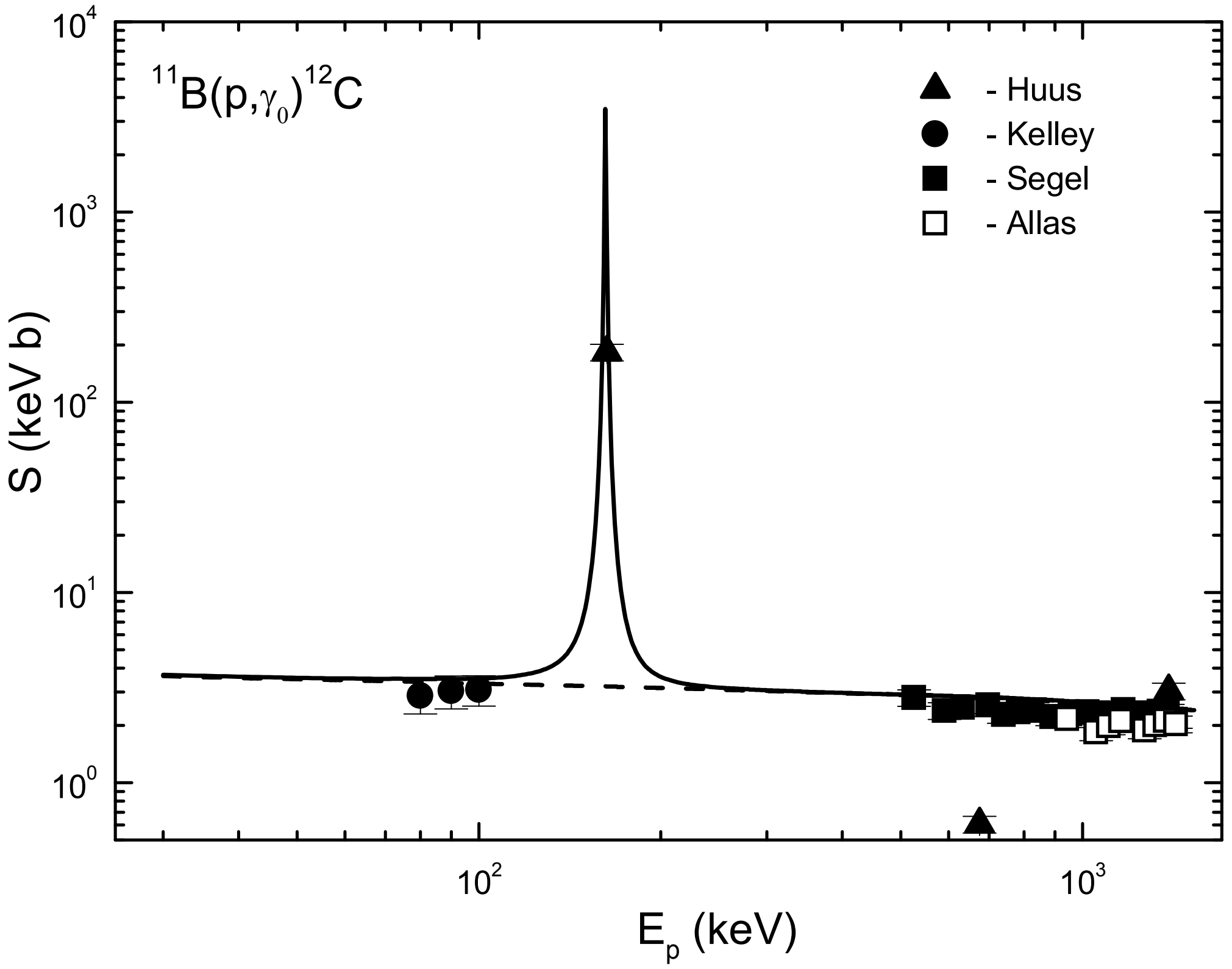

Fig. 13. The astrophysical *S*-factor of the proton radiative capture on $^{11}$Be to the GS in the energy range 50–1.5 $10^3$ keV. Experiment: black triangles (▲) – Ref. 88, dots (●) – Ref. 7, open squares (□) –Ref. 90, black squares (■) – Ref. 89.

Note, that calculated cross section at 163 keV turned to be equal 20 μb what is higher comparing the experimental value also. Meanwhile, the width of potential of Eq. (38) that equals 0.8 keV is considerably larger the proton width of 0.02 keV from Ref. 9. Possibly this mismatch in resonance widths reveals in a rate of decreasing of the calculated cross sections.

Corresponding results for the astrophysical *S*-factor are presented in Fig. 13. Experimental data for the *S*-factor have been recalculated from the total cross sections with using of the exact mass values of point-like particles. Let us remark, that below 100 keV the *S*-factor is practically constant and has a value of 3.5(1) keV b in average. Situation with the resonant behavior of the astrophysical factor is completely analogous what corresponding cross sections reveal. It should be noted that it is not feasible yet to obtain a potential with the width of 0.02 keV. At a time, it is reasonable to encourage experimentalists for new measurements of this channel basing on the modern methods, as the majority of available data refer to 50-60ies.

## 6. Proton-capture reaction $^{14}$C(p, γ)$^{15}$N

### 6.1. *Structure of states in the p$^{14}$C system*

Going to the analysis of the cluster states in the $N^{14}A$ system with the formation of the nuclei $^{15}$N or $^{15}$O in the GS, let us note that the classification of the orbital states of $^{14}$C in the n$^{13}$C channel or $^{14}$N in the p$^{13}$C channel according to Young tableaux was considered by us earlier in Refs. 50, 58, 66, 91, 92, 93, 94 and 95. We regard the results of the classification of the GSs of $^{15}$O and $^{15}$N by orbital symmetry in the considered channels as qualitative, because there are no complete tables of Young tableaux productions for systems with more than eight nucleons,[65] which have been used in earlier similar calculations.[38,39,96] At the same time, simply based on such a classification, we succeeded in describing the available experimental data on the radiative capture of neutrons on $^{13}$C, $^{14}$C and $^{14}$N,[50,58,66,91] and also capture of protons on $^{13}$C.[92] This is why the classification procedure by orbital symmetry given above was used here for the determination of the number of FSs and ASs in partial intercluster potentials and, consequently, to the specified number of nodes of the WFs of the relative motion of the clusters.

Furthermore, we will suppose that for $^{14}A$ it is possible to assume the orbital Young tableau in the form {4442};[50,58,66,67,91,92] therefore, for the $N^{14}A$ system, we have $\{1\} \times \{4442\} \rightarrow \{5442\} + \{4443\}$ in the frame of 1*p*-shell.[65,69] The first of the obtained tableaux is compatible with orbital moments $L = 0$, and 2, and is forbidden because it contains five nucleons in the *s*-shell. The second tableau is allowed and is compatible with orbital moments $L = 1$.[69] As mentioned before, the absence of tables of Young tableaux productions for when the number of particles is 14 and 15 prevents the exact classification of the cluster states in the considered system of particles. However, qualitative estimations of the possible Young tableaux for orbital states allow us to detect the existence of the FS in the $^2S$ wave and the absence of FS for the $^2P$ states. The same structure of FSs and ASs in the different partial waves allows us to construct the potentials of intercluster interactions required for the calculations of the astrophysical *S*-factors, in this case for the proton radiative capture reaction on $^{14}$C.

Thus, by limiting our consideration to only the lowest partial waves with orbital moments $L = 0$, and 1, it could be said that for the n$^{14}$C system (for $^{14}$C it is known $J^{\pi},T = 0^{+},1$), the forbidden and allowed states exist in the $^{2}S_{1/2}$ wave potential. The last of them corresponds to the GS of $^{15}$C with $J^{\pi} = 1/2^{+}$ and is at the binding energy of the n$^{14}$C system of -1.21809 MeV (c.m.).[97] At the same time the potentials of the $^{2}P$ waves of elastic scattering do not have FSs. Considering the p$^{14}$C system let us note that there is the FS in the potential of the $^{2}S_{1/2}$ wave, and in the $^{2}P_{1/2}$ wave there is only AS which corresponds to the GS of $^{15}$N with $J^{\pi} = 1/2^{-}$ and is at the binding energy of the p$^{14}$C system of -10.2074 MeV.[97]

In the case of the n$^{14}$N system (for $^{14}$N we have $J^{\pi},T = 1^{+}0$) there is the bound FS in the potentials of the $S$ scattering wave, and the $^{2}P_{1/2}$ wave has only the AS, which corresponds to the GS of $^{15}$N with $J^{\pi} = 1/2^{-}$ and is at the binding energy of the n$^{14}$N system of -10.8333 MeV.[97] For the p$^{14}$N system we obtain the similar results – there is the FS in the potentials of the $S$ scattering wave, and the $^{2}P_{1/2}$ wave has only the AS, which corresponds to the GS of $^{15}$O with $J^{\pi} = 1/2^{-}$ and is at the binding energy of the p$^{14}$N system of -7.2971 MeV.[97]

Now let us consider the whole spectrum of resonance states in the p$^{14}$C system, i.e., states at positive energies. There are no resonance levels at the energies lower than 1 MeV in the spectra of $^{15}$N for the p$^{14}$C channel, which would have the width value more than 1 keV, however at 1.5 MeV (l.s.) the very wide resonance at $J^{\pi} = 1/2^{+}$ with the width of 405 keV in c.m. is observed. Furthermore, the $^{2}S_{1/2}$ potential with the FS that describes this resonance will be constructed, and the potentials of the $^{2}P$ scattering waves can be equalized to zero, because they have no FSs. As it was mentioned before, the GS of $^{15}$N in the p$^{14}$C channel is the $^{2}P_{1/2}$ wave, therefore it is possible to consider the $E1$ transition from the resonance $^{2}S_{1/2}$ scattering wave at 1.5 MeV to the doublet $^{2}P_{1/2}$ GS of $^{15}$N:

1. $^{2}S_{1/2} \to {}^{2}P_{1/2}$.

The $M1$ transitions to the GS from the doublet $^{2}P$ scattering waves, which potentials simply equal to zero, are possible in principle

2. $$\begin{matrix} ^{2}P_{1/2} \to {}^{2}P_{1/2} \\ ^{2}P_{3/2} \to {}^{2}P_{1/2} \end{matrix}$$

And also the $E1$ transition from the $^{2}D_{3/2}$ wave to the $^{2}P_{1/2}$ GS

3. $^{2}D_{3/2} \to {}^{2}P_{1/2}$.

However, as it was obtained later as a result of our calculations, the transitions 2 and 3, in comparison with the process 1, lead to very small cross sections and they are not taken into account in future calculations.

At first, carry out the standard phase shift analysis of the differential cross sections of the elastic p$^{14}$C scattering, because all intercluster potentials for describing of the proton radiative capture process on $^{14}$C will be based on the phase shifts of the elastic p$^{14}$C scattering.

### 6.2. *Phase shifts and potentials of the elastic scattering*

The excitation functions from Ref. 98, measured at 90°, 125°, 141° and 165° in the energy range from 0.6 to 2.3 MeV (l.s.), are shown in Figs. 14a,b,c,d by dots. These data are used by us further for carrying out of the phase shift analysis and extracting of the resonance form of the $^2S_{1/2}$ scattering phase at 1.5 MeV. The results of the present analysis are shown in Figs. 15a,b,c,d by dots, and the solid lines in Fig. 14 show cross sections calculated with the obtained scattering phase shifts. About 120 first points, cited in Ref. 98, in the stated above energy range were used in this analysis. In addition, it was obtained that for description of the cross sections in the excitation functions, at least at the energies up to 2.2–2.3 MeV, there is no need to take into account $^2P$ or $^2D$ scattering waves, i.e., their values simply can be equalized to zero. Since only one point in the cross sections of excitation functions is considered for each energy and angle, therefore the value of $\chi^2$ at all energies and angles usually lays at the level $10^{-2}$–$10^{-10}$ and taking into account another partial scattering phase shifts already does not lead to its decrease.

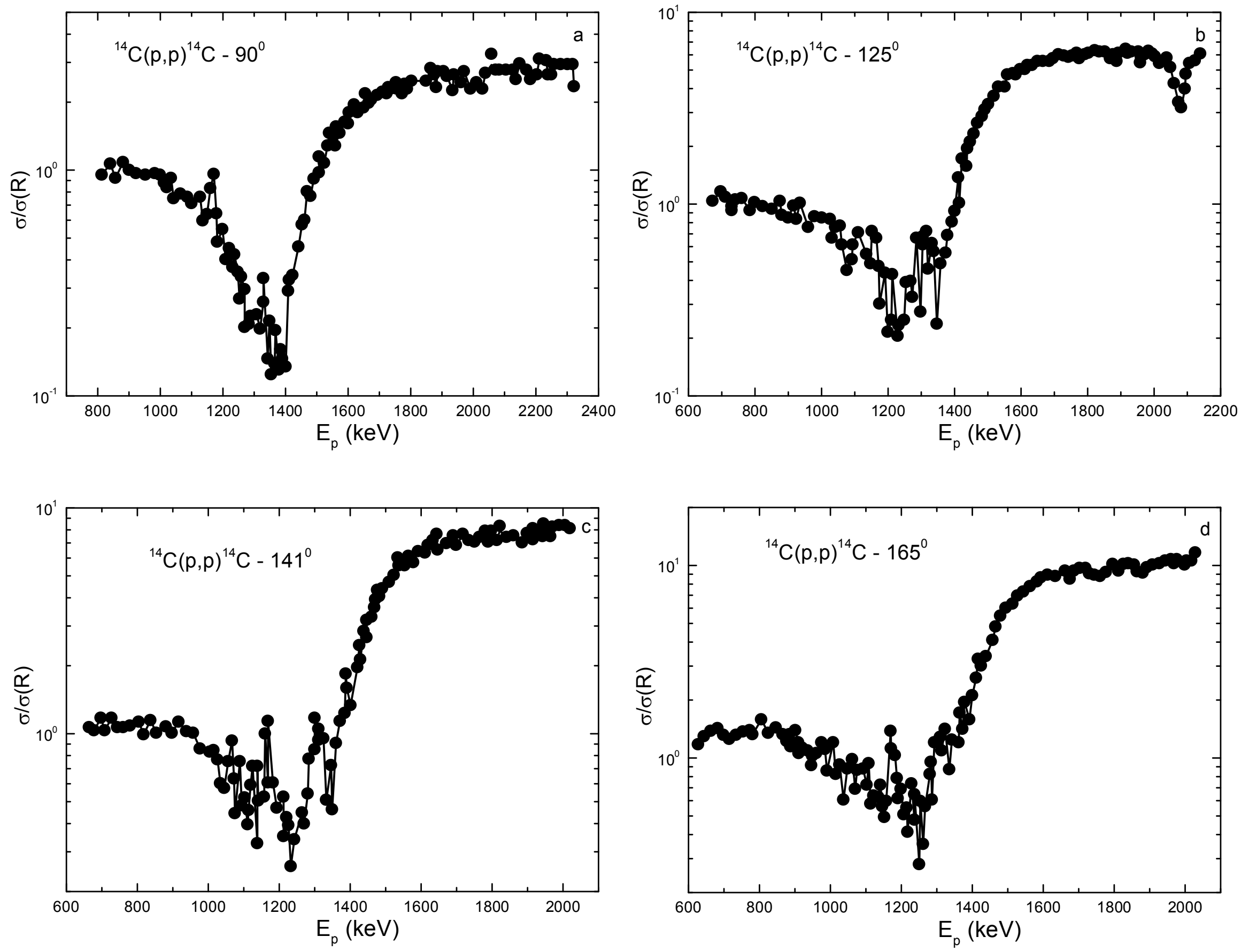

Figs. 14a,b,c,d. The excitation functions in the elastic p$^{14}$C scattering in the range of the $^2S_{1/2}$ resonance.[98] The solid line – their approximation on the basis of the obtained scattering phase shifts.

The resonance energy, as it is seen in Fig. 15, obtained from the excitation function at 90° is at the interval of 1535–1562 keV for which the phase shift value lays within limits of 87°–93° with the value of 90° at 1554 keV. The resonance energy obtained from the excitation function at 125° is at the interval of 1551–1575 keV for which the phase shift value lays within limits of 84°–93°. The resonance energy obtained from the excitation function at 141° is at the interval of 1534–1611 keV for which the phase shift value lays within limits of 84°–90° with the value of 90° at 1534, 1564 and 1611 keV. The resonance energy obtained from the excitation function at 165° is at the interval of 1544–1563 keV for which the phase shift value lays within limits of 87°–91°. For so noticeable scatter of values, it can be said only that the resonance value lays within limits of 1534–1611 that is, in general, agree with data of Ref. 97, where the resonance energy value of 1509 keV is given.

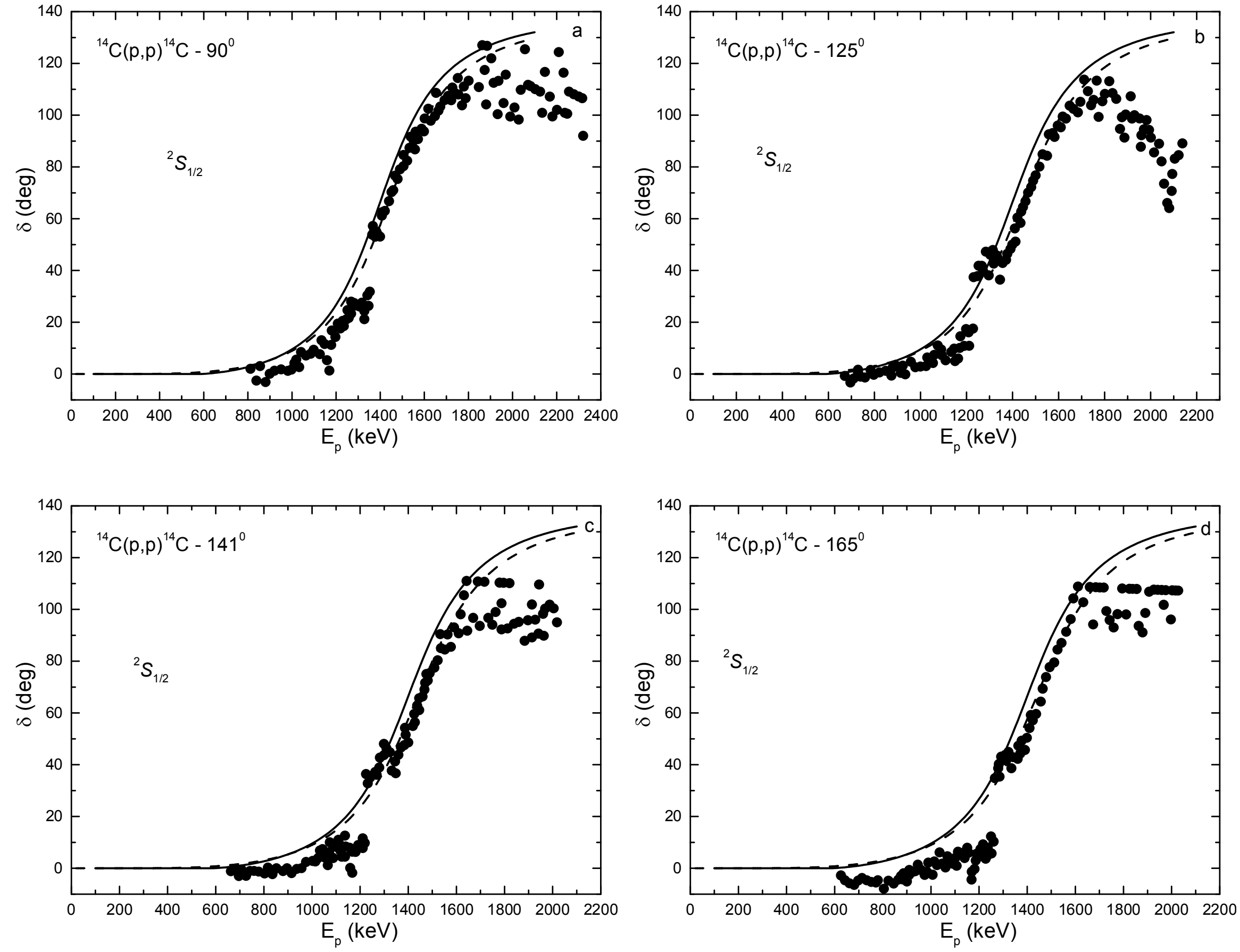

Figs. 15a,b,c,d. The $^2S_{1/2}$ elastic phase shift of the p$^{14}$C scattering at low energies, obtained on the basis of the excitation functions, shown in Figs. 14a,b,c,d. Points – results of our phase shift analysis, carried out on the basis of data from Ref. 98, lines – calculation of the phase shift with the potentials given in the text.

Let us note that work of Ref. 98 mentions that the detailed analysis of the resonances, including 1.5 MeV, was not carried out, because it was done earlier in works of Refs. 99 and 100 on the basis of the proton capture reaction on $^{14}$C. Here, as one can see from the results of the phase shift analysis, the resonance energy at 1.5 MeV slightly overestimated. However, as it was seen in Fig. 14, the data spread on cross sections in excitation functions is too large for clear conclusion about the energy of the resonance. Apparently, the additional and more modern measurements of the elastic scattering cross sections are required, in order to on the basis of these data to perform more unambiguous conclusion about the resonance energy at 1.5 MeV.

For description of the obtained $^2S_{1/2}$ scattering phase in the phase shift analysis it is possible to use simple Gaussian potential of the form of Eq. (9) with FSs and parameters

$$V_0 = 5037.0 \text{ MeV}, \ \alpha = 12.0 \text{ fm}^{-2}, \tag{42}$$

which leads to the scattering phase shifts with the resonance at 1500 keV (l.s.) and with the width of 530 keV (c.m.) that is in a good agreement with the available experimental data of Ref. 97. The parameters of this potential were matched to reproduce in general the resonance data exactly from Ref. 97, which were obtained in works of Refs. 99 and 100. The phase shift of this potential is shown in Figs. 15a,b,c,d by the solid lines and at the resonance energy reaches the value of 90(1)°. The energy behavior of the scattering phase shift of this potential correctly describes obtained in the phase shift analysis scattering phases in whole, taking into account the shift of the resonance energy approximately at 30–50 keV relative to results of Ref. 97. The calculated phase shift line for this potential is parallel to points, obtained in our phase shift analysis, for all scattering angles.

For more accurate description of the obtained in the phase shift analysis data the next potential is needed

$$V_0 = 5035.5 \text{ MeV}, \ \alpha = 12.0 \text{ fm}^{-2}. \tag{43}$$

It leads to the resonance energy of 1550 keV, its width of 575 keV, and the calculation results of the $^2S_{1/2}$ phase shift are shown in all Figs. 15 by the dashed line. As one can see this line appreciably better reproduce results of the carried out here phase shift analysis.

It should be noted over again that the potential is constructed completely unambiguously, if the number of FSs is given (in this case, it is equal to unit), according to the known energy of the resonance level in spectra of any nucleus[97] and its width. In other words, it is not possible to find another combination of the parameters $V$ and $\alpha$, which could be possible to describe the resonance energy of level and its width correctly. The depth of such potential unambiguously determines the resonance location, i.e., resonance energy of the level, and its width $\alpha$ specifies the certain width of this resonance state, which have to correspond to experimental observable values.[97]

For the $^2P_{1/2}$ GS potential of $^{15}$N without FSs in the cluster p$^{14}$C channel the following parameters were found:

$$V_0 = 221.529718 \text{ MeV}, \ \alpha = 0.6 \text{ fm}^{-2}. \tag{44}$$

It allows to find the value of $R_m = 2.52$ fm for the mass radius, the value of $R_{ch} = 2.47$ fm for the charge radius, the binding energy of -10.207400 MeV at the accuracy of the FDM[74] of $10^{-6}$ MeV. The phase shift of such potential decreases smoothly and at 2 MeV approximately equals 179°, and for the AC in the dimensionless form of Eq. (10)

(see Ref. 51), the value of 1.80(1) was obtained at the distance interval 3–10 fm. The mentioned AC error is obtained by its averaging over the given distance interval. The value of 2.56(5) fm (see Ref. 97) was used for radius of $^{14}$C, for radius of $^{15}$N it is known the value of 2.612(9),[97] the proton radius is equal to 0.8775(51).[87] Note once more that we failed in finding AC data in this channel, obtained in another works by independent methods.

### 6.3. *The total capture cross sections*

Going to the description of the obtained results, let us note that the experimental data for total cross section of the proton radiative capture on $^{14}$C to the GS of $^{15}$N or for the astrophysical *S*-factor were measured in Ref. 14 for the energy range 260-740 keV – these results are shown by squares in Figs. 16 and 17 and in numerical form were taken by us from the data base of Ref. 101. The capture cross section to the GS with potential of Eq. (44) for the *E*1 transition from the resonance $^2S$ scattering wave with potential of Eq. (42) was considered for their description. The calculation results of the total cross section and the astrophysical *S*-factor are shown in Figs. 16 and 17 by the solid lines, respectively. They practically completely describe the experimental data at all measured in Ref. 14 energies. Meanwhile, it is good to see that in Fig. 17 the calculated line is in the band of experimental errors and ambiguities of available measurements.[14]

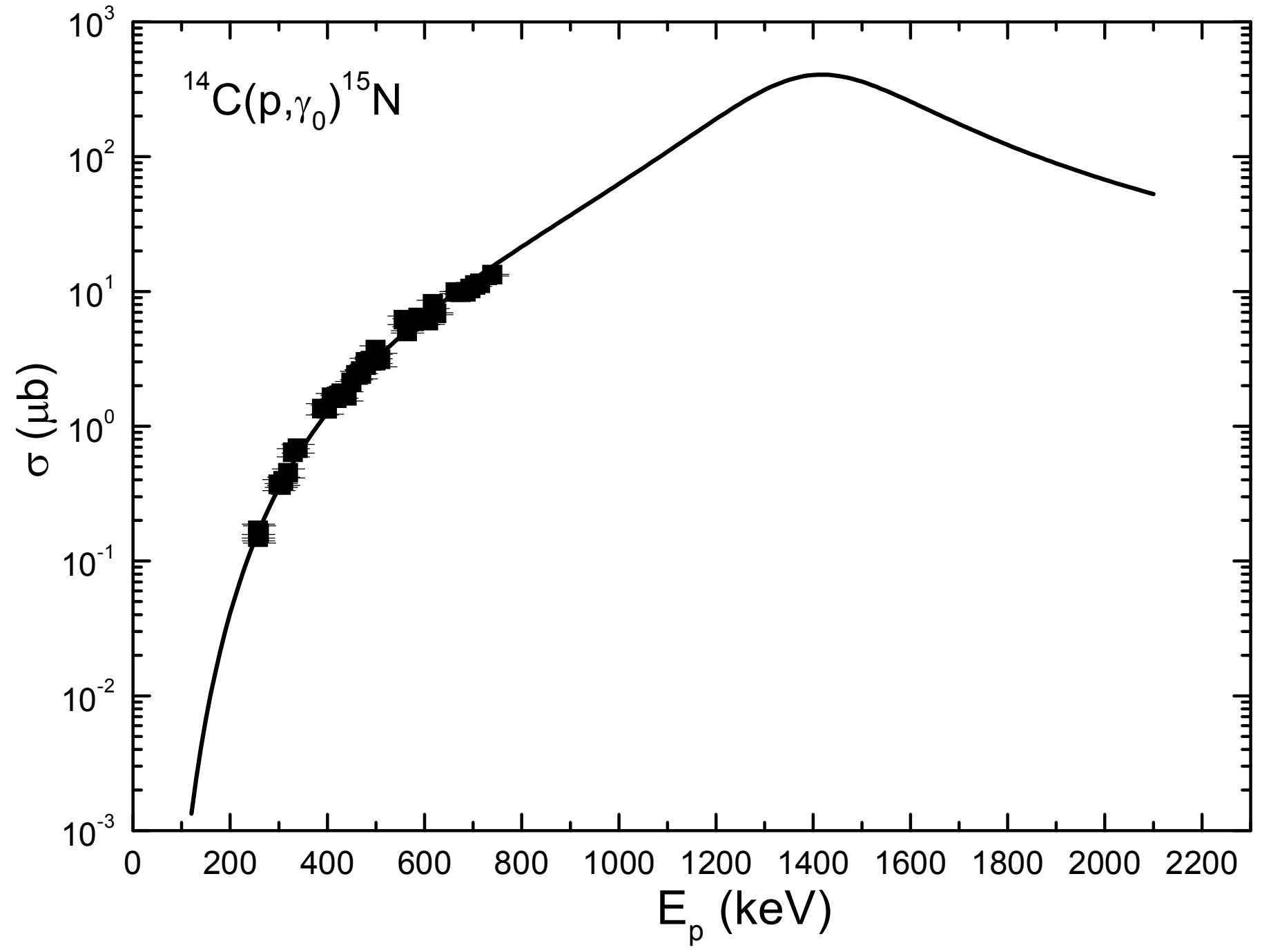

Fig. 16. The total cross section of the proton capture on $^{14}$C to the GS of $^{15}$N. The experimental data are from Ref. 14.

It should be noted here that if scattering potential of Eq. (42) was constructed exclusively based on the characteristics of the resonance at 1.5 MeV, then because of the absence of the AC value in the p$^{14}$C channel the GS potential of Eq. (44) is determined so that to describe available experimental data of the total capture cross sections[14] to the best advantage. However, this potential leads to the quite reasonable radius value of $^{15}N$, and the AC value obtained with it could be used further for comparison with the results of other works. Moreover, the studies of such capture reactions in the used model allow one to extract the AC value, though its accuracy hardly exceeds 10-20%.

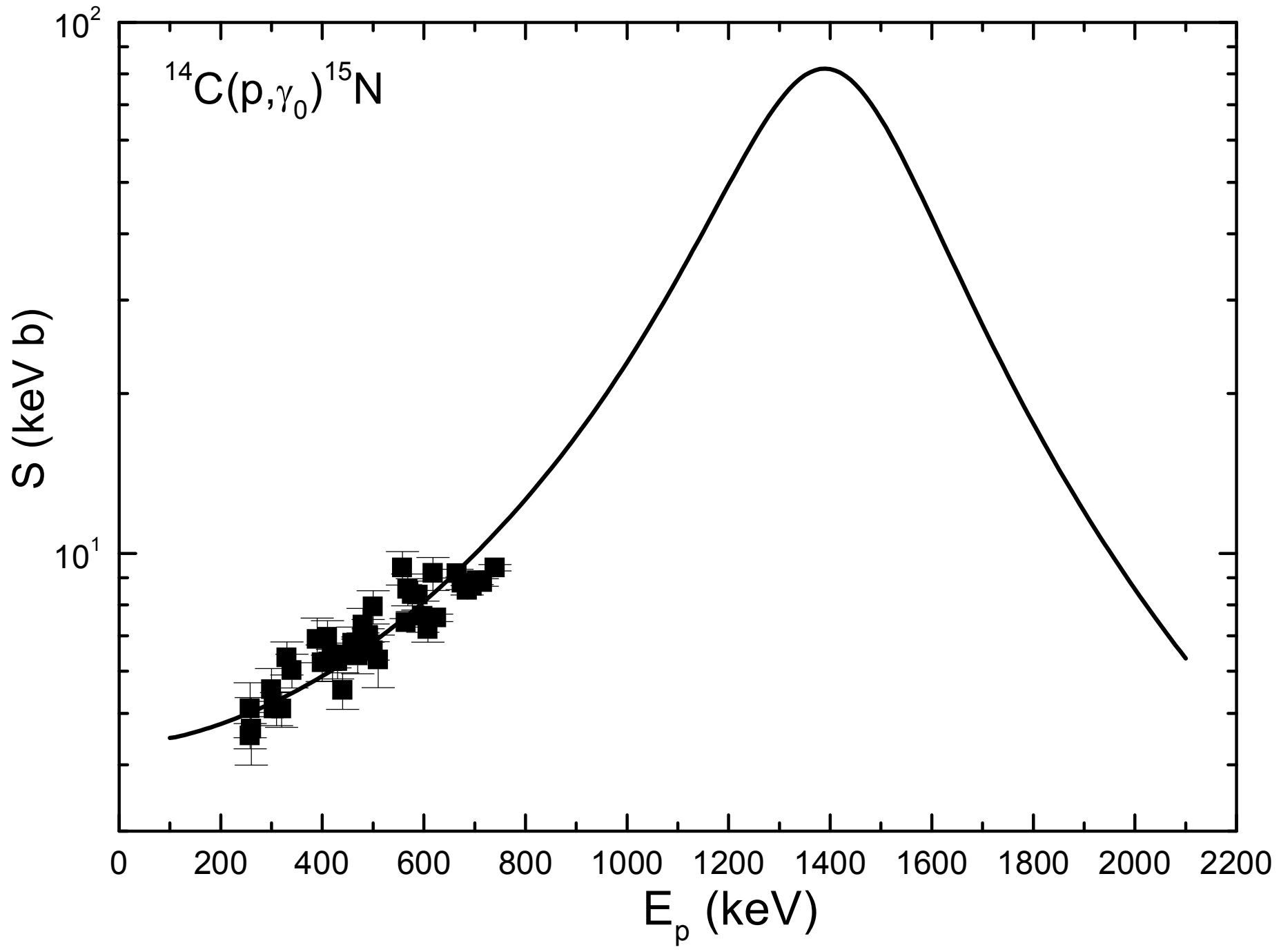

Fig. 17. The astrophysical *S*-factor of the proton capture on $^{14}C$ to the GS of $^{15}N$. The experimental data are from work Ref. 14.

The calculated values of the *S*-factor at the resonance energy also can be used in future for comparison with modern measurements of this reaction. But the measurement spread on the *S*-factor[14] and the experimental errors of these measurements, for example, in the range 258-260 keV reach 30%. Therefore, the calculation accuracy of the maximum *S*-factor value, which in these calculations equals 81-82 keV b at $E_p$= 1390 keV, will be approximately the same, although the phase shift resonance of potential of Eq. (42) is at 1500 keV. The value of the calculated *S*-factor remains almost constant and equals 4.5(1) keV b in the region of low energies, notably 100-140 keV. Apparently, just that very value can be considered as the *S*-factor at zero energy for the considered here proton radiative capture reaction on $^{14}C$.

## 7. Proton-capture reaction $^{15}N(p, \gamma)^{16}O$

### 7.1. *Structure of states in the $p^{15}N$ system*

At first note that the classification of the orbital states of $^{15}N$ according to Young tableaux was qualitatively considered in Refs. 38, 39, 67, 68, 95, 96, and 102. Therefore, we have $\{1\} \times \{4443\} \rightarrow \{5443\} + \{4444\}$ for the $p^{15}N$ system in the frame of 1$p$-shell.[69] The first of the obtained tableaux compatible with the orbital moment $L = 1$ and is forbidden as it contains five cells in the first row,[69] and the second tableau is allowed and compatible with the orbital moment $L = 0$.[69] Thereby, if to limit by the consideration of only lowest partial waves with orbital moment, it could be said that there is bound forbidden state in the $P$ wave potential, and the $S$ wave has no forbidden state. The allowed bound state in the $P$ wave corresponds to the GS of $^{16}$O and is at the binding energy of the p$^{15}$N system of -12.1276 MeV.[103] Because the moment of $^{15}$N is equal to $J^{\pi}T = 1/2^{-}1/2$ Ref. 97 and for $^{16}$O we have $J^{\pi}T = 0^{+}0$, then its GS in the p$^{15}$N channel can be the $^{3}P_{0}$ state (representation in $^{(2S+1)}L_{J}$).

We regard the results on the classification of $^{16}$O by orbital symmetry in the p$^{15}$N channel as the qualitative one as there are no complete tables of Young tableaux productions for the systems with a number of nucleons more than eight,[65] which have been used by us in earlier similar calculations.[38,39,67,68,93,95,96,104] At the same time, just on the basis of such classification, we succeeded with description of available experimental data on the radiative capture of protons and neutrons on $^{13}$C.[82,92,105] That is why here we will use similar classification of cluster states, which gives the certain number of forbidden and allowed states in different partial intercluster potentials. The number of such states determines the number of nodes of the wave function of cluster relative motion with the certain orbital moment $L$.[38,39,67,68,95,96]

### 7.2. *Interaction potentials and structure of resonance states*

In the first place we have considered $E1$ transitions from the resonance $^{3}S_{1}$ scattering wave without FS at the energies up to 1.5 MeV to the triplet GS $^{3}P_{0}$ of $^{16}$O in the p$^{15}$N channel with one bound FS, for considering the total cross section of the radiative proton capture on $^{15}$N to the GS, as in the previous Refs. 38, 39, 67, 68, 95, and 96. There are two resonances at the energies up to 1.1 MeV, which can be compared with the $^{3}S_{1}$ scattering wave.

1. The first resonance is at the energy of 335(4) keV with the width of 110(4) keV (l.s.) and has the moment $J^{\pi},T = 1^{-}0$ (see Table 16.22 in Ref. 103) – it can be caused by the triplet $^{3}S_{1}$ scattering state.

2. The second resonance is at the energy of 710(7) keV with the width of 40(40) keV (l.s.) and has the moment $J^{\pi},T = 0^{-}1$. It can be caused by the $^{1}S_{0}$ scattering state, however there is no resonance in the capture cross section for it.

3. The third resonance is at the energy of 1028(10) keV with the width of 140(10) keV (l.s.) and has the moment $J^{\pi},T = 1^{-}1$ (see Table 16.22 in Ref. 103) – it also can be caused by the triplet $^{3}S_{1}$ scattering state.

These levels have the next energies in the center-of-mass system: The first resonance is at the energy of 312(2) keV with the width of 91(6) keV – it corresponds to the ES of $^{16}$O at the energy of 12.440(2) MeV. The third resonance is at the energy of 962(8) keV with the width of 130(5) keV – it corresponds to the ES of $^{16}$O at the energy of 13.090(8) MeV (see Table 16.13 in Ref. 103). We do not consider the second resonance, because, as it was mentioned, it does not give obvious contribution to the total radiative capture cross sections.

Carrying out the calculations of the total radiative capture cross sections the nuclear part of the p$^{15}$N intercluster potential is usually presented in the Gaussian form of Eq.(9).[38,39,67,68,95,96]

Immediately note that for the potential of the resonance $^3S_1$ waves without FSs two potentials were obtained, which correspond to two resonances of different width at different energies of 335 keV and 1028 keV. The parameters of the first potential:

$$V_{S1} = 1.0857 \text{ MeV}, \ \alpha_{P1} = 0.003 \text{ fm}^{-2} \quad (45)$$

lead to the resonance energy of 335.0(1) keV and its width of 138(1) keV (l.s.), and the scattering phase shift is shown in Fig. 18 by the solid line. The relative accuracy of calculation of the $^3S_1$ scattering phase in these calculations is equal to $10^{-3}$ approximately and for the energy of 335 keV this potential leads to the value of the phase shift of 90.0(1)°.

The next parameters were obtained for the second potential of the $^3S_1$ wave

$$V_{S1} = 105.059 \text{ MeV}, \ \alpha_{P1} = 1.0 \text{ fm}^{-2}. \quad (46)$$

It leads to the resonance at 1028.0(5) keV and its width is equal to 140(1) keV (l.s.), and the scattering phase shift is shown in Fig. 18 by the dashed line – for resonance energy this potential also leads to the phase shift value of 90.0(1)°.

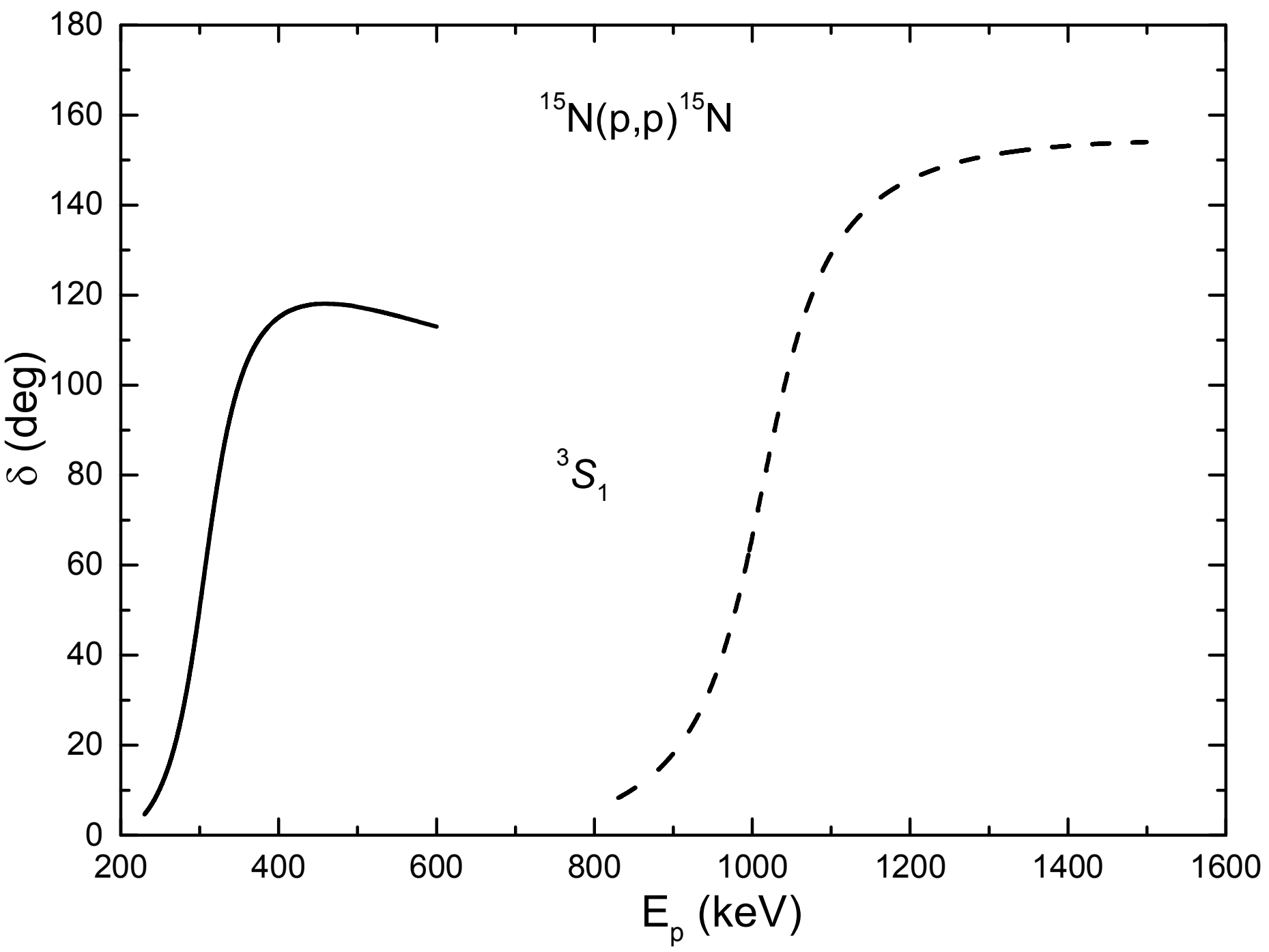

Fig. 18. The phase shifts of the $p^{15}N$ elastic scattering in the $^3S_1$ wave

Here, we must draw attention[38,39,67,68,95,96] that the potential with the given number of BSs is constructed completely unambiguously using the energy values of the resonance level in spectra of $^{16}O$ and its width. It is impossible to find other parameters $V_0$ and $\alpha$, which can be able correctly reproduce the level resonance energy and its width, if the specified number of FSs and ASs are given, which in this case equals zero. The depth of this potential unambiguously determines the location of the resonance, i.e., the level resonance energy $E_r$, and its width gives specific width $\Gamma_r$ of this resonance state. The error of parameters of such potential is determined by the measurement error of the level $E_r$ and its width $\Gamma_r$.[38,39,67,68,95,96] However, here we must note that it is impossible, for the present, to construct the unified $^3S_1$ potential, which would contain two, noted above, resonances with different energies and widths. Therefore, the calculated total cross section for these resonances will consist of two parts – the first with the potential of Eq. (45) and the second for the interaction of Eq. (46), also both of these parts are the same $E1$ transition from the $^3S_1$ scattering wave to the $^3P_0$ GS of $^{16}$O.

There are no resonance levels with $J = 0^+,1^+,2^+$ and widths more than 10 keV (see Table 16.22 in Ref. 103) in the spectra of the elastic p$^{15}$N scattering at the energies up to 1.05 MeV. Therefore, it is possible to use the parameter values for the potentials of the nonresonance $^3P$ waves with one bound FS based on the assumption that in the considered energy range, i.e., up to 1.5 MeV, their phase shifts are equal to zero. The next parameters were obtained for such potentials

$$V_P = 14.4 \text{ MeV}, \ \alpha_P = 0.025 \text{ fm}^{-2}. \tag{47}$$

The calculation of the $P$ phase shifts with this potential at the energy up to 1.5 MeV leads to their values from 180° to 179°. The unified Levinson theorem[37,106,107] was used for determination of the phase shift values at zero energy, therefore phase shifts of the potential with one bound FS have to start from 180°. There is the resonance with $J = 2^+$ (see Ref. 103) in the spectrum of $^{16}$O at the energy of 1.050(150) MeV, however the width values are not given for it, and the next $J = 1^+$ resonance with the width of 68(3) keV is located higher than 1.5 MeV.

Furthermore, we will construct the potential with FS of the $^3P_0$ state, which has to correctly reproduce the binding energy of the GS of $^{16}$O with $J^\pi T = 0^+0$ in the p$^{15}$N channel at -12.1276 MeV (see Ref. 103) and reasonably describe the mean square radius of $^{16}$O, which experimental value equals 2.710(15) fm,[103] when the experimental radius of $^{15}$N equals 2.612(9) fm.[97] The charged and the mass radius of proton are equal to 0.8775(51) fm were used in these calculations.[87] Consequently, the next parameters for the potential of the GS of $^{16}$O in the p$^{15}$N channel were obtained:

$$V_{G.S.} = 1057.99470 \text{ MeV}, \ \alpha_{G.S.} = 1.2 \text{ fm}^{-2}. \tag{48}$$

The potential leads to the binding energy of -12.12760 MeV at the FDM accuracy of $10^{-5}$ MeV, the mean square charged radius of 2.52 fm and the mass radius of 2.57 fm. The value of 1.94(1) at the range of 2–10 fm was obtained for the AC, written in the dimensionless form of Eq. (10).[51] The error of the constant is determined by its averaging over the noted above range interval. The phase shift of this potential is in the range from 180° to 179° at the energy up to 1.5 MeV.

The value of 192(26) fm$^{-1}$ for this AC is given in Ref. 108 that after division on the spectrofactor value of 2.1 and the antisymmetrization coefficient of 16 (see Ref. 109) gives 5.71(77) fm$^{-1}$ or 2.39(88) fm$^{-1/2}$. Its recalculation to the dimensionless value at $\sqrt{2k_0} = 1.22$ gives 1.96(72) – this value is in a good agreement with the constant obtained above. The recalculation of the AC to the dimensionless value is needed because another definition of the AC were used in these works, that is from Eq. (17), which differs from using here by factor $\sqrt{2k_0}$. In addition, the AC from Ref. 108 contains the spectrofactor equals, evidently, 2.1 and the antisymmetrization coefficient, obtained in the review of Ref. 76.

Give here the second variant of the GS potential with the similar parameters, but slightly another AC

$$V_{G.S.} = 976.85193 \text{ MeV}, \ \alpha_{G.S.} = 1.1 \text{ fm}^{-2}. \quad (49)$$

It leads to the same binding energy, does not change the values of mean square charged and mass radii, and for the AC in the dimensionless form of Ref. 51 at the range of 2–9 fm the value of 2.05(1) was obtained. The phase shift of this potential at the energy up to 1.5 MeV lies at the same interval of values as for the potential of Eq. (48).

Earlier it was shown once and again in Refs. 38, 39, 67, 68, 95, and 96 that the additional control of calculation of the binding energy GS or BS on the basis of the variational method (VM)[74,110] leads to the results coinciding with the FDM with the given determination accuracy of the binding energy of two-cluster system, therefore here we already not used the VM for the verification of this binding energy.

### 7.3. *The total proton capture cross sections on $^{15}N$*

Going to the direct consideration of the results of the $M1$ and the $E1$ transitions to the GS of $^{16}$O, note that we succeeded[111,112,113] to find the experimental data for the total cross section of the process of the proton capture on the $^{15}$N in the energy range from 80 keV up to 1.5 MeV, which will be considered furthermore – these results are shown in Figs. 19a and 19b. The first part of the cross section of the $E1$ transition $^3S_1 \to {}^3P_0$ to the GS, calculated with the potentials of the scattering state of Eq. (45) and the GS of Eq. (48), is shown in Fig. 19a by the dashed line, and the dotted line shows the cross section of the $E1$ transition for the scattering potential of Eq. (46) and the GS of Eq. (48). In addition, we considered the $M1$ transition of the form $^3P_1 \to {}^3P_0$ with the potentials of the scattering state of Eq. (47) and the GS of Eq. (48). The results of these calculations are shown in Fig. 19a by the dot-dashed line and the total summed cross section for the referred above capture processes to the ground state is shown by the solid line. The $S$-factor is equal to 39.50(5) keV b and practically constant at the energy range of 50–60 keV. The similar calculations are given in Fig. 19b for the GS potential of Eq. (49), these results differ from the previous one only in the low energy range. This potential of the GS leads to the $S$-factor values of 43.35(5) keV b at the energy range of 50–60 keV. The values of the $S$-factor at these energies, owing to their weak changes, can be considered, evidently, as the $S(0)$-factor for zero energy.

It can be noted that if the parameters of the resonance $^3S_1$ potential are fixed according to the resonance of phase shift relatively unambiguously and for the bound state they are chosen on the basis of the description of the bound state characteristics, therefore for the $^3P_1$ potential of Eq. (47) with the FS which leads to zero scattering phases, another parameter values are possible. However, only the given above parameters of Eq. (47) allow one to obtain the acceptable results of calculations for the transition from the $^3P_1$ wave to the GS, which are shown in Fig. 19a and 19b by the dot-dashed lines. Now, it is impossible to draw a unique conclusion about the form and depth of such scattering potential.

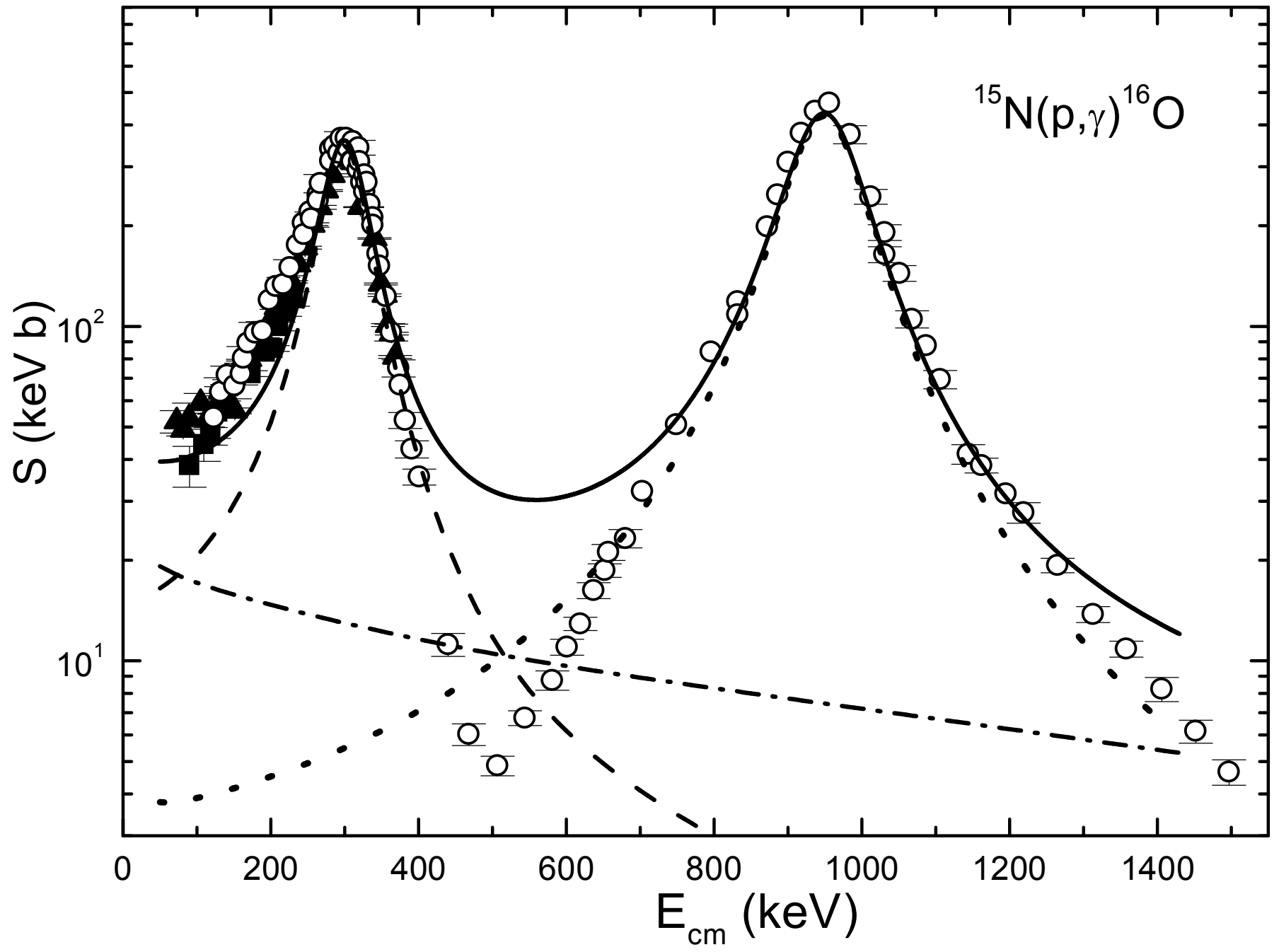

Fig. 19a. The astrophysical *S*-factor of the proton radiative capture on $^{15}N$ to the GS. The experimental data: ▲ – Ref. 111, ■ – Ref. 112, ○ – Ref. 113. Lines – the calculation of the total cross sections for transitions to the GS.

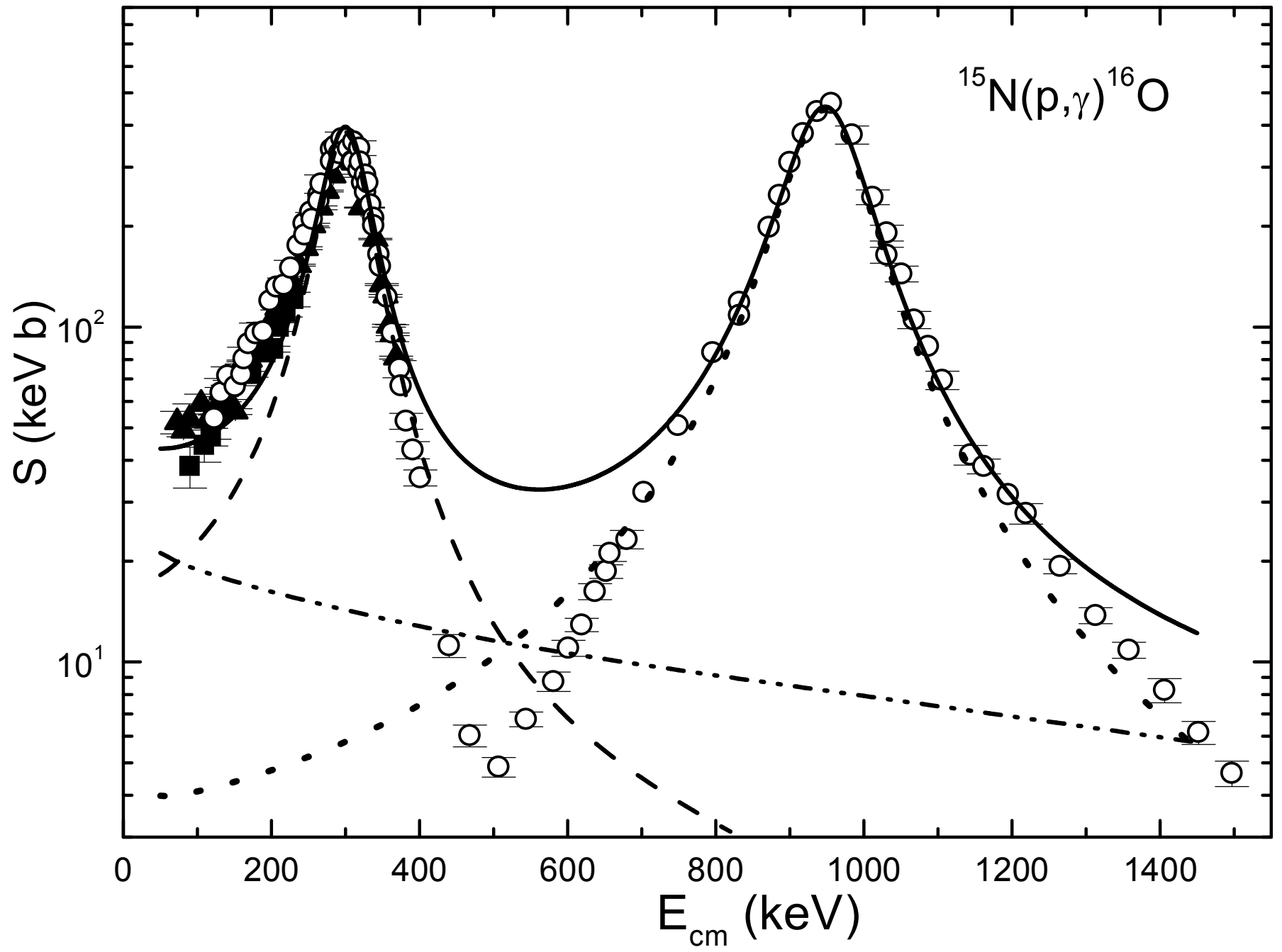

Fig. 19b. The same as in Fig. 19a.

In addition, the medium-energy region, where the minimum of cross sections at 500 keV is observed, is described badly. It is connected with the inefficient abrupt decrease of the cross section after the first resonance at the energy above 400 keV (the dashed line in Figs. 19a and 19b) and with the inefficient abrupt rising of it before the second resonance at the energies up to 600–650 keV (the dotted line in Figs. 19a and 19b). It is also connected with the value of the cross section for the *M*1 transition, which is, for this energy region, more than minimal experimental value of 4.9 keV b at 507 keV.[113] And here, evidently, the results of the phase shift analysis of angular distribution of the elastic p$^{15}$N scattering are needed for the construction of the correct potentials of two resonances in the $^3S_1$ wave.

## 8. Conclusion

As can be seen from the listed results, the obvious assumptions about the methods of construction of the n$^{10}$B interaction potentials, if they have FSs, allow one to obtain acceptable results on the description of the available experimental data for the total cross section of the neutron capture on $^{10}$B.[73,78,80] at the energy range from 25 meV to 61 keV. The possibility to describe all considered experimental data both by capture cross section and according to the GS characteristics, allows us to fix parameters of the GS potential closely enough in the form of Eq. (16). The summed cross sections at the resonance energy of 0.475 MeV, equals 4.5 μb at the width of the resonance of 193 keV and 13.7 μb at the width of 32 keV.

Within the MPCM based on the deep attractive potentials with forbidden states and coordinated with the spectrum of resonance levels we succeeded to convey properly the behavior of the experimental cross sections of the n$^{11}$B radiative capture onto the GS of $^{12}$B in the energy range $10^{-5}$–$10^2$ keV. Furthermore, within the *E*1 and *M*1 transitions the total cross sections have been described at resonance energies as a whole. Potentials for the GS and the first ES are in conformity with basic characteristics of $^{12}$B in the n$^{11}$B channel, namely binding energy, charge radius and asymptotic constant.

It was demonstrated that quite transparent assumptions on a way of construction of the p$^{11}$B interaction potentials with forbidden states[39] make it possible to reproduce reasonably well available experimental data of Refs. 7, 88, 89, and 90 for the radiative capture to the GS of $^{12}$C in the energy region from 80 meV up to 1500 keV, as well as claim that the energy dependence of the observed cross sections occurs due to $^3S_1 \to ^3P_0$ and $^3P_2 \to ^3P_0$ transitions. It should be noted, that there are no experimental data on the asymptotic constants for the GS, so our estimations might be regarded as a preliminary proposal.

The MPCM based on the deep attractive potentials with forbidden states and coordinated with the spectrum of resonance levels mares it possible to convey properly the behavior of the experimental cross sections of the proton radiative capture on $^{11}$B onto the GS of $^{12}$C in rather wide energy range. Furthermore, the total cross sections have been described at resonance energies as a whole. Potentials for the GS and the first ES are in conformity with basic characteristics of $^{12}$C in the p$^{11}$B channel, namely binding energy, charge radius and asymptotic constants.

The resonance $^2S_{1/2}$ phase shift of the p$^{14}$C elastic scattering at energies from 0.6 MeV to 2.3 MeV was found as a result of the carried out phase shift analysis of the experimental differential cross sections in excitation functions.[98] The resonance energy of the phase shift is in a quite agreement with the level spectrum of $^{15}$N in the $p^{14}C$ channel.[97] The results of the carried out phase shift analysis, i.e., phase shift of the elastic $p^{14}C$ scattering and the data of resonances of $^{15}N$,[97] allow one to parametrize intercluster interaction potentials for scattering processes in the resonance $^2S_{1/2}$ wave. These potentials, by-turn, can be used further for carrying out certain calculations for different astrophysical problems, partially considered, for example, in Ref. 38, 39, and 96.

It should be noted that it is not enough data of Ref. 98 for carrying out of the accurate phase shift analysis and the another additional data on differential cross sections, for example, angular distributions in the resonance region are needed. Such data can allow one to determine the location of the $^2S_{1/2}$ resonance in the region of 1.5 MeV and more accurately determine its width exactly on the basis of the elastic scattering phase shifts.

Furthermore, the measured cross section or the astrophysical $S$-factor of the proton capture reaction on $^{14}C$ are succeeded to correctly describe on the basis only of the $E1$ transition from the resonance $^2S_{1/2}$ scattering wave with FS to the $^2P_{1/2}$ GS without FS of $^{15}N$, considered in the two-body $p^{14}C$ model. The carrying out in future more detailed measurements of total cross sections of this reaction, especially, in the resonance region at 1.5 MeV allows one to draw, apparently, more concrete conclusions about the quality of description of the considered cross sections of the reaction of the proton capture on $^{14}C$ in the framework of the MPCM.

The intercluster potentials of the bound state, constructed on the basis of the quite obvious requirements for description of the binding energy, the mean square radii of $^{16}O$ and the AC values in the $p^{15}N$ channel, and also the scattering potentials describing the resonances allow one to reproduce generally correctly the available experimental data for the total cross sections of the proton radiative capture on $^{15}N$ at low energies.[113] Meanwhile, all $p^{15}N$ potentials using here are constructed on the basis of the given above classification of the FSs and the ASs according to Young tableaux.

However, it is difficult to do defined and final conclusions if there are no results on the phase shift analysis, carried out on the basis of differential cross sections of the $p^{15}N$ elastic scattering. Such data are needed, approximately, in the range of 0.3–1.5 MeV, however the results at the energies lower 1 MeV are absent up to now. Meanwhile, the available data on the differential cross sections[114] were obtained in the form of excitation function only at two angles and only above the energy of 0.97 MeV. Therefore, furthermore it is desirable to carry out the detailed measurement of the differential cross sections of the elastic scattering in the energy range from 0.1–0.3 MeV up to 1.3–1.5 MeV. Such data have to contain angular distributions in the range of two resonances at 335 keV and 1028 keV (see Ref. 103) at angles from 30º to 170º.

Thereby, the MPCM again confirms, as already done in 26 reactions (see Table 1) presented in Refs. 38, 39, 58, 66, 67, 68, 91, 93, 94, 102, 115, 116, its ability to describe correctly the cross sections of the processes such as the radiative capture of neutral and charged particles on light nuclei at thermal and astrophysical energies. As this occurs, such results are obtained using the potentials matched with the resonance scattering phases, or with the level spectra of the final nucleus and the BS characteristics of the considered nuclei and some basic principles of the construction of such potentials were checked partially in the three-body calculations.[117]

Table 1. The characteristics of nuclei and cluster systems, and references to works in which they were considered.[+)]

| No. | Nucleus ($J^\pi$,$T$) | Cluster channel | $T_z$ | $T$ | Refs. |
|---|---|---|---|---|---|
| 1. | $^3$He (1/2$^+$,1/2) | p$^2$H | +1/2 + 0 = +1/2 | 1/2 | 38, 67, 68, 93 |
| 2. | $^3$H (1/2$^+$,1/2) | n$^2$H | –1/2 + 0 = –1/2 | 1/2 | 39, 58, 66 |
| 3. | $^4$He (0$^+$,0) | p$^3$H | +1/2 – 1/2 = 0 | 0 + 1 | 38, 68 |
| 4. | $^6$Li (1$^+$,0) | $^2$H$^4$He | 0 + 0 = 0 | 0 | 38, 68 |
| 5. | $^7$Li (3/2$^-$,1/2) | $^3$H$^4$He | –1/2 + 0 = –1/2 | 1/2 | 38, 68 |
| 6. | $^7$Be (3/2$^-$,1/2) | $^3$He$^4$He | +1/2 + 0 = +1/2 | 1/2 | 38, 68 |
| 7. | $^7$Be (3/2$^-$,1/2) | p$^6$Li | +1/2 + 0 = +1/2 | 1/2 | 38, 67, 68 |
| 8. | $^7$Li (3/2$^-$,1/2) | n$^6$Li | –1/2 + 0 = –1/2 | 1/2 | 39, 58, 66 |
| 9. | $^8$Be (0$^+$,0) | p$^7$Li | +1/2 – 1/2 = 0 | 0 + 1 | 38, 67, 68 |
| 10. | $^8$Li (2$^+$,1) | n$^7$Li | –1/2 – 1/2 = –1 | 1 | 39, 58, 66, 94 |
| 11. | $^{10}$B (3$^+$,0) | p$^9$Be | +1/2 – 1/2 = 0 | 0 + 1 | 38, 68 |
| 12. | $^{10}$Be (0$^+$,1) | n$^9$Be | –1/2 – 1/2 = –1 | 1 | 91, 116 |
| 13. | $^{11}$B(3/2$^-$,1/2) | n$^{10}$B | –1/2 + 0 = –1/2 | 1/2 | Present work |
| 14. | $^{12}$C (0$^+$,0) | p$^{11}$B | +1/2 – 1/2 = 0 | 0 | Present work |
| 15. | $^{12}$B (1$^+$,1) | n$^{11}$B | –1/2 – 1/2 = –1 | 1 | Present work |
| 16. | $^{13}$N (1/2$^-$,1/2) | p$^{12}$C | +1/2 + 0 = +1/2 | 1/2 | 38, 67, 68 |
| 17. | $^{13}$C (1/2$^-$,1/2) | n$^{12}$C | –1/2 + 0 = –1/2 | 1/2 | 39, 58, 66 |
| 18. | $^{14}$N (1$^+$,0) | p$^{13}$C | +1/2 – 1/2 = 0 | 0 + 1 | 67, 68 |
| 19. | $^{14}$C (0$^+$,1) | n$^{13}$C | –1/2 – 1/2 = –1 | 1 | 39, 58, 66 |
| 20. | $^{15}$C (1/2$^+$,3/2) | n$^{14}$C | –1/2 – 1 = –3/2 | 3/2 | 39, 91, 115 |
| 21. | $^{15}$N (1/2$^-$,1/2) | p$^{14}$C | +1/2 – 1 = –1/2 | 1/2 | Present work |
| 22. | $^{15}$N (1/2$^-$,1/2) | n$^{14}$N | –1/2 + 0 = –1/2 | 1/2 | 39, 91, 115 |
| 23. | $^{16}$N (2$^-$,1) | n$^{15}$N | –1/2 – 1/2 = –1 | 1 | 39, 91, 102 |
| 24. | $^{16}$O (0$^+$,0) | p$^{15}$N | +1/2 – 1/2 = 0 | 0 | Present work |
| 25. | $^{16}$O (0$^+$,0) | $^4$He$^{12}$C | 0 + 0 = 0 | 0 | 38, 68 |
| 26. | $^{17}$O (5/2$^+$,1/2) | n$^{16}$O | –1/2 + 0 = –1/2 | 1/2 | 39, 58, 91 |

[+)] $T$ – isospin and $T_z$ – its projection, $J^\pi$ – total moment and parity.

## Acknowledgments

The work was performed under grant No. 0151/GF2 "Studying of the thermonuclear processes in the primordial nucleosynthesis of the Universe" of the Ministry of Education and Science of the Republic of Kazakhstan.

In conclusion, the authors express their deep gratitude to Prof. Akram Mukhamedzhanov (Cyclotron Institute Texas A&M University) and Prof. Rakhim Yarmukhamedov (INP, Tashkent, Uzbekistan) for provision of the information on the ACs in different channels.

## References

1. M. Heil, F. Käppeler, M. Wiescher and A. Mengoni, *Astrophys. J.* **507**, 997 (1998).
2. V. Guimaraes and C. A. Bertulani, *AIP Conf. Proc.* **1245**, 30 (2010); available online at: arXiv:0912.0221v1 [nucl-th] 1 Dec 2009.
3. M. Igashira and T. Ohsaki, *Sci. Tech. Adv. Materials* **5**, 567 (2004); available online at: http://iopscience.iop.org/1468-6996/5/5-6/A06
4. Y. Nagai, T. Shima, T. S. Suzuki, H. Sato, T. Kikuchi, T. Kii, M. Igashira and T. Ohsaki, *Hyperfine Interactions* **103**, 43 (1996).
5. Z. H. Liu, C. J. Lin, H. Q. Zhang, Z. C. Li, J. S. Zhang, Y. W. Wu, F. Yang, M. Ruan, J. C. Liu, S. Y. Li and Z. H. Peng, *Phys. Rev. C* **64**, 034312 (2001).
6. D. Clayton, *Isotopes in the Cosmos. Hydrogen to Gallium* (Cambridge University Press, Cambridge, 2003).
7. J. H. Kelley, R. S. Canon, S. J. Gaff, R. M. Prior, B. J. Rice, E. C. Schreiber, M. Spraker, D. R. Tilley, E. A. Wulf and H. R. Weller, *Phys. Rev. C* **62**, 025803 (2000).
8. B. D. Anderson, M. R. Dwarakanath, J. S. Schweitzer and A. V. Nero, *Nucl. Phys. A*. **233**, 286 (1974).
9. F. Ajzenberg-Selove, *Nucl. Phys. A*. **506**, 1 (1990).
10. F. E. Cecil, D. Ferg, H. Liu, J. C. Scorby, J. A. McNeil and P. D. Kunz, *Nucl. Phys. A*. **539**, 75 (1992).
11. J. H. Applegate and C. J. Hogan, *Phys. Rev. D* **31**, 3037 (1985).
12. J. H. Applegate, C. J. Hogan and R. J. Scherer, *Astrophys. J.* **329**, 572 (1988).
13. R. A. Malaney and W. A. Fowler, *Astrophys. J.* **333**, 14 (1988).
14. J. Görres, S. Graff, M. Wiescher, R. E. Azuma, C. A. Barnes, H. W. Becker and T. R. Wang, *Nucl. Phys. A* **517**, 329 (1990).
15. L. Yaffe and W. H. Stevens. *Phys. Rev.* **79**, 893 (1950).
16. W. A. Fowler, G. E. Caughlan and B. A. Zimmermann, *Ann. Rev. Astron. Astrophys.* **13**, 69 (1975).
17. C. A. Barnes, D. D. Clayton and D. N. Schramm, *Essays in Nuclear Astrophysics Presented to William A. Fowler* (Cambridge University Press, Cambridge, 1982).
18. E. G. Adelberger *et al*., *Rev. Mod. Phys.* **83**, 195 (2011).
19. C. Rolfs, W. S. Rodney, *Nucl. Phys. A* **235**, 450 (1974).
20. G. Imbriani *et al*., *Astron. Astrophys.* **420**, 625 (2004).
21. G. Imbriani *et al*., *Eur. Phys. J. A* **25**, 455 (2005).
22. R. C. Runkle, A. E. Champagne, C. Angulo, C. Fox, C. Iliadis, R. Longland and J. Pollanen, *Phys. Rev. Lett.* **94**, 082503 (2005).
23. K. Esmakhanova, N. Myrzakulov, G. Nugmanova, Ye. Myrzakulov, L. M. Chechin and R. Myrzakulov, *Int. J. Mod. Phys. D* **20**, 2419 (2011).
24. L. M. Chechin, *Chinese Physics Letters* **23**, 2344 (2006).
25. L. M. Chechin, *Astronomy Reports* **54**, 719 (2010).
26. M. White, D. Scott and J. Silk, *Ann. Rev. Astron. & Astrophys.* **32**, 319 (1994).
27. L. M. Chechin and Sh. R. Myrzakul, *Rus. Phys. J.* **52**, 286 (2009).
28. T. Omarov and L. M. Chechin, *Gen. Relativ. Grav.* **31**, 443 (1999).
29. K. Wildermuth and Y. C. Tang, *A unified theory of the nucleus* (Vieweg, Branschweig, 1977).
30. T. Mertelmeir and H. M. Hofmann, *Nucl. Phys. A* **459**, 387 (1986).

31. J. Dohet-Eraly, *Microscopic cluster model of elastic scattering and bremsstrahlung of light nuclei* (Université Libre De Bruxelles, Bruxelles, 2013); http://theses.ulb.ac.be/ETD-db/collection/available/ULBetd-09122013-100019/unrestricted/these_Jeremy_Dohet-Eraly.pdf.
32. J. Dohet-Eraly and D. Baye, *Phys. Rev. C* **84**, 014604 (2011).
33. P. Descouvemont and M. Dufour, *Microscopic cluster model*, in: *Clusters in Nuclei*, Second edition by C. Beck. (Springer-Verlag, Berlin Heidelberg, 2012).
34. P. Descouvemont, *Microscopic cluster models. I.*; available online at: http://www.nucleartheory.net/Talent_6_Course/TALENT_lectures/pd_microscopic_1.pdf.
35. A. V. Nesterov, F. Arickx, J. Broeckhove and V. S. Vasilevsky, *Phys. Part. Nucl.* **41**, 716 (2010).
36. A. V. Nesterov, V. S. Vasilevsky and T. P. Kovalenko, *Ukr. J. Phys.* **58**, 628 (2013).
37. O. F. Nemets, V. G. Neudatchin, A. T. Rudchik, Yu. F. Smirnov and Yu. M. Tchuvil'sky, *Nucleon Association in Atomic Nuclei and the Nuclear Reactions of the Many Nucleon Transfers* (in Russian) (Naukova dumka, Kiev, 1988).
38. S. B. Dubovichenko, *Thermonuclear processes of the Universe* (NOVA Sci. Publ., New-York, 2012); available online at: https://www.novapublishers.com/catalog/product_info.php?products_id=31125
39. S. B. Dubovichenko, *Selected methods of nuclear astrophysics*, Third Edition, revised and enlarged (in Russian) (Lambert Acad. Publ. GmbH&Co. KG, Saarbrucken, Germany, 2013); available online at: https://www.lap-publishing.com/catalog/details/store/us/book/978-3-659-34710-8/Избранные-методы-ядерной-астрофизики
40. S. B. Dubovichenko and D. M. Zazulin, *Rus. Phys. J.* **53**, 458 (2010).
41. S. B. Dubovichenko. *Rus. Phys. J.* **55**, 561 (2012).
42. S. B. Dubovichenko. *Rus. Phys. J.* **51**, 1136 (2008).
43. S. B. Dubovichenko. *Phys. Atom. Nucl.* **71**, 65 (2008).
44. S. B. Dubovichenko. *Rus. Phys. J.* **52**, 294 (2009).
45. S. B. Dubovichenko. *Phys. Atom. Nucl.* **75**, 285 (2012).
46. S. B. Dubovichenko. *Rus. Phys. J.* **55**, 992 (2013).
47. F. Nikitiu. *Phase shifts analysis* (Mir, Moscow. 1983) (in Russian).
48. P. E. Hodgson, *The optical model of elastic scattering* (Clarendon Press, Oxford, 1963).
49. C. Angulo, M. Arnould, M. Rayet, P. Descouvemont, D. Baye, C. Leclercq-Willain, A. Coc, S. Barhoumi, P. Aguer, C. Rolfs, R. Kunz, J. W. Hammer, A. Mayer, T. Paradellis, S. Kossionides, C. Chronidou, K. Spyrou, S. Degl'Innocenti, G. Fiorentini, B. Ricci, S. Zavatarelli, C. Providencia, H. Wolters, J. Soares, C. Grama, J. Rahighi, A. Shotter and M. Lamehi Rachti, *Nucl. Phys. A* **656**, 3 (1999).
50. S. B. Dubovichenko and A. V. Dzhazairov-Kakhramanov, In *The Universe Evolution, Astrophysical and Nuclear Aspects*. Edited by I. Strakovsky and L. Blokhintsev (NOVA Sci. Publ., New-York, 2013), pp. 49-108.
51. G. R. Plattner and R. D. Viollier, *Nucl Phys A* **365**, 8 (1981).
52. S. B. Dubovichenko, A. V. Dzhazairov-Kakhramanov, *Phys. Atom. Nucl.* **58**, 579 (1995).
53. S. B. Dubovichenko and A. V. Dzhazairov-Kakhramanov, *Phys. Part. Nucl.* **28**, 615 (1997).

54. I. Ajzenberg and V. Grajner, *Mechanisms of nuclear excitation* (Atomizdat, Moscow, 1973) (in Russian).
55. D. A. Varshalovich, A. N. Moskalev and V. K. Khersonskii, *Quantum Theory of Angular Momentum* (World Scientific, Singapore, 1989).
56. Fundamental Physical Constants 2010, (NIST Ref. on Constants, Units and Uncertainty), proton magnetic moment, http://physics.nist.gov/cgi-bin/cuu/Value?mup#mid
57. M. P. Avotina and A. V. Zolotavin, *Moments of the ground and excited states of nuclei* (Atomizdat, Moscow, 1979).
58. S. B. Dubovichenko, *Phys. Part. Nucl.* **44**, 803 (2013).
59. Fundamental Physical Constants 2010, (NIST Ref. on Constants, Units and Uncertainty), neutron mass, http://physics.nist.gov/cgi-bin/cuu/Value?mnu#mid
60. Fundamental Physical Constants 2010, (NIST Ref. on Constants, Units and Uncertainty), proton mass, http://physics.nist.gov/cgi-bin/cuu/Value?mpu#mid
61. Nuclear Wallet Cards database ($^{10}$B isotope), (2012), http://cdfe.sinp.msu.ru/cgi-bin/gsearch_ru.cgi?z=5&a=10
62. Nuclear Wallet Cards database ($^{11}$B isotope), (2012), http://cdfe.sinp.msu.ru/cgi-bin/gsearch_ru.cgi?z=5&a=11
63. Nuclear Wallet Cards database ($^{14}$C isotope), (2012), http://cdfe.sinp.msu.ru/cgi-bin/gsearch_ru.cgi?z=6&a=14
64. Nuclear Wallet Cards database ($^{15}$N isotope), (2012), http://cdfe.sinp.msu.ru/cgi-bin/gsearch_ru.cgi?z=7&a=15
65. C. Itzykson and M. Nauenberg, *Rev. Mod. Phys.* **38**, 95 (1966).
66. S. B. Dubovichenko, A. V. Dzhazairov-Kakhramanov and N. A. Burkova, *Int. Jour. Mod. Phys. E* **22**, 1350028 (2013).
67. S. B. Dubovichenko and A. V. Dzhazairov-Kakhramanov, *Int. J. Mod. Phys. E* **21**, 1250039 (2012).
68. S. B. Dubovichenko and Yu. N. Uzikov, *Phys. Part. Nucl.* **42**, 251 (2011).
69. V. G. Neudatchin and Yu. F. Smirnov, *Nucleon associations in light nuclei* (Nauka, Moscow, 1969) (in Russian).
70. D. R. Tilley, J. H. Kelley, J. L. Godwin, D. J. Millener, J. E. Purcell, C. G. Sheu and H. R. Weller, *Nucl. Phys. A* **745**, 155 (2004).
71. G. P. Lamaze, R. A. Schrack and O. A. Wasson, *Nucl. Sci. Eng.* **68**, 183 (1978).
72. J. H. Kelley, E. Kwan, J. E. Purcell, C. G. Sheu and H. R. Weller, *Nucl. Phys. A* **880**, 88 (2012).
73. M. Igashira, K. Masuda, S. Mizuno, M. Mizumachi, T. Ohsaki, T. Suzuki, Y. Nagai and H. Kitazawa, Proceedings of the Conference “Measurement, Calculation and Evaluation of Photon Production Data”, (International Atomic Energy Agency, Bologna. 14-17 November 1994) P.269.
74. S. B. Dubovichenko, *Calculation methods of nuclear characteristics* (in Russian) Second Edition, revised and enlarged (Lambert Acad. Publ. GmbH&Co. KG, Saarbrucken, Germany, 2012); available online at: https://www.lap-publishing.com/catalog/details//store/ru/book/978-3-659-21137-9/методы-расчета-ядерных-характеристик
75. E. I. Dolinskii, A. M. Mukhamedzhanov and R. Yarmukhamedov, *Direct nuclear reactions on light nuclei with the emission of neutrons* (FAN, Tashkent, 1978).

76. L. D. Blokhintsev, I. Borbei and E. I. Dolinskii, Fiz. Elem. Chast. At. Yad. **8**, 1189 (1977).
77. R. Yarmukhamedov, *Determination of ANC for n$^{10}$B channel in $^{11}$B nucleus* (2013, private communication).
78. G. A. Bartholomew and P. J. Campion, *Can. J. Phys.* **35**, 1347 (1957).
79. S. F. Mughabghab, *Atlas of Neutron Resonances*, *Resonance Parameters and Thermal Cross Sections, Z = 1-100*, Fifth Edition, (Elsevier, Amsterdam, 2006).
80. R. B. Firestone, M. Krticka, D. P. McNabb, B. Sleaford, U. Agvaanluvsan, T. Belgya and Zs. Revay, *AIP Conf. Proc.* **1005**, 26 (2008).
81. C. J. Lin, H. Q. Zhang, Z. H. Liu, W. Y. Wu, F. Yang and M. Ruan, *Phys. Rev. C.* **68**, 047601 (2003).
82. S. B. Dubovichenko, A. V. Dzhazairov-Kakhramanov and N. A. Burkova, *Jour. Nucl. Part. Phys.* **2(2)** 6 (2012).
83. N. V. Afanasyeva, S. B. Dubovichenko and A. V. Dzhazairov-Kakhramanov, *Jour. of Nucl. Ener. Sci. & Power Generation Technology* 2013, 2:3, http://dx.doi.org/10.4172/2325-9809.1000112; available online at: arXiv: 1212.1765 [nucl-th].
84. Nuclear Reaction Database (EXFOR): http://cdfe.sinp.msu.ru/exfor/index.php
85. F. P. Mooring *et al.*, *Argonne National Laboratory* **6877**, 5 (1964).
86. W. L. Imhof, R. G. Johnson, F. J. Vaughn and M. Walt, *Phys. Rev.* **125**, 1334 (1962).
87. Fundamental Physical Constants 2010, (NIST Ref. on Constants, Units and Uncertainty), proton rms charge radius, http://physics.nist.gov/cgi-bin/cuu/Value?rp#mid
88. T. Huus and R. B. Day, *Phys. Rev.* **91**, 599 (1953).
89. R. E. Segel, S. S. Hanna, R. G. Allas, *Phys. Rev.* **139**, B818 (1965).
90. R. G. Allas, S. S. Hanna, L. Meyer-Schützmeister and R. E. Segel, *Nucl. Phys.* **58**, 122 (1964).
91. S. B. Dubovichenko, A. V. Dzhazairov-Kakhramanov and N. V. Afanasyeva, *Int. Jour. Mod. Phys. E* **22**, 1350075 (2013).
92. S. B. Dubovichenko, *Phys. Atom. Nucl.* **75**, 173 (2012).
93. S. B. Dubovichenko and A. V. Dzhazairov-Kakhramanov, *Euro. Phys. Jour. A* **39**, 139 (2009).
94. S. B. Dubovichenko and A. V. Dzhazairov-Kakhramanov, *Ann. der Phys.* **524**, 850 (2012).
95. S. B. Dubovichenko and A. V. Dzhazairov-Kakhramanov, In *The Big Bang: Theory, Assumptions and Problems*. Edited by J. R. O'Connel and A. L. Hale (NOVA Sci. Publ., New-York, 2012) pp. 1-60.
96. S. B. Dubovichenko, *Thermonuclear processes of the Universe*, Series "Kazakhstan space research" (A-tri, Almaty, 2011) **7**, p. 402; (in Russian) available online at: http://xxx.lanl.gov/abs/1012.0877
97. F. Ajzenberg-Selove, *Nucl. Phys. A.* **523** 1 (1991).
98. J. D. Henderson, E. L. Hudspeth and W. R. Smith, *Phys. Rev.* **172**, 1058 (1968).
99. G. A. Bartholomew, F. Brown, H. E. Gove, A. E. Litherland and E. B. Paul, *Can. J. Phys.* **33**, 441 (1955).
100. G. A. Bartholomew, A. E. Litherland, E. B. Paul and H. E. Gove. *Can. J. Phys.* **34**, 147 (1956).
101. Nuclear Reaction Database (EXFOR): http://cdfe.sinp.msu.ru/cgi-

bin/exfV3.cgi?entry=&ztarg=6&atarg=14&inpart=P&inpart1=&outpart=G&outpart1=&zfin=&afin=&sf=1&sf=2&sf1=&sf1=&sf1=&sf1=&sf1=&min=&max=&code=&year=&author=&institute=&n=50

102. S. B. Dubovichenko, *Rus. Phys. J.* **56**, 494 (2013).
103. F. Ajzenberg-Selove, *Nucl. Phys. A* **565**, 1 (1993).
104. S. B. Dubovichenko, *Rus. Phys. J.* **55**, 138 (2012).
105. S. B. Dubovichenko, A. V. Dzhazairov-Kakhramanov, N. A. Burkova, *Jour. Nucl. Part. Phys.* **3**, 108 (2013).
106. V. G. Neudatchin, A. A. Sakharuk and Yu. F. Smirnov, *Sov. J. Part. Nucl.* **23**, 210 (1992).
107. V. G. Neudatchin, V. I. Kukulin, V. N. Pomerantsev and A. A. Sakharuk, *Phys. Rev. C* **45**, 1512 (1992).
108. A. M. Mukhamedzhanov, M. L. Cognata, and V. Kroha, *Phys. Rev. C* **83**, 044604 (2011).
109. A. M. Mukhamedzhanov, Private communication (2012).
110. S. B. Dubovichenko, *Calculation methods of nuclear characteristics* (Complex, Almaty, 2006); http://arxiv.org/abs/1006.4947
111. A. Caciolli *et al.*, arXiv:1107.4514v1 [astro-ph.SR] 22 Jul 2011.
112. D. Bemmerer *et al.*, *Jour. Phys. G* **36**, 045202 (2009).
113. P. J. LeBlanc *et al.*, *Phys. Rev. C* **82**, 055804 (2010).
114. S. Bashkin, R. R. Carlson, R. A. Douglas, *Phys. Rev.* **114**, 1543 (1958).
115. S. B. Dubovichenko, *Phys. Atom. Nucl.* **76**, 841 (2013).
116. S. B. Dubovichenko, *JETP* **117**, 649 (2013).
117. S. B. Dubovichenko, *JETP* **113**, 221 (2011).